# Analysis of Two-State Folding Using Parabolic Approximation III: Non-Arrhenius Kinetics of FBP28 WW Part-I

**AUTHOR NAME:** Robert S. Sade

**AUTHOR ADDRESS:** Vinkensteynstraat 128, 2562 TV, Den Haag, Netherlands

**AUTHOR EMAIL ADDRESS:** robert.sade@gmail.com

**AUTHOR AFFILIATION:** Independent Researcher

# ABSTRACT

A model which treats the denatured and the native conformers as being confined to harmonic Gibbs energy wells has been used to analyse the non-Arrhenius behaviour of spontaneously-folding fixed two-state systems. The results demonstrate that when pressure and solvent are constant: (*i*) a two-state system is physically defined only for a finite temperature range; (*ii*) irrespective of the primary sequence, the 3-dimensional structure of the native conformer, the residual structure in the denatured state, and the magnitude of the folding and unfolding rate constants, the equilibrium stability of a two-state system is a maximum when its denatured conformers bury the least amount of solvent accessible surface area (SASA) to reach the activated state; (*iii*) the Gibbs barriers to folding and unfolding are not always due to the incomplete compensation of the activation enthalpies and entropies; (*iv*) the difference in heat capacity between the reaction-states is due to both the size of the solvent-shell and the non-covalent interactions; (*v*) the position of the transition state ensemble along the reaction coordinate (RC) depends on the choice of the RC; and (*vi*) the atomic structure of the transiently populated reaction-states cannot be inferred from perturbation-induced changes in their energetics.

# INTRODUCTION

It was shown elsewhere, henceforth referred to as Papers I and II, that the equilibrium and kinetic behaviour of spontaneously-folding fixed two-state systems can be analysed by a treatment that is analogous to that given by Marcus for electron transfer.[1-3] In this framework termed the parabolic approximation, the Gibbs energy functions of the denatured state ensemble (DSE) and the native state ensemble (NSE) are represented by parabolas whose curvature is given by their temperature-invariant force constants, $\alpha$ and $\omega$, respectively. The temperature-invariant mean length of the reaction coordinate (RC) is given by $m_{D-N}$ and is identical to the separation between the vertices of the DSE and the NSE-parabolas along the abscissa. Similarly, the position of the transition state ensemble (TSE) relative to the DSE and the NSE are given by $m_{TS-D(T)}$ and $m_{TS-N(T)}$, respectively, and are identical to the separation between the *curve-crossing* and the vertices of the DSE and the NSE-parabolas, respectively. The Gibbs energy of unfolding at equilibrium, $\Delta G_{D-N(T)}$, is identical to the separation between the vertices of the DSE and the NSE-parabolas along the ordinate. Similarly, the Gibbs activation energy for folding ($\Delta G_{TS-D(T)}$) and unfolding ($\Delta G_{TS-N(T)}$) are identical to the separation between the *curve-crossing* and the vertices of the DSE and the NSE-parabolas along the ordinate, respectively.

The purpose of this article is to use the framework described in Papers I and II to analyse the non-Arrhenius behaviour of the 37-residue FBP28 WW domain, at an unprecedented range and resolution.[4]

# EQUATIONS

The expressions for the position of the TSE relative to the vertices of the DSE and the NSE Gibbs parabolas are given by

$$m_{TS-D(T)} = \frac{\omega m_{D-N} - \sqrt{\lambda\omega + \Delta G_{D-N(T)}(\omega-\alpha)}}{(\omega-\alpha)} = \frac{\omega m_{D-N} - \sqrt{\varphi}}{(\omega-\alpha)} \qquad (1)$$

$$m_{TS-N(T)} = \frac{\sqrt{\lambda\omega + \Delta G_{D-N(T)}(\omega-\alpha)} - \alpha m_{D-N}}{(\omega-\alpha)} = \frac{\sqrt{\varphi} - \alpha m_{D-N}}{(\omega-\alpha)} \qquad (2)$$

where the discriminant $\varphi = \lambda\omega + \Delta G_{D-N(T)}(\omega - \alpha)$, and $\lambda = \alpha(m_{D-N})^2$ is the *Marcus reorganization energy* for two-state protein folding. The expressions for the activation energies for folding and unfolding are given by

$$\Delta G_{TS-D(T)} = \alpha\left(m_{TS-D(T)}\right)^2 = \frac{\alpha\left(\omega m_{D-N} - \sqrt{\varphi}\right)^2}{(\omega - \alpha)^2} = \lambda\beta^2_{T(fold)(T)} \tag{3}$$

$$\Delta G_{TS-N(T)} = \omega\left(m_{TS-N(T)}\right)^2 = \frac{\omega\left(\sqrt{\varphi} - \alpha m_{D-N}\right)^2}{(\omega - \alpha)^2} = \frac{\omega}{\alpha}\lambda\beta^2_{T(unfold)(T)} \tag{4}$$

where the parameters $\beta_{T(fold)(T)}$ $(= m_{TS-D(T)}/m_{D-N})$ and $\beta_{T(unfold)(T)}$ $(= m_{TS-N(T)}/m_{D-N})$ are according to Tanford's framework.[5] The expressions for the rate constants for folding ($k_{f(T)}$) and unfolding ($k_{u(T)}$), and $\Delta G_{D-N(T)}$ are given by

$$k_{f(T)} = k^0 \exp\left(-\frac{\alpha\left(\omega m_{D-N} - \sqrt{\varphi}\right)^2}{RT(\omega - \alpha)^2}\right) = k^0 \exp\left(-\frac{\lambda\beta^2_{T(fold)(T)}}{RT}\right) \tag{5}$$

$$k_{u(T)} = k^0 \exp\left(-\frac{\omega\left(\sqrt{\varphi} - \alpha m_{D-N}\right)^2}{RT(\omega - \alpha)^2}\right) = k^0 \exp\left(-\frac{\omega\lambda\beta^2_{T(unfold)(T)}}{\alpha RT}\right) \tag{6}$$

$$\Delta G_{D-N(T)} = \Delta G_{TS-N(T)} - \Delta G_{TS-D(T)} = \lambda\left(\frac{\omega}{\alpha}\beta^2_{T(unfold)(T)} - \beta^2_{T(fold)(T)}\right) \tag{7}$$

where, $k^0$ is the temperature-invariant prefactor with units identical to those of the rate constants (s$^{-1}$), $R$ is the gas constant, $T$ is the absolute temperature. If the temperature-dependence of $\Delta G_{D-N(T)}$ and the values of $\alpha$, $\omega$, and $m_{D-N}$ are known for any two-state system at constant pressure and solvent conditions (see **Methods**), the temperature-dependence of the *curve-crossing* relative to the ground states may be readily ascertained. The temperature-dependence of *curve-crossing* is central to this analysis since all other parameters can be readily derived by manipulating the same using standard kinetic and thermodynamic relationships.

The activation entropies for folding ($\Delta S_{\text{TS-D}(T)}$) and unfolding ($\Delta S_{\text{TS-N}(T)}$) are given by the first derivatives of $\Delta G_{\text{TS-D}(T)}$ and $\Delta G_{\text{TS-N}(T)}$ functions with respect to temperature

$$\Delta S_{\text{TS-D}(T)} = -\alpha \frac{\partial \left(m_{\text{TS-D}(T)}\right)^2}{\partial T} = -2\alpha\, m_{\text{TS-D}(T)} \frac{\partial m_{\text{TS-D}(T)}}{\partial T} = \frac{\alpha\, m_{\text{TS-D}(T)} \Delta C_{p\text{D-N}}}{\sqrt{\varphi}} \ln\left(\frac{T_S}{T}\right) \quad (8)$$

$$\Delta S_{\text{TS-N}(T)} = -\omega \frac{\partial \left(m_{\text{TS-N}(T)}\right)^2}{\partial T} = -2\omega\, m_{\text{TS-N}(T)} \frac{\partial m_{\text{TS-N}(T)}}{\partial T} = \frac{\omega\, m_{\text{TS-N}(T)} \Delta C_{p\text{D-N}}}{\sqrt{\varphi}} \ln\left(\frac{T}{T_S}\right) \quad (9)$$

where $T_S$ is the temperature at which the entropy of unfolding at equilibrium is zero ($\Delta S_{\text{D-N}(T)} = 0$) and $\Delta C_{p\text{D-N}}$ is the temperature-invariant difference in heat capacity between the DSE and the NSE.[6] The activation enthalpies for folding ($\Delta H_{\text{TS-D}(T)}$) and unfolding ($\Delta H_{\text{TS-N}(T)}$) may be readily obtained by recasting the Gibbs equation: $\Delta H_{(T)} = \Delta G_{(T)} + T\Delta S_{(T)}$, or from the temperature-dependence of $k_{f(T)}$ and $k_{u(T)}$ to give

$$\Delta H_{\text{TS-D}(T)} = \alpha\left(m_{\text{TS-D}(T)}\right)^2 \left(1 + \frac{T\Delta C_{p\text{D-N}}}{m_{\text{TS-D}(T)}\sqrt{\varphi}} \ln\left(\frac{T_S}{T}\right)\right) = RT^2 \left(\frac{\partial \ln k_{f(T)}}{\partial T}\right) \quad (10)$$

$$\Delta H_{\text{TS-N}(T)} = \omega\left(m_{\text{TS-N}(T)}\right)^2 \left(1 + \frac{T\Delta C_{p\text{D-N}}}{m_{\text{TS-N}(T)}\sqrt{\varphi}} \ln\left(\frac{T}{T_S}\right)\right) = RT^2 \left(\frac{\partial \ln k_{u(T)}}{\partial T}\right) \quad (11)$$

The difference in heat capacity between the DSE and the TSE (i.e., for the partial unfolding reaction $[TS] \rightleftharpoons D$) is given by

$$\Delta C_{p\text{D-TS}(T)} = \frac{\alpha}{2\varphi\sqrt{\varphi}}\left[m_{\text{TS-D}(T)} 2\varphi \Delta C_{p\text{D-N}} + \omega m_{\text{D-N}} T \left(\Delta S_{\text{D-N}(T)}\right)^2\right] \quad (12)$$

Similarly, the difference in heat capacity between the TSE and the NSE (for the partial unfolding reaction $N \rightleftharpoons [TS]$) is given by

$$\Delta C_{p\text{TS-N}(T)} = \frac{\omega}{2\varphi\sqrt{\varphi}}\left[m_{\text{TS-N}(T)} 2\varphi \Delta C_{p\text{D-N}} - \alpha m_{\text{D-N}} T \left(\Delta S_{\text{D-N}(T)}\right)^2\right] \quad (13)$$

The reader may refer to Papers I and II for the derivations.

## RESULTS AND DISCUSSION

As mentioned earlier and discussed in sufficient detail in Papers I and II, the analysis we are going to perform has an explicit requirement for a minimal experimental dataset which are: (*i*) an experimental chevron obtained at constant temperature, pressure and solvent conditions (except for the denaturant); (*ii*) an equilibrium thermal denaturation curve obtained under constant pressure, and in solvent conditions identical to those in which the chevron was acquired but without the denaturant, using either calorimetry or spectroscopy; and (*iii*) the calorimetrically determined $\Delta C_{pD-N}$ value (i.e., the slope of the linear regression of a plot of model-independent $\Delta H_{D-N(T_m)}$ *vs* $T_m$, where $\Delta H_{D-N(T_m)}$ is the enthalpy of unfolding at the midpoint of thermal denaturation, $T_m$; see Fig. 4 in Privalov, 1989).[7] Fitting the chevron to a modified chevron-equation using non-linear regression yields the values of $m_{D-N}$, the force constants $\alpha$ and $\omega$, and the prefactor $k^0$ ($k^0$ is assumed to be temperature-invariant; see **Methods** in Paper I). Fitting a spectroscopic sigmoidal equilibrium thermal denaturation curve using standard two-state approximation (van't Hoff analysis using temperature-invariant $\Delta C_{pD-N}$) yields van't Hoff $\Delta H_{D-N(T_m)}$ and $T_m$ and enables the temperature-dependence of $\Delta H_{D-N(T)}$, $\Delta S_{D-N(T)}$ and $\Delta G_{D-N(T)}$ functions to be ascertained across a wide temperature regime (Eqs. (A1)-(A3), **Figure 1** and **Figure 1−figure supplement 1**).[6] Once the values of $m_{D-N}$, the force constants, the prefactor, and the temperature dependence of $\Delta G_{D-N(T)}$ are known, the rest of the analysis is fairly straightforward. The values of all the reference temperatures that appear in this article are given in **Table 1**.

## Temperature-dependence of $m_{TS-D(T)}$ and $m_{TS-N(T)}$

Substituting the expression for the temperature-dependence of $G_{D-N(T)}$ (Eq. (A3), **Figure 1**) in Eqs. (1) and (2) enables the temperature-dependence of the *curve-crossing* relative to the DSE and the NSE to be ascertained (**Figure 2**; substituted expressions not shown). Because by postulate the force constants, $\Delta C_{pD-N}$, and $m_{D-N}$ are temperature-invariant for any given primary sequence that folds in a two-state manner at constant pressure and solvent conditions, we get from inspection of Eqs. (1) and (2) that the discriminant $\varphi$, and $\sqrt{\varphi}$ must be a maximum when $\Delta G_{D-N(T)}$ is a maximum. Because $\Delta G_{D-N(T)}$ is a maximum at $T_S$ (the temperature at which the entropy of unfolding at equilibrium, $\Delta S_{D-N(T)}$, is zero),[6] a corollary is that φ and $\sqrt{\varphi}$ must be a maximum at $T_S$; and any deviation in the temperature from $T_S$ will only lead to their decrease. Consequently, $m_{TS\text{-}D(T)}$ and $\beta_{T(fold)(T)}\,(=m_{TS\text{-}D(T)}/m_{D\text{-}N})$ are always a minimum, and $m_{TS\text{-}N(T)}$ and $\beta_{T(unfold)(T)}\,(=m_{TS\text{-}N(T)}/m_{D\text{-}N})$ are always a maximum at $T_S$. This gives rise to two further corollaries: Any deviation in the temperature from $T_S$ can only lead to: (*i*) an increase in $m_{TS\text{-}D(T)}$ and $\beta_{T(fold)(T)}$; and (*ii*) a decrease in $m_{TS\text{-}N(T)}$ and $\beta_{T(unfold)(T)}$ (**Figure 2** and **Figure 2−figure supplement 1**). In other words, when $T = T_S$, the TSE is the least native-like in terms of the SASA (solvent accessible surface area), and any deviation in temperature causes the TSE to become more native-like. A further consequence of $m_{TS\text{-}D(T)}$ being a minimum at $T_S$ is that if for a two-state-folding primary sequence there exists a chevron with a well-defined linear folding arm at $T_S$, then $m_{TS\text{-}D(T)} > 0$ and $\beta_{T(fold)(T)} > 0$ for all temperatures (**Figure 2A** and **Figure 2−figure supplement 1A**). Since the *curve-crossing* is physically undefined for φ < 0 owing to there being no real roots, the maximum theoretically possible value of $m_{TS\text{-}D(T)}$ will occur when φ = 0 and is given by: $m_{TS\text{-}D(T)}\big|^{max}_{T=T_\alpha,T_\omega} = \omega m_{D\text{-}N}/(\omega-\alpha) > m_{D\text{-}N}$ where $T_\alpha$ and $T_\omega$ are the temperature limits such that for $T < T_\alpha$ and $T > T_\omega$, a two-state system is not physically defined (see Paper II). Because $m_{D\text{-}N} = m_{TS\text{-}D(T)} + m_{TS\text{-}N(T)}$ for a two-state system, and $m_{D\text{-}N}$ is temperature-invariant by postulate, the theoretical minimum of $m_{TS\text{-}N(T)}$ is given by: $m_{TS\text{-}N(T)}\big|^{min}_{T=T_\alpha,T_\omega} = -\alpha m_{D\text{-}N}/(\omega-\alpha) < 0$. Now, since $m_{TS\text{-}N(T)}$ is a maximum and positive at $T_S$ but its minimum is negative, a consequence is that $m_{TS\text{-}N(T)} = \beta_{T(unfold)(T)} = 0$ at two unique temperatures, one in the ultralow ($T_{S(\alpha)}$) and the other in the high ($T_{S(\omega)}$) temperature regime, and negative for $T_\alpha \leq T < T_{S(\alpha)}$ and $T_{S(\omega)} < T \leq T_\omega$ (**Figure 2B** and **Figure 2−figure supplement 1B**). Obviously, $m_{TS\text{-}D(T)} = m_{D\text{-}N}$ and $\beta_{T(fold)(T)}$ is unity at $T_{S(\alpha)}$ and $T_{S(\omega)}$. To summarize, unlike $m_{TS\text{-}D(T)}$ and $\beta_{T(fold)(T)}$ which are positive for all temperatures and a minimum at $T_S$, $m_{TS\text{-}N(T)}$ and $\beta_{T(unfold)(T)}$ are a maximum at $T_S$, zero at $T_{S(\alpha)}$ and $T_{S(\omega)}$, and negative for $T_\alpha \leq T < T_{S(\alpha)}$ and $T_{S(\omega)} < T \leq T_\omega$.

The predicted Leffler-Hammond shift, which must be valid for any two-state system, is in agreement with the experimental data on the temperature-dependent behaviour of other two-state systems (Table 1 in Dimitriadis et al., 2004; Table 1 in Taskent et al., 2008; Fig. 5C in Otzen and Oliveberg, 2004),[8-12] with the rate at which the *curve-crossing* shifts with stability (relative to the vertex of the DSE-parabola) being given by $\partial m_{TS-D(T)} / \partial \Delta G_{D-N(T)} = -1/2\sqrt{\varphi}$. Importantly, just as the Leffler-Hammond movement is rationalized in physical organic chemistry using Marcus theory,[13-15] we can similarly rationalize these effects in protein folding using parabolic approximation (**Figures 3**, **4**, and **Figure 4−figure supplement 1**). When $T = T_S$, $\Delta G_{TS-D(T)}$ is a minimum, and $\Delta G_{D-N(T)}$ and $\Delta G_{TS-N(T)}$ are both a maximum; and any increase or decrease in the temperature relative to $T_S$ leads to a decrease in $\Delta G_{TS-N(T)}$, and an increase in $\Delta G_{TS-D(T)}$, consequently, leading to a decrease in $\Delta G_{D-N(T)}$ (**Figures 1**, **3B** and **5**). Naturally at $T_c$ and $T_m$, $\Delta G_{TS-D(T)} = \Delta G_{TS-N(T)}$, $k_{f(T)} = k_{u(T)}$, and $\Delta G_{D-N(T)} = 0$ (**Figure 3C**). The reason why $m_{TS-D(T)} = m_{D-N}$, and $m_{TS-N(T)} = 0$ at $T_{S(\alpha)}$ and $T_{S(\omega)}$ is apparent from **Figures 4A**, **4C** and **Figure 4−figure supplement 1A**: The right arm of the DSE-parabola intersects the vertex of the NSE-parabola leading to $\Delta G_{TS-D(T)} = \alpha(m_{TS-D(T)})^2 = \alpha(m_{D-N})^2 = \lambda$, $\Delta G_{TS-N(T)} = \omega(m_{TS-N(T)})^2 = 0$, and $\Delta G_{D-N(T)} = -\lambda$. Importantly, in contrast to unfolding which can become barrierless at $T_{S(\alpha)}$ and $T_{S(\omega)}$, folding is barrier-limited at all temperatures, with the absolute minimum of $\Delta G_{TS-D(T)}$ occurring at $T_S$; and any deviation in the temperature from $T_S$ will only lead to an increase in $\Delta G_{TS-D(T)}$ (**Figure 5A**). Thus, a corollary is that if folding is barrier-limited at $T_S$ (i.e., the chevron has a well-defined linear folding arm with a finite slope at $T_S$), then a protein that folds *via* two-state mechanism can never spontaneously (i.e., unaided by ligands, co-solvents etc.) switch to a downhill mechanism (Type 0 scenario according to the Energy Landscape Theory; see Fig. 6 in Onuchic et al., 1997), no matter what the temperature, and irrespective of how fast or slow it folds. Although unfolding is barrierless at $T_{S(\alpha)}$ and $T_{S(\omega)}$, it is once again barrier-limited for $T_\alpha \leq T < T_{S(\alpha)}$ and $T_{S(\omega)} < T \leq T_\omega$, with the *curve-crossing* occurring to the right of the vertex of the NSE-parabola (**Figures 4A**, **4B**, **Figure 4−figure supplement 1B** and **5B**), such that $m_{TS-D(T)} > m_{D-N}$, $m_{TS-N(T)} < 0$, $\beta_{T(fold)(T)} > 1$ and $\beta_{T(unfold)(T)} < 0$ (**Figure 2** and **Figure 2−figure supplement 1**).

To summarize, for any two-state folder, unfolding is *conventional barrier-limited* for $T_{S(\alpha)} < T < T_{S(\omega)}$ and the position of the TSE or the *curve-crossing* occurs in between the vertices of the DSE and the NSE parabolas. As the temperature deviates from $T_S$, the SASA of the TSE becomes progressively native-like, with a concomitant increase and a decrease in $\Delta G_{TS-D(T)}$ and $\Delta G_{TS-N(T)}$, respectively. When $T = T_{S(\alpha)}$ and $T_{S(\omega)}$, the *curve-crossing* occurs precisely at the vertex of the NSE-parabola, the SASA of the TSE is identical to that of the NSE, and unfolding is barrierless; and for $T_\alpha \leq T < T_{S(\alpha)}$ and $T_{S(\omega)} < T \leq T_\omega$, unfolding is once again barrier-limited but falls under the *Marcus-inverted-regime* with the *curve-crossing* occurring on the right-arm of the NSE-parabola, leading to the SASA of the NSE being greater than that of the TSE (i.e., the TSE is more compact than the NSE). Importantly, for $T < T_\alpha$ and $T > T_\omega$, the TSE cannot be physically defined owing to $\sqrt{\varphi}$ being mathematically undefined for $\varphi < 0$. A consequence is that $k_{f(T)}$ and $k_{u(T)}$ become physically undefined, leading to $\Delta G_{\text{D-N}(T)} = RT \ln\left(k_{f(T)}/k_{u(T)}\right)$ being physically undefined, such that all of the conformers will be confined to a single Gibbs energy well, which is the DSE, and the protein will cease to function.[16] Thus, from the view point of the physics of phase transitions, $T_\alpha \leq T \leq T_\omega$ denotes the *coexistence temperature-range* where the DSE and the NSE, which are in a dynamic equilibrium, will coexist as two distinct phases; and for $T < T_\alpha$ and $T > T_\omega$ there will be a single phase, which is the DSE, with $T_\alpha$ and $T_\omega$ being the limiting temperatures for coexistence, or phase boundary temperatures from the view point of the DSE.[17-23] This is roughly analogous to the operating temperature range of a logic circuit such as a microprocessor; and just as this range is a function of its constituent material, the physically definable temperature range of a two-state system is a function of the primary sequence when pressure and solvent are constant, and importantly, can be modulated by a variety of *cis*-acting and *trans*-acting factors (see Paper-I). The limit of equilibrium stability below which a two-state system becomes physically undefined is given by: $\Delta G_{\text{D-N}(T)}\big|_{T=T_\alpha,T_\omega} = -\lambda\omega/(\omega-\alpha)$. Consequently, the physically meaningful range of equilibrium stability for a two-state system is given by: $\Delta G_{\text{D-N}(T_S)} + \left[\lambda\omega/(\omega-\alpha)\right]$, where $\Delta G_{\text{D-N}(T_S)}$ is the stability at $T_S$ and is apparent from inspection of **Figure 5−figure supplement 1**. This is akin to the stability range over which Marcus theory is physically realistic (see Kresge, 1973, page 494).[24]

Because by postulate $m_{\text{D-N}}$, $m_{\text{TS-D}(T)}$ and $m_{\text{TS-N}(T)}$ are true proxies for $\Delta\text{SASA}_{\text{D-N}}$, $\Delta\text{SASA}_{\text{D-TS}(T)}$ and $\Delta\text{SASA}_{\text{TS-N}(T)}$, respectively (see Paper I), we have three fundamentally important corollaries that must hold for all two-state systems at constant pressure and solvent conditions: (*i*) the Gibbs barrier to folding is the least when the denatured conformers bury the least amount of SASA to reach the TSE (**Figure 5−figure supplement 2A**); (*ii*) the Gibbs barrier to unfolding is the greatest when the native conformers expose the greatest amount of SASA to reach the TSE (**Figure 5−figure supplement 2B**); and (*iii*) equilibrium stability is the greatest when the conformers in the DSE are displaced the least from the mean of their ensemble along the SASA-RC to reach the TSE (the *principle of least displacement*; **Figure 5−figure supplement 1**).

## Temperature-dependence of the folding, unfolding, and the observed rate constants

Inspection of **Figures 6A** and **Figure 6−figure supplement 1A** demonstrates that Eq. (5) makes a remarkable prediction that $k_{f(T)}$ has a non-linear dependence on temperature. Starting from the lowest temperature ($T_\alpha$) at which a two-state system is physically defined, $k_{f(T)}$ initially increases with an increase in the temperature and reaches a maximal value at $T = T_{H(TS-D)}$ where $\partial \ln k_{f(T)}/\partial T = \Delta H_{TS-D(T)}/RT^2 = 0$; and any further increase in temperature beyond this point will cause a decrease in $k_{f(T)}$ until the temperature $T_\omega$ is reached, such that for $T > T_\omega$, $k_{f(T)}$ is undefined. Inspection of **Figures 6B** and **Figure 6−figure supplement 1B** demonstrates that the temperature-dependence of $k_{u(T)}$ is far more complex: Starting from $T_\alpha$, $k_{u(T)}$ increases with temperature for the regime $T_\alpha \leq T < T_{S(\alpha)}$ (the low-temperature *Marcus-inverted-regime*), reaches a maximum when $T = T_{S(\alpha)}$ ($k_{u(T)} = k^0$; the first extremum of $k_{u(T)}$), and decreases with further rise in temperature for the regime $T_{S(\alpha)} < T < T_{H(TS-N)}$ such that when $T = T_{H(TS-N)}$, $k_{u(T)}$ is a minimum (the second extremum of $k_{u(T)}$). And for $T_{H(TS-N)} < T < T_{S(\omega)}$, an increase in temperature will lead to an increase in $k_{u(T)}$, eventually leading to its saturation at $T = T_{S(\omega)}$ ($k_{u(T)} = k^0$; the third extremum of $k_{u(T)}$), and decreases with further rise in temperature for $T_{S(\omega)} < T \leq T_\omega$ (the high-temperature *Marcus-inverted-regime*). Thus, in contrast to $k_{f(T)}$ which has only one extremum, $k_{u(T)}$ is characterised by three extrema where $\partial \ln k_{u(T)}/\partial T = \Delta H_{TS-N(T)}/RT^2 = 0$, and may be rationalized from the temperature-dependence of $m_{TS-D(T)}$ and $m_{TS-N(T)}$, the Gibbs barrier heights for folding and unfolding, and the intersection of the DSE and the NSE Gibbs parabolas (**Figures 2-5** and their figure supplements). We will show in subsequent publications that the inverted behaviour at very low and high temperatures is not common to all fixed two-state systems and depends on the mean and variance of the Gaussian distribution of the SASA of the conformers in the DSE and the NSE.

Since the ultimate test of any hypothesis is experiment, the most important question now is how well do the calculated rate constants compare with experiment? Although Nguyen et al. have investigated the non-Arrhenius behaviour of the FBP28 WW, they find that the behaviour of its wild type is erratic, with its folding being three-state for $T < T_m$ and two-state for $T > T_m$ (Fig. 3A in Nguyen et al., 2003). Consequently, non-Arrhenius data for the wild type FBP28 WW are lacking. Incidentally, this atypical behaviour is probably artefactual since the protein aggregates and forms fibrils under the experimental conditions in which the measurements were made (see Figs. 2, 3 and 6 in Ferguson et al., 2003).[25,26] Nevertheless, data for ΔNΔC Y11R W30F, a variant of FBP28 WW are available between ~ 298 and ~357 K (Fig. 4A in Nguyen et al., 2003). Now since the relaxation time constants for the fast phase of wild type FBP28 WW (~ 30 μs at 39.5 °C and < 15 μs at 65 °C, page 3950, Fig. 3A, Nguyen et al., 2003) are very similar to those of ΔNΔC Y11R W30F (~ 28 μs at 40 °C and 11 μs at 65 °C, page 3952), a reasonable approximation is that the temperature-dependence of $k_{f(T)}$ and $k_{u(T)}$ of the wild type and the mutant must be similar. Consequently, the temperature-dependence of the rate constants for the wild type FBP28 WW calculated using parabolic approximation must be very similar to the data for ΔNΔC Y11R W30F reported by Nguyen et al. The remarkable agreement between the said datasets is readily apparent from a comparison of Fig. 4A of Nguyen et al., and **Figure 6−figure supplement 2**, and serves an important test of the hypothesis.

Since the temperature-dependence of $k_{f(T)}$ and $k_{u(T)}$ across a wide temperature range is known, the variation in the observed rate constant ($k_{obs(T)}$) with temperature may be readily ascertained using (see **Appendix**)

$$\ln k_{obs(T)} = \ln \left[ k^0 \exp\left( -\frac{\alpha(\omega m_{D-N} - \sqrt{\varphi})^2}{RT(\omega-\alpha)^2} \right) + k^0 \exp\left( -\frac{\omega(\sqrt{\varphi} - \alpha m_{D-N})^2}{RT(\omega-\alpha)^2} \right) \right] \qquad (14)$$

Inspection of **Figure 7** demonstrates that $\ln(k_{obs(T)})$ vs temperature is a smooth 'W-shaped' curve, with $k_{obs(T)}$ being dominated by $k_{f(T)}$ around $T_{H(TS-N)}$, and by $k_{u(T)}$ for $T < T_c$ and $T > T_m$, which is precisely why the kinks in $\ln(k_{obs(T)})$ occur around these temperatures. It is easy to see that at $T_c$ or $T_m$, $k_{f(T)} = k_{u(T)} \Rightarrow k_{obs(T)} = 2k_{f(T)} = 2k_{u(T)}$, $\Delta G_{D-N(T)} = RT \ln(k_{f(T)}/k_{u(T)}) = 0$ or $\Delta G_{TS-D(T)} = \Delta G_{TS-N(T)}$ (**Figures 3C** and **Figure 7−figure supplement 1**). In other words, for a two-state system, $T_c$ and $T_m$ determined at equilibrium must be identical to the temperatures at which $k_{f(T)}$ and $k_{u(T)}$ intersect. This is a consequence of the *principle of microscopic reversibility*, i.e., the equilibrium and kinetic stabilities must be identical for a two-state system at all temperatures.[27] It is precisely for this reason that the value of the prefactor in the Arrhenius expressions for the rate constants must be identical for both the folding and the unfolding reactions at all temperatures (Eqs. (5) and (6)). The steep increase in $k_{obs(T)}$ for $T < T_c$ and $T > T_m$ is due to the $\Delta G_{TS-N(T)}$ approaching zero as described earlier. The argument that the shapes of the curves must be conserved across two-state systems applies not only to the temperature-dependence of $m_{TS-D(T)}$, $m_{TS-N(T)}$, $\Delta G_{TS-D(T)}$ and $\Delta G_{TS-N(T)}$ described so far, but to the rest of the state functions that will be described in this article (see Paper-I).

An important conclusion that we may draw from these data is the following: Because we have assumed a temperature-invariant prefactor and yet find that the kinetics are non-Arrhenius, it essentially implies that one does not need to invoke a *super-Arrhenius temperature-dependence of the configurational diffusion constant* to explain the non-Arrhenius behaviour of proteins.[28-32] Instead, as long as the enthalpies and the entropies of unfolding/folding at equilibrium display a large variation with temperature, and equilibrium stability is a non-linear function of temperature, both $k_{f(T)}$ and $k_{u(T)}$ will have a non-linear dependence on temperature. This leads to two corollaries: (*i*) since the large variation in equilibrium enthalpies and entropies of unfolding, including the pronounced curvature in $\Delta G_{D-N(T)}$ of proteins with temperature is due to the large and positive $\Delta C_{pD-N}$, "*non-Arrhenius kinetics can be particularly acute for reactions that are accompanied by large changes in the heat capacity*"; and (*ii*) because the change in heat capacity upon unfolding is, to a first approximation, proportional to the change in SASA that accompanies it, and since the change in SASA upon unfolding/folding increases with chain-length,[33,34] "*non-Arrhenius kinetics, in general, can be particularly pronounced for large proteins, as compared to very small proteins and peptides*."

**Temperature-dependence of activation enthalpies**

Inspection of **Figure 8** demonstrates that for the partial folding reaction $D \rightleftharpoons [TS]$: (*i*) $\Delta H_{TS-D(T)} > 0$ for $T_\alpha \leq T < T_{H(TS-D)}$; (*ii*) $\Delta H_{TS-D(T)} < 0$ for $T_{H(TS-D)} < T \leq T_\omega$; and (*iii*) $\Delta H_{TS-D(T)} = 0$ for $T = T_{H(TS-D)}$. Thus, the activation of the denatured conformers to the TSE is enthalpically: (*i*) unfavourable for $T_\alpha \leq T < T_{H(\text{TS-D})}$; (*ii*) favourable for $T_{H(\text{TS-D})} < T \leq T_\omega$; and (*iii*) neutral when $T = T_{H(\text{TS-D})}$. Consequently, at $T_{H(\text{TS-D})}$, $\Delta G_{\text{TS-D}(T)}$ is purely due to the difference in entropy between the DSE and the TSE ($\Delta G_{\text{TS-D}(T)} = -T\Delta S_{\text{TS-D}(T)}$) with $k_{f(T)}$ being given by

$$k_{f(T)}\Big|_{T=T_{H(\text{TS-D})}} = k^0 \exp\left(\frac{\Delta S_{\text{TS-D}(T)}}{R}\right)\Bigg|_{T=T_{H(\text{TS-D})}} = k^0 \exp\left(\frac{\alpha\, m_{\text{TS-D}(T)}\Delta C_{p\text{D-N}}}{R\sqrt{\varphi}}\ln\left(\frac{T_S}{T}\right)\right)\Bigg|_{T=T_{H(\text{TS-D})}} \quad (15)$$

Because $k_{f(T)}$ is a maximum at $T_{H(\text{TS-D})}$ ($\partial \ln k_{f(T)}/\partial T = 0$), a corollary is that *"for a two-state folder at constant pressure and solvent conditions, if the prefactor is temperature-invariant, then $k_{f(T)}$ will be a maximum when the Gibbs barrier to folding is purely entropic."* This statement is valid only if the prefactor is temperature-invariant. Now since $\Delta G_{\text{TS-D}(T)} > 0$ for all temperatures (**Figure 5A** and **Table 1**), it is imperative that $\Delta S_{\text{TS-D}(T)} < 0$ at $T_{H(\text{TS-D})}$ (see activation entropy for folding).

Unlike the $\Delta H_{\text{TS-D}(T)}$ function which changes its algebraic sign only once across the entire temperature range over which a two-state system is physically defined, the behaviour of $\Delta H_{\text{TS-N}(T)}$ function is far more complex (**Figure 9**): (*i*) $\Delta H_{\text{TS-N}(T)} > 0$ for $T_\alpha \leq T < T_{S(\alpha)}$ and $T_{H(\text{TS-N})} < T < T_{S(\omega)}$; (*ii*) $\Delta H_{\text{TS-N}(T)} < 0$ for $T_{S(\alpha)} < T < T_{H(\text{TS-N})}$ and $T_{S(\omega)} < T \leq T_\omega$; and (*iii*) $\Delta H_{\text{TS-N}(T)} = 0$ at $T_{S(\alpha)}$, $T_{H(\text{TS-N})}$, and $T_{S(\omega)}$. Consequently, we may state that the activation of native conformers to the TSE is enthalpically: (*i*) unfavourable for $T_\alpha \leq T < T_{S(\alpha)}$ and $T_{H(\text{TS-N})} < T < T_{S(\omega)}$; (*ii*) favourable for $T_{S(\alpha)} < T < T_{H(\text{TS-N})}$ and $T_{S(\omega)} < T \leq T_\omega$; and (*iii*) neutral at $T_{S(\alpha)}$, $T_{H(\text{TS-N})}$, and $T_{S(\omega)}$. If we reverse the reaction-direction, the algebraic signs invert leading to a change in the interpretation. Thus, for the partial folding reaction $[TS] \rightleftharpoons N$, the flux of the conformers from the TSE to the NSE is enthalpically: (*i*) favourable for $T_\alpha \leq T < T_{S(\alpha)}$ and $T_{H(\text{TS-N})} < T < T_{S(\omega)}$ ($\Delta H_{\text{N-TS}(T)} < 0$); (*ii*) unfavourable for $T_{S(\alpha)} < T < T_{H(\text{TS-N})}$ and $T_{S(\omega)} < T \leq T_\omega$ ($\Delta H_{\text{N-TS}(T)} > 0$); and (*iii*) neither favourable nor unfavourable at $T_{S(\alpha)}$, $T_{H(\text{TS-N})}$, and $T_{S(\omega)}$ (**Figure 9−figure supplement 1A**). Note that the term "flux" implies "diffusion of the conformers from one reaction state to the other on the Gibbs energy surface," and as such is an "operational definition."

Importantly, although $\partial \ln k_{u(T)}/\partial T = 0 \Rightarrow \Delta H_{\text{TS-N}(T)} = 0$ at $T_{S(\alpha)}$, $T_{H(\text{TS-N})}$, and $T_{S(\omega)}$, the behaviour of the system at $T_{S(\alpha)}$ and $T_{S(\omega)}$ is distinctly different from that at $T_{H(\text{TS-N})}$: While

$m_{TS-N(T)} = \Delta G_{TS-N(T)} = \Delta H_{TS-N(T)} = \Delta S_{TS-N(T)} = 0$, $m_{TS-D(T)} = m_{D-N}$, $\Delta G_{TS-D(T)} = \Delta G_{N-D(T)} = \lambda$, and $k_{u(T)} = k^0$ at $T_{S(\alpha)}$ and $T_{S(\omega)}$ (note that if both $\Delta G_{TS-N(T)}$ and $\Delta H_{TS-N(T)}$ are zero, then $\Delta S_{TS-N(T)}$ must also be zero, see activation entropies), $k_{u(T)}$ is a minimum ($k_{u(T)} \ll k^0$) with the Gibbs barrier to unfolding being purely entropic ($\Delta G_{TS-N(T)} = -T\Delta S_{TS-N(T)}$) at $T_{H(TS-N)}$. Consequently, we may write

$$k_{u(T)}\Big|_{T=T_{H(TS-N)}} = k^0 \exp\left(\frac{\Delta S_{TS-N(T)}}{R}\right)\Big|_{T=T_{H(TS-N)}} = k^0 \exp\left(\frac{\omega m_{TS-N(T)}\Delta C_{pD-N}}{R\sqrt{\varphi}}\ln\left(\frac{T}{T_S}\right)\right)\Big|_{T=T_{H(TS-N)}} \quad (16)$$

Thus, a corollary is that *"for two-state system at constant pressure and solvent conditions, if the prefactor is temperature-invariant, then $k_{u(T)}$ will be a minimum when the Gibbs barrier to unfolding is purely entropic."* Since $\Delta G_{TS-N(T)} > 0$ at $T_{H(TS-N)}$ (**Figure 5B** and **Table 1**), it is imperative that $\Delta S_{TS-N(T)}$ be negative at $T_{H(TS-N)}$ (see activation entropy for unfolding).

The criteria for two-state folding from the viewpoint of enthalpy are the following: (*i*) the condition that $\Delta H_{D-N(T)} = \Delta H_{TS-N(T)} - \Delta H_{TS-D(T)}$ must be satisfied at all temperatures; (*ii*) the intersection of $\Delta H_{TS-D(T)}$ and $\Delta H_{TS-N(T)}$ functions calculated directly from the temperature-dependence of the experimentally determined $k_{f(T)}$ and $k_{u(T)}$, respectively, must be identical to the independently estimated $T_H$ from equilibrium thermal denaturation experiments; and (*iii*) the condition that $T_{H(TS-N)} < T_H < T_S < T_{H(TS-D)}$ must be satisfied. A corollary of the last statement is that both $\Delta H_{TS-D(T)}$ and $\Delta H_{TS-N(T)}$ functions must be positive at the point of intersection. These aspects are readily apparent from **Figure 9−figure supplement 1B** and **Figure 9−figure supplement 2**.

## Temperature-dependence of activation entropies

Inspection of **Figure 10** shows that for the partial folding reaction $D \rightleftharpoons [TS]$, $\Delta S_{TS-D(T)}$ which is positive at low temperature, decreases in magnitude with an increase in temperature and becomes zero at $T_S$, where the SASA of the TSE is the least native-like, $\Delta G_{TS-D(T)}$ is a minimum ($\partial \Delta G_{TS-D(T)}/\partial T = -\Delta S_{TS-D(T)} = 0$) and $\Delta G_{D-N(T)}$ is a maximum ($\partial \Delta G_{D-N(T)}/\partial T = -\Delta S_{D-N(T)} = 0$; **Figures 1, 2, 5A, Figure 10−figure supplements 1** and **2**); and any further increase in temperature beyond this point causes $\Delta S_{TS-D(T)}$ to become negative. Thus, the activation of denatured conformers to the TSE is entropically: (*i*) favourable for $T_\alpha \leq T < T_S$; (*ii*) unfavourable for $T_S < T \leq T_\omega$; and (*iii*) neutral when $T = T_S$. At $T_S$ the Gibbs barrier to folding is purely due to the difference in enthalpy between the DSE and the TSE with $k_{f(T)}$ being given by

$$k_{f(T)}\Big|_{T=T_S} = k^0 \exp\left(-\frac{\Delta G_{\text{TS-D}(T)}}{RT}\right)\Big|_{T=T_S} = k^0 \exp\left(-\frac{\Delta H_{\text{TS-D}(T)}}{RT}\right)\Big|_{T=T_S} \tag{17}$$

Inspection of **Figure 11** demonstrates that the behaviour of the $\Delta S_{\text{TS-N}(T)}$ function is far more complex than the $\Delta S_{\text{TS-D}(T)}$ function: (*i*) $\Delta S_{\text{TS-N}(T)} > 0$ for $T_\alpha \leq T < T_{S(\alpha)}$ and $T_S < T < T_{S(\omega)}$; (*ii*) $\Delta S_{\text{TS-N}(T)} < 0$ for $T_{S(\alpha)} < T < T_S$ and $T_{S(\omega)} < T \leq T_\omega$; and (*iii*) $\Delta S_{\text{TS-N}(T)} = 0$ at $T_{S(\alpha)}$, $T_S$, and $T_{S(\omega)}$. Consequently, we may state that the activation of native conformers to the TSE is entropically: (*i*) favourable for $T_\alpha \leq T < T_{S(\alpha)}$ and $T_S < T < T_{S(\omega)}$; (*ii*) unfavourable for $T_{S(\alpha)} < T < T_S$ and $T_{S(\omega)} < T \leq T_\omega$; and (*iii*) neutral at $T_{S(\alpha)}$, $T_S$, and $T_{S(\omega)}$. If we reverse the reaction-direction (**Figure 11−figure supplement 1A**), the algebraic signs invert leading to a change in the interpretation. Consequently, we may state that for the partial folding reaction $[TS] \rightleftharpoons N$, the flux of the conformers from the TSE to the NSE is entropically: (*i*) unfavourable for $T_\alpha \leq T < T_{S(\alpha)}$ and $T_S < T < T_{S(\omega)}$ ($\Delta S_{\text{N-TS}(T)} < 0$); (*ii*) favourable for $T_{S(\alpha)} < T < T_S$ and $T_{S(\omega)} < T \leq T_\omega$ ($\Delta S_{\text{N-TS}(T)} > 0$); and (*iii*) neutral at $T_{S(\alpha)}$, $T_S$, and $T_{S(\omega)}$.

At $T = T_S$, the Gibbs barrier to unfolding is purely due to the difference in enthalpy between the TSE and the NSE ($\Delta G_{\text{TS-N}(T)} = \Delta H_{\text{TS-N}(T)}$) with $k_{u(T)}$ being given by

$$k_{u(T)}\Big|_{T=T_S} = k^0 \exp\left(-\frac{\Delta G_{\text{TS-N}(T)}}{RT}\right)\Big|_{T=T_S} = k^0 \exp\left(-\frac{\Delta H_{\text{TS-N}(T)}}{RT}\right)\Big|_{T=T_S} \tag{18}$$

Although $\Delta S_{\text{TS-N}(T)} = 0 \Rightarrow S_{\text{TS}(T)} = S_{\text{N}(T)}$ at $T_{S(\alpha)}$, $T_S$, and $T_{S(\omega)}$, the underlying thermodynamics is fundamentally different at $T_S$ as compared to $T_{S(\alpha)}$ and $T_{S(\omega)}$. While both $\Delta G_{\text{TS-N}(T)}$ and $m_{\text{TS-N}(T)}$ are positive and a maximum, and $\Delta G_{\text{TS-N}(T)}$ is purely enthalpic at $T_S$ ($\Delta G_{\text{TS-N}(T)} = \Delta H_{\text{TS-N}(T)}$), at $T_{S(\alpha)}$ and $T_{S(\omega)}$ we have $m_{\text{TS-N}(T)} = 0 \Rightarrow \Delta G_{\text{TS-N}(T)} = \omega(m_{\text{TS-N}(T)})^2 = 0 \Rightarrow \Delta H_{\text{TS-N}(T)} = 0$, and $\Delta G_{\text{N-D}(T)} = \Delta G_{\text{TS-D}(T)} = \lambda$; and because $\Delta G_{\text{TS-N}(T)} = 0$ at $T_{S(\alpha)}$ and $T_{S(\omega)}$, the rate constant for unfolding will reach an absolute maximum for that particular solvent and pressure at these two temperatures. To summarize, while at $T_S$ we have $G_{TS(T)} \gg G_{N(T)}$, $S_{D(T)} = S_{TS(T)} = S_{N(T)}$, and $k_{u(T)} \ll k^0$, when $T = T_{S(\alpha)}$ and $T_{S(\omega)}$, we have $G_{TS(T)} = G_{N(T)}$, $H_{TS(T)} = H_{N(T)}$, $S_{TS(T)} = S_{N(T)}$, and $k_{u(T)} = k^0$ (**Figure 11−figure supplements 2** and **3**). Thus, a fundamentally important conclusion that we may draw from these relationships is that "*if two reaction-states on the folding pathway of a two-state system have identical SASA and Gibbs energy under identical environmental conditions, then their absolute enthalpies and entropies must be identical.*" This must hold irrespective of whether or not the two reaction-states have identical, similar or dissimilar structures. We will revisit this scenario when we discuss the heat capacities of activation and the inapplicability of the Hammond postulate to protein folding reactions.

The criteria for two-state folding from the viewpoint of entropy are the following: (*i*) the condition that $\Delta S_{D-N(T)} = \Delta S_{TS-N(T)} - \Delta S_{TS-D(T)}$ must be satisfied at all temperatures; (*ii*) the intersection of $\Delta S_{TS-D(T)}$ and $\Delta S_{TS-N(T)}$ functions calculated directly from the slopes of the temperature-dependent shift in the *curve-crossing* relative to the DSE and the NSE, respectively, must be identical to the independently estimated $T_S$ from equilibrium thermal denaturation experiments (**Figure 11−figure supplements 1B**, **4** and **5**); and (*iii*) both $\Delta S_{TS-D(T)}$ and $\Delta S_{TS-N(T)}$ functions must independently be equal to zero at $T_S$.

**Temperature-dependence of the Gibbs activation energies**

Although the general features of the temperature-dependence of $\Delta G_{TS-D(T)}$ and $\Delta G_{TS-N(T)}$ were described earlier (**Figure 5** and its figure supplements), it is instructive to discuss the same in terms of their constituent enthalpies and entropies.

The determinants of $\Delta G_{TS-D(T)}$ in terms of its activation enthalpy and entropy may be readily deduced by partitioning the entire temperature range over which the two-state system is physically defined ($T_\alpha \leq T \leq T_\omega$) into three distinct regimes using four unique reference temperatures: $T_\alpha$, $T_S$, $T_{H(TS-D)}$, and $T_\omega$ (**Figure 12** and **Figure 12−figure supplement 1**). (1) For $T_\alpha \leq T < T_S$, the activation of conformers from the DSE to the TSE is entropically favoured ($T\Delta S_{TS-D(T)} > 0$) but is more than offset by the endothermic activation enthalpy ($\Delta H_{TS-D(T)} > 0$), leading to incomplete compensation and a positive $\Delta G_{TS-D(T)}$ ($\Delta H_{TS-D(T)} - T\Delta S_{TS-D(T)} > 0$). When $T = T_S$, $\Delta G_{TS-D(T)}$ is a minimum (its lone extremum), and is purely due to the endothermic enthalpy of activation ($\Delta G_{TS-D(T)} = \Delta H_{TS-D(T)} > 0$). (2) For $T_S <$ $T < T_{H(TS-D)}$, the activation of denatured conformers to the TSE is enthalpically and entropically disfavoured ($\Delta H_{TS-D(T)} > 0$ and $T\Delta S_{TS-D(T)} < 0$) leading to a positive $\Delta G_{TS-D(T)}$. (3) In contrast, for $T_{H(TS-D)} < T \leq T_\omega$, the favourable exothermic activation enthalpy ($\Delta H_{TS-D(T)} < 0$) is more than offset by the unfavourable entropy of activation ($T\Delta S_{TS-D(T)} < 0$), leading once again to a positive $\Delta G_{TS-D(T)}$. When $T = T_{H(TS-D)}$, $\Delta G_{TS-D(T)}$ is purely due to the negative change in the activation entropy or the *negentropy* of activation ($\Delta G_{TS-D(T)} = -T\Delta S_{TS-D(T)} > 0$), $\Delta G_{TS-D(T)}/T$ is a minimum, and $k_{f(T)}$ is a maximum (their lone extrema; see Massieu-Planck functions below). An important conclusion that we may draw from these analyses is the following: While it is true that for the temperature regimes $T_\alpha \leq T < T_S$ and $T_{H(TS-D)} < T \leq T_\omega$, $\Delta G_{TS-D(T)}$ is due to the incomplete compensation of the opposing activation enthalpy and entropy, this is clearly not the case for $T_S < T < T_{H(TS-D)}$ where both these two state functions are unfavourable and complement each other to generate a positive Gibbs activation barrier.

Similarly, the determinants of $\Delta G_{TS-N(T)}$ in terms of its activation enthalpy and entropy may be readily divined by partitioning the entire temperature range into five distinct regimes using six unique reference temperatures: $T_\alpha$, $T_{S(\alpha)}$, $T_{H(TS-N)}$, $T_S$, $T_{S(\omega)}$, and $T_\omega$ (**Figure 13** and **Figure 13−figure supplement 1**). (1) For $T_\alpha \leq T < T_{S(\alpha)}$, which is the ultralow temperature *Marcus-inverted-regime* for unfolding, the activation of the native conformers to the TSE is entropically favoured ($T\Delta S_{TS-N(T)} > 0$) but is more than offset by the unfavourable enthalpy of activation ($\Delta H_{TS-N(T)} > 0$) leading to incomplete compensation and a positive $\Delta G_{TS-N(T)}$ ($\Delta H_{TS-N(T)} - T\Delta S_{TS-N(T)} > 0$). When $T = T_{S(\alpha)}$, $\Delta S_{TS-N(T)} = \Delta H_{TS-N(T)} = 0 \Rightarrow \Delta G_{TS-N(T)} = 0$. The first extrema of $\Delta G_{TS-N(T)}$ and $\Delta G_{TS-N(T)}/T$ (which are a minimum), and the first extremum of $k_{u(T)}$ (which is a maximum, $k_{u(T)} = k^0$) occur at $T_{S(\alpha)}$. (2) For $T_{S(\alpha)} < T < T_{H(TS-N)}$, the activation of the native conformers to the TSE is enthalpically favourable ($\Delta H_{TS-N(T)} < 0$) but is more than offset by the unfavourable negentropy of activation ($T\Delta S_{TS-N(T)} < 0$) leading to $\Delta G_{TS-N(T)} > 0$. When $T = T_{H(TS-N)}$, $\Delta H_{TS-N(T)} = 0$ for the second time, and the Gibbs barrier to unfolding is purely due to the negentropy of activation ($\Delta G_{TS-N(T)} = -T\Delta S_{TS-N(T)} > 0$). The second extrema of $\Delta G_{TS-N(T)}/T$ (which is a maximum) and $k_{u(T)}$ (which is a minimum) occur at $T_{H(TS-N)}$. (3) For $T_{H(TS-N)} < T < T_S$, the activation of the native conformers to the TSE is entropically and enthalpically unfavourable ($\Delta H_{TS-N(T)} > 0$ and $T\Delta S_{TS-N(T)} < 0$) leading to $\Delta G_{TS-N(T)} > 0$.

When $T = T_S$, $\Delta S_{TS-N(T)} = 0$ for the second time, and the Gibbs barrier to unfolding is purely due to the endothermic enthalpy of activation ($\Delta G_{TS-N(T)} = \Delta H_{TS-N(T)} > 0$). The second extremum of $\Delta G_{TS-N(T)}$ (which is a maximum) occurs at $T_S$. (4) For $T_S < T < T_{S(\omega)}$, the activation of the native conformers to the TSE is entropically favourable ($T\Delta S_{TS-N(T)} > 0$) but is more than offset by the endothermic enthalpy of activation ($\Delta H_{TS-N(T)} > 0$) leading to incomplete compensation and a positive $\Delta G_{TS-N(T)}$. When $T = T_{S(\omega)}$, $\Delta S_{TS-N(T)} = \Delta H_{TS-N(T)} = 0$ for the third and the final time, and $\Delta G_{TS-N(T)} = 0$ for the second and final time. The third extrema of $\Delta G_{TS-N(T)}$ and $\Delta G_{TS-N(T)}/T$ (which are a minimum), and the third extremum of $k_{u(T)}$ (which is a maximum, $k_{u(T)} = k^0$) occur at $T_{S(\omega)}$. (5) For $T_{S(\omega)} < T \leq T_\omega$, which is the high-temperature *Marcus-inverted-regime* for unfolding, the activation of the native conformers to the TSE is enthalpically favourable ($\Delta H_{TS-N(T)} < 0$) but is more than offset by the unfavourable negentropy of activation ($T\Delta S_{TS-N(T)} < 0$), leading to $\Delta G_{TS-N(T)} > 0$. Once again we note that although the Gibbs barrier to unfolding is due to the incomplete compensation of the opposing enthalpies and entropies of activation for the temperature regimes $T_\alpha \leq T < T_{S(\alpha)}$, $T_{S(\alpha)} < T < T_{H(TS-N)}$, $T_S < T < T_{S(\omega)}$, and $T_{S(\omega)} < T \leq T_\omega$, both the enthalpy and the entropy of activation are unfavourable and collude to generate the Gibbs barrier to unfolding for the temperature regime $T_{H(TS-N)} < T < T_S$. Thus, a fundamentally important conclusion that we may draw from this analysis is that "*the Gibbs barriers to folding and unfolding are not always due to the incomplete compensation of the opposing enthalpy and entropy.*"

In a *protein folding* scenario where the activated conformers diffuse on the Gibbs energy surface to reach the NSE, the algebraic signs of the state functions invert leading to a change in the interpretation (**Figure 13−figure supplements 2** and **3**). Thus, for the partial folding reaction $[TS] \rightleftharpoons N$: (1) For $T_\alpha \leq T < T_{S(\alpha)}$, the flux of the conformers from the TSE to the NSE is entropically disfavoured ($T\Delta S_{TS-N(T)} > 0 \Rightarrow T\Delta S_{N-TS(T)} < 0$) but is more than compensated by the favourable change in enthalpy ($\Delta H_{TS-N(T)} > 0 \Rightarrow \Delta H_{N-TS(T)} < 0$), leading to $\Delta G_{N-TS(T)} < 0$. (2) For $T_{S(\alpha)} < T < T_{H(TS-N)}$, the flux of the conformers from the TSE to the NSE is enthalpically unfavourable ($\Delta H_{TS-N(T)} < 0 \Rightarrow \Delta H_{N-TS(T)} > 0$) but is more than compensated by the favourable change in entropy ($T\Delta S_{TS-N(T)} < 0 \Rightarrow T\Delta S_{N-TS(T)} > 0$) leading to $\Delta G_{N-TS(T)} < 0$. When $T = T_{H(TS-N)}$, the flux is driven purely by the positive change in entropy ($\Delta G_{\text{N-TS}(T)} = -T\Delta S_{\text{N-TS}(T)} < 0$). (3) For $T_{H(\text{TS-N})} < T < T_S$, the flux of the conformers from the TSE to the NSE is entropically and enthalpically favourable ($\Delta H_{\text{N-TS}(T)} < 0$ and $T\Delta S_{\text{N-TS}(T)} > 0$) leading to $\Delta G_{\text{N-TS}(T)} < 0$. When $T = T_S$, the flux is driven purely by the exothermic change in enthalpy ($\Delta G_{\text{N-TS}(T)} = \Delta H_{\text{N-TS}(T)} < 0$). (4) For $T_S < T < T_{S(\omega)}$, the flux of the conformers from the TSE to the NSE is entropically unfavourable ($T\Delta S_{\text{TS-N}(T)} > 0 \Rightarrow T\Delta S_{\text{N-TS}(T)} < 0$) but is more than compensated by the exothermic change in enthalpy ($\Delta H_{\text{TS-N}(T)} > 0 \Rightarrow \Delta H_{\text{N-TS}(T)} < 0$) leading to $\Delta G_{\text{N-TS}(T)} < 0$. (5) For $T_{S(\omega)} < T \leq T_\omega$, the flux of the conformers from the TSE to the NSE is enthalpically unfavourable ($\Delta H_{\text{TS-N}(T)} < 0 \Rightarrow \Delta H_{\text{N-TS}(T)} > 0$) but is more than compensated by the favourable change in entropy ($T\Delta S_{\text{TS-N}(T)} < 0 \Rightarrow T\Delta S_{\text{N-TS}(T)} > 0$), leading to $\Delta G_{\text{N-TS}(T)} < 0$.

Thus, the criteria for two-state folding from the viewpoint of Gibbs energy are the following: (*i*) the condition that $\Delta G_{\text{D-N}(T)} = \Delta G_{\text{TS-N}(T)} - \Delta G_{\text{TS-D}(T)}$ must be satisfied at all temperatures; (*ii*) the cold and heat denaturation temperatures estimated from equilibrium thermal denaturation must be identical to independently determined temperatures at which $k_{f(T)}$ and $k_{u(T)}$ are identical, i.e., the temperatures at which $\Delta G_{\text{TS-D}(T)}$ and $\Delta G_{\text{TS-N}(T)}$ functions intersect must be identical to the temperatures at which $\Delta H_{\text{D-N}(T)} - T\Delta S_{\text{D-N}(T)} = \Delta G_{\text{D-N}(T)} = 0$. The basis for these relationships, as mentioned earlier, is the *principle of microscopic reversibility*;[27] (*iii*) $\Delta G_{\text{TS-D}(T)}$ and $\Delta G_{\text{TS-N}(T)}$ must be a minimum and a maximum, respectively, at $T_S$; and (*iv*) the condition that $T_{H(\text{TS-N})} < T_H < T_S < T_{H(\text{TS-D})}$ must be satisfied. A far more detailed explanation in terms of chain and desolvation entropies and enthalpies is given in the accompanying article.

**Massieu-Planck functions**

The Massieu-Planck function, $\Delta G/T$, or its equivalent $-R\ln K$ ($K$ is the equilibrium constant) predates the Gibbs energy function by a few years and is especially useful when analysing temperature-dependent changes in protein behaviour (see Schellman, 1997, on the use of Massieu-Planck functions to analyse protein folding, and why the use of $\Delta G$ *versus* $T$ curves can sometimes lead to ambiguous conclusions).[6,35] Comparison of **Figure 6−figure supplement 1A** and **Figure 14A** demonstrates that although $\Delta G_{\text{TS-D}(T)}$ is a minimum at $T_S$ (**Figure 5A**), $k_{f(T)}$ will be a maximum not at $T_S$ but instead at $T_{H(TS-D)}$ where the Massieu-Planck activation potential for folding ($\Delta G_{TS-D(T)}/T \equiv -R\ln K_{TS-D(T)}$) is a minimum, and is readily apparent if we recast the Arrhenius expression for $k_{f(T)}$ in terms of the equilibrium constant for the partial folding reaction $D \rightleftharpoons [TS]$.

$$k_{f(T)} = k^0 \exp\left(-\frac{\cancel{RT} \ln K_{TS-D(T)}}{\cancel{RT}}\right) = k^0 K_{TS-D(T)} \tag{19}$$

Eq. (19) shows that the rate determining $K_{TS-D(T)}$ ($[TS]/[D]$) or the population of activated conformers relative to those that nestle at the bottom of the denatured Gibbs energy well is a maximum not at $T_S$ but at $T_{H(TS-D)}$ (**Figure 14−figure supplement 1A**). Similarly, comparison of **Figure 6−figure supplement 1B** and **Figure 14B** shows that although $\Delta G_{TS-N(T)}$ is a maximum at $T_S$ (**Figure 5B**), the minimum in $k_{u(T)}$ will occur not at $T_S$ but instead at $T_{H(TS-N)}$ where the Massieu-Planck activation potential for unfolding ($\Delta G_{TS-N(T)}/T \equiv -R\ln K_{TS-N(T)}$) is a maximum (Eq. (20)).

$$k_{u(T)} = k^0 \exp\left(-\frac{\cancel{RT} \ln K_{TS-N(T)}}{\cancel{RT}}\right) = k^0 K_{TS-N(T)} \tag{20}$$

Thus, for the partial unfolding reaction $N \rightleftharpoons [TS]$, the rate determining $K_{TS-N(T)}$ ($[TS]/[N]$) or the population of activated conformers relative to those at the bottom of the native Gibbs basin is a minimum not at $T_S$ but at $T_{H(TS-N)}$ (**Figure 14−figure supplement 1B**). Similarly, we see that although the $\Delta G_{N-D(T)}$ is a minimum or the most negative at $T_S$ (**Figure 1−figure supplement 1**), $K_{N-D(T)}$ ($[N]/[D]$) is a maximum not at $T_S$ but at $T_H$ where $\Delta H_{N-D(T)} = 0$ and $k_{f(T)}/k_{u(T)}$ is a maximum (**Figure 14−figure supplement 2A**).[6] Because the ratio of the solubilities of any two reaction-states is identical to the equilibrium constant, we may state that for any two-state folder at constant pressure and solvent conditions: (*i*) the solubility of the TSE as compared to the DSE is the greatest when the Gibbs barrier to folding is purely entropic, and this occurs precisely at $T_{H(TS-D)}$ (**Figure 14−figure supplement 3A**); (*ii*) the solubility of the TSE as compared to the NSE is the least when the Gibbs barrier to unfolding is purely entropic and occurs precisely at $T_{H(TS-N)}$ (**Figure 14−figure supplement 3B**); (*iii*) the solubilities of the TSE and the NSE are identical at $T_{S(\alpha)}$ and $T_{S(\omega)}$ where $\Delta S_{TS-N(T)} = \Delta H_{TS-}$ $_{\text{N}(T)} = \Delta G_{\text{TS-N}(T)} = 0$, and $k_{u(T)} = k^0$ (**Figure 14−figure supplement 3B**); and (*iv*) the solubility of the NSE as compared to the DSE is the greatest when the net flux of the conformers from the DSE to the NSE is driven purely by the difference in entropy between these two reaction-states and occurs precisely at $T_H$ (**Figure 14−figure supplement 2B**). The notion that "certain aspects of the temperature-dependent protein behaviour are greatly simplified when the Massieu-Planck functions are used in preference to the Gibbs energy" is readily apparent from inspection of **Figure 14−figure supplements 4** and **5**: While the natural logarithms of $k_{f(T)}$ and $k_{u(T)}$ have a complex dependence on their respective Gibbs barriers, a simple linear relationship exists between the rate constants and their respective Massieu-Planck functions.

## Temperature-dependence of $\Delta C_{p\text{D-TS}(T)}$ and $\Delta C_{p\text{TS-N}(T)}$

In order to provide a rational explanation for the temperature-dependence of the $\Delta C_{p\text{D-TS}(T)}$ and $\Delta C_{p\text{TS-N}(T)}$ functions, it is instructive to first discuss the inter-relationships between $\Delta \text{SASA}_{\text{D-N}}$, $m_{\text{D-N}}$, and $\Delta C_{p\text{D-N}}$. According to the "*liquid-liquid transfer*" model (LLTM) the greater heat capacity of the DSE as compared to the NSE (i.e., $\Delta C_{p\text{D-N}} > 0$ and substantial) is predominantly due to anomalously high heat capacity and low entropy of water that surrounds the exposed non-polar residues in the DSE (referred to as "microscopic icebergs" or "clathrates"; see references in Baldwin, 2014).[36] Because the size of the solvation shell depends on the SASA of the non-polar solute, it naturally follows that the change in the heat capacity must be proportional to the change in the non-polar SASA that accompanies a reaction. Consequently, protein unfolding reactions which are accompanied by large changes in non-polar SASA lead to large and positive changes in the heat capacity.[33,37,38] Because the denaturant *m* values are also directly proportional to the change in SASA that accompanies protein unfolding reactions, the expectation is that $m_{\text{D-N}}$ and $\Delta C_{p\text{D-N}}$ values must also be proportional to each other: The greater the $m_{\text{D-N}}$ value, the greater is the $\Delta C_{p\text{D-N}}$ value and *vice versa* (Figs. 2, 3 and 5 in Myers et al., 1995). However, since the residual structure in the DSEs of proteins under folding conditions is both sequence and solvent-dependent (i.e., the SASAs of the DSEs two proteins of identical chain lengths but dissimilar primary sequences need not necessarily be the same even under identical solvent conditions),[39,40] and because we do not yet have reliable theoretical or experimental methods to accurately quantify the SASA of the DSEs of proteins under folding conditions (i.e., the values are model-dependent),[41-43] the data scatter in plots that show correlation between the experimentally determined $m_{\text{D-N}}$ or $\Delta C_{p\text{D-N}}$ values (which reflect the true $\Delta\text{SASA}_{\text{D-N}}$) and the calculated values of $\Delta\text{SASA}_{\text{D-N}}$ can be significant (Fig. 2 in Myers et al., 1995, and Fig. 3 in Robertson and Murphy, 1997). Now, since the solvation shell around the DSEs of large proteins is relatively greater than that of small proteins even when the residual structure in the DSEs under folding conditions is taken into consideration, large proteins on average expose relatively greater amount of non-polar SASA upon unfolding than do small proteins; consequently, both $m_{\text{D-N}}$ and $\Delta C_{p\text{D-N}}$ values also correlate linearly with chain-length, albeit with considerable scatter since chain length, owing to the residual structure in the DSEs, is unlikely to be a true descriptor of the SASA of the DSEs of proteins under folding conditions (note that the scatter can also be due to certain proteins having anomalously high or low number of non-polar residues). The point we are trying to make is the following: Because the native structures of proteins are relatively insensitive to small variations in pH and co-solvents,[44] and since the number of ways in which foldable polypeptides can be packed into their native structures is relatively limited (as inferred from the limited number of protein folds, see SCOP: www.mrc-lmb.cam.ac.uk and CATH: www.cathdb.info databases), one might find a reasonably good correlation between chain lengths and the SASAs of the NSEs of proteins of differing primary sequences under varying solvents (Fig. 1 in Miller et al., 1987).[45,46] However, since the SASAs of the DSEs under folding conditions, owing to residual structure are variable, until and unless we find a way to accurately simulate the DSEs of proteins, and if and only if these theoretical methods are sensitive to point mutations, changes in pH, co-solvents, temperature and pressure, it is almost impossible to arrive at a universal equation that will describe how the $\Delta\text{SASA}_{\text{D-N}}$ under folding conditions will vary with chain length, and by logical extension, how $m_{\text{D-N}}$ and $\Delta C_{p\text{D-N}}$ will vary with SASA or chain length. Nevertheless, if we consider a single two-state-folding primary sequence under constant pressure and solvent conditions and vary the temperature, and if the properties of the solvent are temperature-invariant (for example, no change in the pH due to the temperature-dependence of the $pK_a$ of the constituent buffer), then the manner in which the $\Delta C_{p\text{D-TS}(T)}$ and $\Delta C_{p\text{TS-N}(T)}$ functions vary with temperature must be consistent with the temperature-dependence of $m_{\text{TS-D}(T)}$ and $m_{\text{TS-N}(T)}$, respectively, and by logical extension, with $\Delta\text{SASA}_{\text{D-TS}(T)}$ and $\Delta\text{SASA}_{\text{TS-N}(T)}$, respectively.

Inspection of **Figures 15** and **Figure 15−figure supplements 1**, **2** and **3** demonstrate that: (*i*) both $\Delta C_{p\text{D-TS}(T)}$ and $\Delta C_{p\text{TS-N}(T)}$ vary with temperature; and (*ii*) their gross features stem primarily from the second derivatives of the temperature-dependence of the *curve-crossing* with respect to the DSE and the NSE. The prediction that the change in heat capacities for the partial unfolding reactions, $N \rightleftharpoons [TS]$ and $[TS] \rightleftharpoons D$, must vary with temperature is due to Eqs. (12) and (13). Although this may not be readily apparent from a casual inspection of the equations, even a cursory examination of **Figures 8** and **9** shows that it is simply not possible for $\Delta C_{p\text{D-TS}(T)}$ and $\Delta C_{p\text{TS-N}(T)}$ functions to be temperature-invariant since the slopes of the $\Delta H_{\text{TS-D}(T)}$ and the $\Delta H_{\text{TS-N}(T)}$ functions are continuously changing with temperature. If we recall that the force constants are temperature-invariant, it becomes readily apparent that the second terms in the brackets on the right-hand-side (RHS) of Eqs. (12) and (13) i.e., $\omega T(\Delta S_{\text{D-N}(T)})^2$ and $\alpha T(\Delta S_{\text{D-N}(T)})^2$, respectively, will be parabolas with a minimum (zero) at $T_S$. This is due to $\Delta S_{\text{D-N}(T)}$ being negative for $T < T_S$, positive for $T > T_S$, and zero for $T = T_S$. Furthermore, since $\varphi$, $\sqrt{\varphi}$ and $m_{\text{TS-N}(T)}$ are a maximum, and $m_{\text{TS-D}(T)}$ a minimum at $T_S$, the expectation is that $\Delta C_{p\text{D-TS}(T)}$ must be a minimum (or $\Delta C_{p\text{TS-D}(T_S)}$ is the least negative), and $\Delta C_{p\text{TS-N}(T)}$ must be a maximum at $T_S$. Thus, for $T = T_S$, Eqs. (12) and (13) become

$$\left. \begin{array}{l} \Delta C_{p\text{D-TS}(T)} = \dfrac{\alpha m_{\text{TS-D}(T)} \Delta C_{p\text{D-N}}}{\sqrt{\varphi}} \\[2mm] \Delta C_{p\text{TS-N}(T)} = \dfrac{\omega m_{\text{TS-N}(T)} \Delta C_{p\text{D-N}}}{\sqrt{\varphi}} \end{array} \right]_{T=T_S} > 0 \Rightarrow \left. \dfrac{\Delta C_{p\text{TS-N}(T)}}{\Delta C_{p\text{D-TS}(T)}} \right|_{T=T_S} = \left. \dfrac{\omega m_{\text{TS-N}(T)}}{\alpha m_{\text{TS-D}(T)}} \right|_{T=T_S} \quad (21)$$

The prediction that the extrema of $\Delta C_{p\text{D-TS}(T)}$ and $\Delta C_{p\text{TS-N}(T)}$ functions must occur at $T_S$ is readily apparent from **Figure 15** and **Figure 15−figure supplement 1B**. Importantly, consistent with the relationship between $m_{\text{D-N}}$ and $\Delta C_{p\text{D-N}}$ values, comparison of these two figures with **Figure 2** and **Figure 2−figure supplement 1** demonstrates that just as $m_{\text{TS-D}(T)}$ and $m_{\text{TS-N}(T)}$ are a minimum and a maximum at $T_S$, respectively, so too are $\Delta C_{p\text{D-TS}(T)}$ and $\Delta C_{p\text{TS-N}(T)}$ functions. This leads to two obvious corollaries: (*i*) the difference in heat capacity between the DSE and the TSE is a minimum when the difference in SASA between the DSE and the TSE is a minimum; and (*ii*) the difference in heat capacity between the TSE and the NSE is a maximum when the difference in SASA between the TSE and the NSE is a maximum. Because $\Delta S_{\text{TS-D}(T)} = \Delta S_{\text{TS-N}(T)} = 0$, $\Delta G_{\text{TS-D}(T)}$ is a minimum, and both $\Delta G_{\text{TS-N}(T)}$ and $\Delta G_{\text{D-N}(T)}$ are a maximum, at $T_S$ (**Figures 1, 5** and **Figure 11−figure supplement 1B**), a fundamentally important conclusion is that the *Gibbs barriers to folding and unfolding are a* *minimum and a maximum, respectively, and equilibrium stability is a maximum, and are all purely enthalpic when* $\Delta C_{p\text{D-TS}(T)}$ *and* $\Delta C_{p\text{TS-N}(T)}$ *are a minimum and a maximum, respectively.*

Inspection of **Figure 15** and **Figure 15−figure supplement 1** demonstrates that unlike $\Delta C_{p\text{D-TS}(T)}$ which is positive across the entire temperature range, $\Delta C_{p\text{TS-N}(T)}$ which is a maximum and positive at $T_S$, decreases with any deviation in temperature from $T_S$, and is zero at $T_{C_p\text{TS-N}(\alpha)}$ and $T_{C_p\text{TS-N}(\omega)}$; consequently, $\Delta C_{p\text{TS-N}(T)} < 0$ for $T_\alpha \leq T < T_{C_p\text{TS-N}(\alpha)}$ and $T_{C_p\text{TS-N}(\omega)} < T \leq T_\omega$. The reason for this behaviour is apparent from inspection of **Figures 9** and **11**: The slope of the $\Delta H_{\text{TS-N}(T)}$ and $\Delta S_{\text{TS-N}(T)}$ functions becomes zero at $T_{C_p\text{TS-N}(\alpha)}$ and $T_{C_p\text{TS-N}(\omega)}$; and any further decrease or increase in temperature, respectively, causes the slope to invert. This can be mathematically shown as follows: Since $m_{\text{TS-N}(T)} = 0$ at $T_{S(\alpha)}$ and $T_{S(\omega)}$, we have $\sqrt{\varphi} = \alpha m_{\text{D-N}}$ and $\varphi = (\alpha m_{\text{D-N}})^2$ at $T_{S(\alpha)}$ and $T_{S(\omega)}$. Substituting these relationships in Eq. (13) leads to

$$\Delta C_{p\text{TS-N}(T)}\Big|_{T=T_{S(\alpha)}, T_{S(\omega)}} = -\frac{\omega \alpha m_{\text{D-N}} T (\Delta S_{\text{D-N}(T)})^2}{2\varphi\sqrt{\varphi}}\Bigg|_{T=T_{S(\alpha)}, T_{S(\omega)}} = -\frac{\omega T}{2}\left(\frac{\Delta S_{\text{D-N}(T)}}{\alpha m_{\text{D-N}}}\right)^2\Bigg|_{T=T_{S(\alpha)}, T_{S(\omega)}} \quad (22)$$

Further, since $\Delta C_{p\text{D-N}} = \Delta C_{p\text{D-TS}(T)} + \Delta C_{p\text{TS-N}(T)}$ for a two-state system, we have

$$\Delta C_{p\text{D-TS}(T)}\Big|_{T=T_{S(\alpha)}, T_{S(\omega)}} = \Delta C_{p\text{D-N}} + \frac{\omega T}{2}\left(\frac{\Delta S_{\text{D-N}(T)}}{\alpha m_{\text{D-N}}}\right)^2\Bigg|_{T=T_{S(\alpha)}, T_{S(\omega)}} > \Delta C_{p\text{D-N}} \quad (23)$$

Because $\Delta C_{p\text{TS-N}(T)} < 0$ at $T_{S(\alpha)}$ and $T_{S(\omega)}$, and the lone extremum of $\Delta C_{p\text{TS-N}(T)}$ (which is algebraically positive and a maximum) occurs at $T_S$, it implies that there will be two unique temperatures at which $\Delta C_{p\text{TS-N}(T)} = 0$, one in the low temperature ($T_{C_p\text{TS-N}(\alpha)}$) such that $T_{S(\alpha)} < T_{C_p\text{TS-N}(\alpha)} < T_S$, and the other in the high temperature regime ($T_{C_p\text{TS-N}(\omega)}$) such that $T_S < T_{C_p\text{TS-N}(\omega)} < T_{S(\omega)}$. Thus, at the these two unique temperatures $T_{C_p\text{TS-N}(\alpha)}$ and $T_{C_p\text{TS-N}(\omega)}$, we have $\Delta C_{p\text{D-TS}(T)} = \Delta C_{p\text{D-N}} \Rightarrow \beta_{\text{H(fold)}(T)} = 1$ and $\beta_{\text{H(unfold)}(T)} = 0$; and for the temperature regimes $T_\alpha \leq T < T_{C_p\text{TS-N}(\alpha)}$ and $T_{C_p\text{TS-N}(\omega)} < T \leq T_\omega$, we have $\Delta C_{p\text{D-TS}(T)} > \Delta C_{p\text{D-N}} \Rightarrow \beta_{\text{H(fold)}(T)} > 1$, and $\Delta C_{p\text{TS-N}(T)} < 0 \Rightarrow \beta_{\text{H(unfold)}(T)} < 0$ (see heat capacity RC below for the definition of $\beta_{\text{H(fold)}(T)}$ and $\beta_{\text{H(unfold)}(T)}$).

Although the prediction that $\Delta C_{p\text{TS-N}(T)}$ must approach zero at very low and high temperatures may not be readily verified by experiment for the low-temperature regime owing to technical difficulty in making a measurement, the prediction for the high-temperature regime is strongly supported by the data on CI2 from the Fersht lab: Despite the temperature-range not being substantial (320 to 340 K), and the data points that define the $\Delta H_{\text{TS-N}(T)}$ function being sparse (7 in total), it is apparent even from a cursory inspection that it is clearly non-linear with temperature (Fig. 5B in Tan et al., 1996).[47] Although Fersht and co-workers have fitted the data to a linear function and reached the natural conclusion that the heat capacity of activation for unfolding is temperature-invariant, they nevertheless explicitly mention that if the non-linearity of $\Delta H_{\text{TS-N}(T)}$ were given due consideration, and the data are fit to an empirical-quadratic instead of a linear function, $\Delta C_{p\text{TS-N}(T)}$ indeed becomes temperature-dependent and is predicted to approach zero at ~ 360 K (see text in page 382 in Tan et al., 1996).[47] Now, since $\Delta C_{p\text{TS-N}(T)} > 0$ and a maximum, and $\Delta C_{p\text{D-TS}(T)}$ is a minimum and positive at $T_S$, and decrease and increase, respectively, with any deviation in temperature from $T_S$, and since $\Delta C_{p\text{TS-N}(T)}$ becomes zero at $T_{C_p\text{TS-N}(\alpha)}$ and $T_{C_p\text{TS-N}(\omega)}$, the obvious mathematical consequence is that $\Delta C_{p\text{D-TS}(T)}$ and $\Delta C_{p\text{TS-N}(T)}$ functions must intersect at two unique temperatures. Because at the points of intersection we have the relationship: $\Delta C_{p\text{D-TS}(T)} = \Delta C_{p\text{TS-N}(T)} = \Delta C_{p\text{D-N}}/2$, a consequence is that $\Delta C_{p\text{TS-N}(T)}$ must be positive at the said temperatures, with the low-temperature intersection occurring between $T_{C_p\text{TS-N}(\alpha)}$ and $T_S$, and the high-temperature intersection between $T_S$ and $T_{C_p\text{TS-N}(\omega)}$. This is readily apparent from inspection of **Figure 15−figure supplement 1B**: Both $\Delta C_{p\text{D-TS}(T)}$ and $\Delta C_{p\text{TS-N}(T)}$ are identical at 214.1 K and 345.9 K. An equivalent interpretation is that at these temperatures, the absolute heat capacity of the TSE is exactly half the algebraic sum of the absolute heat capacities of the DSE and the NSE. As we shall show in subsequent publications, the intersection of various state functions is a source of interesting relationships that may be used as constraints in simulations (see also **Figure 9−figure supplement 2**).

**The position of the TSE along the heat capacity RC**

Inspection and comparison of **Figure 2−figure supplement 1** and **Figure 15−figure supplement 1B** demonstrates that although the manner in which the $\Delta C_{p\text{D-TS}(T)}$ and $\Delta C_{p\text{TS-N}(T)}$ functions vary with temperature is consistent with the relationship between $m_{\text{D-N}}$ and $\Delta C_{p\text{D-N}}$ values, there is nevertheless an intriguing anomaly that is at odds with the LLTM for heat capacity. If we consider the partial folding reaction $D \rightleftharpoons [TS]$, it is readily apparent from these figures that although the denatured conformer diffuses $> \sim 70\%$ along the normalized SASA-RC to reach the TSE for 240 K $< T <$ 320 K, $\Delta C_{pD\text{-}TS(T)} \ll \Delta C_{pTS\text{-}N(T)}$ throughout this regime. Conversely, if we consider the total unfolding reaction $N \rightleftharpoons D$, a large fraction of $\Delta C_{pD\text{-}N}$ is accounted for not by the second-half of the unfolding reaction ($[TS] \rightleftharpoons D$) but by the first-half ($N \rightleftharpoons [TS]$), despite the native conformer diffusing less than ~30% along the SASA-RC to reach the TSE. To put things into perspective, we will need to normalize the heat capacities of activation. Adopting Leffler's framework for the relative sensitivities of the activation and equilibrium enthalpies in response to a perturbation in temperature,[48] we may write

$$\beta_{H(\text{fold})(T)} = \frac{\partial \Delta H_{TS\text{-}D(T)}/\partial T}{\partial \Delta H_{N\text{-}D(T)}/\partial T} = \frac{\Delta C_{pTS\text{-}D(T)}}{\Delta C_{pN\text{-}D}} \equiv \frac{\Delta C_{pD\text{-}TS(T)}}{\Delta C_{pD\text{-}N}} \tag{24}$$

$$\beta_{H(\text{unfold})(T)} = \frac{\partial \Delta H_{TS\text{-}N(T)}/\partial T}{\partial \Delta H_{D\text{-}N(T)}/\partial T} = \frac{\Delta C_{pTS\text{-}N(T)}}{\Delta C_{pD\text{-}N}} \tag{25}$$

where $\beta_{H(\text{fold})(T)} = \beta_{S(\text{fold})(T)}$ and $\beta_{H(\text{unfold})(T)} = \beta_{S(\text{unfold})(T)}$ (see Paper-II) are classically interpreted to be a measure of the position of the TSE along the heat capacity RC.[49] Naturally, for a two-state system the algebraic sum of $\beta_{H(\text{fold})(T)}$ and $\beta_{H(\text{unfold})(T)}$ is unity. Recasting Eqs. (24) and (25) in terms of (12) and (13) gives

$$\beta_{H(\text{fold})(T)} = \frac{\alpha}{2\varphi\sqrt{\varphi}\Delta C_{pD\text{-}N}} \left[ m_{TS\text{-}D(T)} 2\varphi \Delta C_{pD\text{-}N} + \omega m_{D\text{-}N} T \left(\Delta S_{D\text{-}N(T)}\right)^2 \right]$$

$$= \frac{\alpha m_{D\text{-}N}}{2\varphi\sqrt{\varphi}\Delta C_{pD\text{-}N}} \left[ \beta_{T(\text{fold})(T)} 2\varphi \Delta C_{pD\text{-}N} + \omega T \left(\Delta S_{D\text{-}N(T)}\right)^2 \right] \tag{26}$$

$$\beta_{H(\text{unfold})(T)} = \frac{\omega}{2\varphi\sqrt{\varphi}\Delta C_{pD\text{-}N}} \left[ m_{TS\text{-}N(T)} 2\varphi \Delta C_{pD\text{-}N} - \alpha m_{D\text{-}N} T \left(\Delta S_{D\text{-}N(T)}\right)^2 \right]$$

$$= \frac{\omega m_{D\text{-}N}}{2\varphi\sqrt{\varphi}\Delta C_{pD\text{-}N}} \left[ \beta_{T(\text{unfold})(T)} 2\varphi \Delta C_{pD\text{-}N} - \alpha T \left(\Delta S_{D\text{-}N(T)}\right)^2 \right] \tag{27}$$

When $T = T_S$, $\Delta S_{D\text{-}N(T)} = 0$ and Eqs. (26) and (27) reduce to

$$\beta_{\text{H(fold)}(T)}\Big|_{T=T_S} = \frac{\alpha m_{\text{TS-D}(T)}}{\sqrt{\varphi}}\Big|_{T=T_S} = \frac{\alpha \beta_{\text{T(fold)}(T)} m_{\text{D-N}}}{\sqrt{\varphi}}\Big|_{T=T_S} \quad (28)$$

$$\beta_{\text{H(unfold)}(T)}\Big|_{T=T_S} = \frac{\omega m_{\text{TS-N}(T)}}{\sqrt{\varphi}}\Big|_{T=T_S} = \frac{\omega \beta_{\text{T(unfold)}(T)} m_{\text{D-N}}}{\sqrt{\varphi}}\Big|_{T=T_S} \quad (29)$$

As explained earlier, because $\Delta C_{p\text{D-N}}$ is temperature-invariant by postulate, and $\Delta C_{p\text{D-TS}(T)}$ is a minimum, and $\Delta C_{p\text{TS-N}(T)}$ is a maximum at $T_S$, $\beta_{\text{H(fold)}(T)}$ and $\beta_{\text{H(unfold)}(T)}$ are a minimum and a maximum, respectively, at $T_S$. How do $\beta_{\text{H(fold)}(T)}$ and $\beta_{\text{H(unfold)}(T)}$ compare with their counterparts, $\beta_{\text{T(fold)}(T)}$ and $\beta_{\text{T(unfold)}(T)}$? This is important because a statistically significant correlation exists between $m_{\text{D-N}}$ and $\Delta C_{p\text{D-N}}$, and both these two parameters independently correlate with $\Delta\text{SASA}_{\text{D-N}}$. Recasting Eqs. (28) and (29) gives

$$\frac{\beta_{\text{H(fold)}(T)}}{\beta_{\text{T(fold)}(T)}}\Big|_{T=T_S} \equiv \frac{\beta_{\text{S(fold)}(T)}}{\beta_{\text{T(fold)}(T)}}\Big|_{T=T_S} = \frac{\alpha m_{\text{D-N}}}{\sqrt{\varphi}}\Big|_{T=T_S} < 1 \quad (30)$$

$$\frac{\beta_{\text{H(unfold)}(T)}}{\beta_{\text{T(unfold)}(T)}}\Big|_{T=T_S} \equiv \frac{\beta_{\text{S(unfold)}(T)}}{\beta_{\text{T(unfold)}(T)}}\Big|_{T=T_S} = \frac{\omega m_{\text{D-N}}}{\sqrt{\varphi}}\Big|_{T=T_S} > 1 \quad (31)$$

Since $m_{\text{TS-N}(T)} > 0$ and a maximum, and $m_{\text{TS-D}(T)} > 0$ and a minimum, respectively, at $T_S$, it is readily apparent from inspection of Eqs. (1) and (2) that $\sqrt{\varphi} > \alpha m_{\text{D-N}}$ and $\omega m_{\text{D-N}} > \sqrt{\varphi}$ at $T_S$. Consequently, we have: $\beta_{\text{T(fold)}(T)}\Big|_{T=T_S} > \beta_{\text{H(fold)}(T)}\Big|_{T=T_S}$ and $\beta_{\text{T(unfold)}(T)}\Big|_{T=T_S} < \beta_{\text{H(unfold)}(T)}\Big|_{T=T_S}$.

In agreement with the predictions of Eqs. (30) and (31), inspection of **Figure 16** demonstrates that although the denatured conformer advances by > ~ 70% along the SASA-RC to reach the TSE when $T = T_S$, it accounts for < ~20% of the total change in $\Delta C_{p\text{D-N}}$ (i.e., $\beta_{\text{T(fold)}(T)}\Big|_{T=T_S} > \beta_{\text{H(fold)}(T)}\Big|_{T=T_S}$), with the rest of the change (> ~ 80%) in heat capacity coming from a mere ~ 30% diffusion of the activated conformer along the SASA-RC to reach the bottom of the native Gibbs basin (i.e., $\beta_{\text{T(unfold)}(T)}\Big|_{T=T_S} < \beta_{\text{H(unfold)}(T)}\Big|_{T=T_S}$). The theoretical prediction that $\beta_{\text{T(fold)}(T)} > \beta_{\text{H(fold)}(T)}$ across a substantial temperature range is supported by the finding by Gloss and Matthews (1998) that the position of the TSE relative to the DSE along the heat capacity RC is consistently lower than the same along the SASA-RC (see also page 178 in Bilsel and Matthews, 2000, and references therein).[50,51]

Now, if we accept the long held premise that the greater heat capacity of the DSE as compared to the NSE is purely or predominantly due to structured water around the exposed non-polar residues in the DSE, then the only way we can explain why $\Delta C_{pD\text{-}TS(T)} \ll \Delta C_{pTS\text{-}N(T)}$ despite $\beta_{T(fold)(T)} > \sim 70\%$ for the partial folding reaction $D \rightleftharpoons [TS]$ is that the non-polar SASA of both the DSE and the TSE are very similar at $T_S$. Because it is physically near-impossible for the denatured conformer to advance by $> \sim 70\%$ along the SASA-RC to reach the TSE, and yet keep the non-polar SASA fairly constant such that $\Delta C_{pD\text{-}TS(T)}$ is just about 20% of $\Delta C_{pD\text{-}N}$, the natural conclusion is that *"the large and positive difference in heat capacity between the DSE and the NSE cannot be only due to the clathrates of water molecules around exposed non-polar residues in the DSE."*[38,52-54] This brings us to two studies on the heat capacities of proteins, one by Sturtevant almost four decades ago, and the other by Lazaridis and Karplus.[55,56] While Sturtevant identified six possible sources of heat capacity which are: (*i*) the hydrophobic effect; (*ii*) electrostatic charges; (*iii*) hydrogen bonds; (*iv*) conformational entropy; (*v*) intramolecular vibrations; and (*vi*) changes in equilibria, and concluded that the most important of these are the *hydrophobic*, *conformational* and *vibrational* effects, Lazaridis and Karplus concluded from their molecular dynamics simulations on truncated CI2 that the heat capacity can have a significantly large and a positive contribution from intra-protein non-covalent interactions. What these two studies essentially imply is that when the pressure and solvent properties are defined and temperature-invariant, the ability of the conformers in a protein reaction-state to absorb thermal energy and yet resist an increase in temperature is dependent on: (*i*) its molecular structure; and (*ii*) the size and the character of its molecular surface (i.e., the relative proportion of polar and non-polar SASA). While the first variable determines the capacity of the reaction-state to absorb thermal energy and distribute it across its various internal modes of motion (the vibrational, rotational, and to some extent, the translational entropy from elements such as the N and C-terminal regions, loops etc. that can flap around in the solvent), the second variable determines not only the size and thickness of the solvent shell but also how tightly or loosely the solvent molecules are bound to the protein surface and to themselves (i.e., the dynamics of water in the solvation shell as compared to bulk water; see Fig. 1 in Frauenfelder et al., 2009), and by extension, the amount of excess thermal energy needed to disrupt the solvent shell as the reaction-states interconvert due to thermal noise.[36,52,57-61] Further discussion on the determinants of heat capacity is beyond the scope of this article and will be addressed elsewhere.

## On the inapplicability of the Hammond postulate to protein folding

Although it is difficult to provide a detailed physical explanation for the temperature-dependence of the heat capacities of activation without deconvoluting the activation enthalpies and entropies into their constituent *chain* and *desolvation* enthalpies and entropies (shown in the accompanying article), it is instructive to give one extreme example to emphasize why both the solvent shell and the non-covalent interactions make a significant contribution to heat capacity (note that as long as the difference in the number of covalent bonds between the reaction-states is zero, to a first approximation, their contribution to the *difference in heat capacity* between the reaction-states can be ignored; see Lecture II in Finkelstein and Ptitsyn, 2002, and references therein).[38,56,62,63]

It was shown earlier that when $T = T_{S(\alpha)}$ and $T_{S(\omega)}$, we have $m_{TS-N(T)} = 0 \Rightarrow \Delta SASA_{TS-N(T)} = 0$, leading to a unique set of relationships: $G_{TS(T)} = G_{N(T)}$, $H_{TS(T)} = H_{N(T)}$, $S_{TS(T)} = S_{N(T)}$, and $k_{u(T)} = k^0$ (**Figures 2B, Figure 2−figure supplement 1B, 4C, 5B, 6B, 9,** and **11**). However, we note from Eq. (22) that $\Delta C_{pTS-N(T)} < 0$ at these two temperatures and is ~ −6.2 kcal.mol$^{-1}$.K$^{-1}$ for FBP28 WW (**Figure 15B**). Since the molar concentration of the TSE is identical to that of the NSE at $T_{S(\alpha)}$ and $T_{S(\omega)}$, what this physically means is that if we were to take a mole of NSE and a mole of TSE and heat them at constant pressure under identical solvent conditions, we will find that the NSE, relative to the TSE, will absorb thermal energy equivalent to ~6.2 calories before both the TSE and the NSE will independently register a 10$^{-3}$ K rise in temperature. Because at these two temperatures the SASA, the Gibbs energy, the enthalpy, and the entropy of the TSE and the NSE are identical, this large difference in heat capacity which is ~15-fold greater than $\Delta C_{pD-N}$ (6.2/0.417 = 14.8) must stem from a complex combination of: (*i*) a difference in the number and kinds of non-covalent interactions;[64] (*ii*) the precise 3D-arrangement of the non-covalent interactions (i.e., the network of interactions) leading to a difference in their fundamental frequencies;[55,56] and (*iii*) the character of the surface exposed to the solvent (i.e., polar *vs* non-polar SASA) between the said reaction-states.[65-67] Thus, a fundamentally important conclusion that we may draw from this behaviour is that "*two reaction-states on a protein folding pathway need not necessarily have the same* *structure even if their interconversion proceeds with concomitant zero net-change in SASA, enthalpy, entropy, and Gibbs energy.*" A corollary is that the reaction-states on a protein folding pathway are distinct entities with respect to both their internal structure and the character of their molecular surface. What this implies is that the Hammond postulate which states that "*if two states, as for example, a transition state and an unstable intermediate, occur consecutively during a reaction process and have nearly the same energy content, their interconversion will involve only a small reorganization of the molecular structures,*"[68] although may be applicable to reactions of small molecules, is inapplicable to protein folding. The inapplicability stems primarily from the profound differences between non-covalent protein folding reactions and covalent reactions of small molecules. In the simplest reactions of small molecules, except for the one or two bonds that are being reconfigured, the rest of the reactant-structure, to a first approximation, usually remains fairly intact as the reaction proceeds (this need not necessarily hold for all simple chemical reactions and probably not for complex reactions). Consequently, if we were to use the bond-length of the bond that is being reconfigured as the RC, and find that the difference in Gibbs energy between any two reaction-states that occur consecutively along the RC are very similar, a reasonable assumption/expectation would be that their structures must be very similar.[69-77] However, such an assumption cannot be valid for protein folding since an incredibly large number of chain and solvent configurations can lead to conformers having exactly the same Gibbs energy. Consequently, it is difficult to imagine how one can infer the structure of the transiently populated protein reaction-states, including the TSEs, to a near-atomic resolution purely from energetics (see Φ-value analysis later).[78-80]

**The position of the TSE along the entropic RC**

The Leffler parameters for the relative sensitivities of the activation and equilibrium Gibbs energies in response to a perturbation in temperature are given by the ratios of the derivatives of the activation and equilibrium Gibbs energies with respect to temperature.[13-15,81] Thus, for the partial folding reaction $D \rightleftharpoons [TS]$, we have

$$\beta_{G(\text{fold})(T)} = \frac{\partial \Delta G_{\text{TS-D}(T)}/\partial T}{\partial \Delta G_{\text{N-D}(T)}/\partial T} = \frac{-\Delta S_{\text{TS-D}(T)}}{-\Delta S_{\text{N-D}(T)}} \equiv \frac{-\Delta S_{\text{TS-D}(T)}}{\Delta S_{\text{D-N}(T)}} \qquad (32)$$

where $\beta_{G(fold)(T)}$ is classically interpreted to be a measure of the position of the TSE relative to the DSE along the entropic RC.[49] Recasting Eq. (32) in terms of (8) and (A4) and rearranging gives

$$\beta_{G(fold)(T)} = \frac{\alpha\, m_{TS-D(T)}\, \cancel{\Delta S_{D-N(T)}}}{\cancel{\Delta S_{D-N(T)}}\, \sqrt{\varphi}} = \frac{\alpha\, m_{TS-D(T)}}{\sqrt{\varphi}} = \frac{\alpha \beta_{T(fold)(T)} m_{D-N}}{\sqrt{\varphi}} \quad (33)$$

Similarly for the partial unfolding reaction $N \rightleftharpoons [TS]$ we have

$$\beta_{G(unfold)(T)} = \frac{\partial \Delta G_{TS-N(T)}/\partial T}{\partial \Delta G_{D-N(T)}/\partial T} = \frac{\Delta S_{TS-N(T)}}{\Delta S_{D-N(T)}} \quad (34)$$

where $\beta_{G(unfold)(T)}$ is a measure of the position of the TSE relative to the NSE along the entropic RC. Substituting Eqs. (9) and (A6) in (34) gives

$$\beta_{G(unfold)(T)} = \frac{\omega\, m_{TS-N(T)}\, \cancel{\Delta S_{D-N(T)}}}{\cancel{\Delta S_{D-N(T)}}\, \sqrt{\varphi}} = \frac{\omega\, m_{TS-N(T)}}{\sqrt{\varphi}} = \frac{\omega \beta_{T(unfold)(T)} m_{D-N}}{\sqrt{\varphi}} \quad (35)$$

Inspection of Eqs. (32) and (34) shows that $\beta_{G(fold)(T)} + \beta_{G(unfold)(T)} = 1$ for any given reaction-direction. Now, since $\Delta S_{D-N(T)} = \Delta S_{TS-D(T)} = \Delta S_{TS-N(T)} = 0$ at $T_S$, $\beta_{G(fold)(T)}$ and $\beta_{G(unfold)(T)}$ will be undefined for $T = T_S$. However, these are removable discontinuities as is apparent from Eqs. (33) and (35); consequently, curves simulated using the latter set of equations will have a hole at $T_S$. If we ignore the hole at $T_S$ to enable a physical description and their comparison to other RCs, the extremum of $\beta_{G(fold)(T)}$ (which is positive and a minimum) and the extremum of $\beta_{G(unfold)(T)}$ (which is positive and a maximum) will occur at $T_S$ (**Figure 17** and **Figure 17−figure supplement 1**) and is a consequence of $m_{TS-D(T)}$ being a minimum, and both $m_{TS-N(T)}$ and $\varphi$ being a maximum, respectively, at $T_S$. This can also be demonstrated by differentiating Eqs. (32) and (34) with respect to temperature (not shown). Comparison of Eqs. (28) and (33), and Eqs. (29) and (35) demonstrate that when $T = T_S$, we have $\beta_{H(fold)(T)} = \beta_{G(fold)(T)}$ and $\beta_{H(unfold)(T)} = \beta_{G(unfold)(T)}$, i.e., the position of the TSE along the heat capacity and entropic RCs are identical at $T_S$, and non-identical for $T \neq T_S$ (**Figure 17**). Further, since $m_{TS-N(T)} = \beta_{T(unfold)(T)} = 0$ at $T_{S(\alpha)}$ and $T_{S(\omega)}$ (**Figure 2B** and **Figure 2−figure supplement 1B**), $\beta_{G(unfold)(T)} \equiv \beta_{T(unfold)(T)} = 0$ and $\beta_{G(fold)(T)} \equiv \beta_{T(fold)(T)} = 1$, and not identical for $T \neq T_{S(\alpha)}$ and $T_{S(\omega)}$; and for $T_\alpha \leq T < T_{S(\alpha)}$ and $T_{S(\omega)} < T \leq T_\omega$ (the ultralow and high temperature *Marcus-inverted-regimes*, respectively), $\beta_{G(fold)(T)}$ and $\beta_{T(fold)(T)}$ are greater than unity, and $\beta_{G(unfold)(T)}$ and $\beta_{T(unfold)(T)}$ are negative (**Figure 18**). Note that although $\beta_{G(fold)(T)}$ is unity at $T_{S(\alpha)}$ and $T_{S(\omega)}$, the structures of the TSE and the NSE cannot be assumed to be identical as explained earlier.

Although it is beyond the scope of this manuscript to perform a large-scale survey of literature for corroborating evidence, the notion that these equations must hold for any two-state folder (as long as they conform to the postulates laid out in Paper-I) is readily apparent from the experimental data of Kelly, Gruebele and colleagues.[25,82-84] However, the reader will note that what Gruebele and coworkers refer to as $\Phi_T(T, P)$ (see Eq. (8) in Crane et al., 2000 and Jäger et al., 2001, Eq. (5) in Ervin and Gruebele, 2002, and Eq. (3) in Nguyen et al., 2003) is equivalent to $\beta_{G(T)}$ in this article. We will reserve the letter $\Phi$ for $\Phi$-value analysis which we will address later.[79] Inspection of Fig. 7a in Crane et al., 2000 demonstrates that $\beta_{G(fold)(T)}$ increases with temperature for $T > T_S$ for both the wild type hYAP WW domain and its mutant W39F (~0.4 at 38 °C and ~0.8 at 78 °C). This pattern is once again repeated for the wild type and several mutants of Pin WW domain (Fig. 8 in Jäger et al., 2001) and more importantly for ΔNΔC Y11R W30F, a variant of FBP28 WW (inset in Fig. 4B in Nguyen et al., 2003). Nevertheless, all is not in agreement since the shapes of their $\beta_{G(fold)(T)}$ curves are distinctly different from what is expected from the formalism discussed in this article. This discrepancy most probably has to do with their use of Taylor expansion with three adjustable parameters to calculate the temperature-dependence of equilibrium stability and the Gibbs activation energies. While it is stated that the use of this non-classical model and the associated adjustable parameters in preference to the physically realistic Schellman formalism (which requires the model-independent calorimetrically determined value of $\Delta C_{pD-N}$)[6] makes little or no difference to the temperature-dependence of equilibrium stability over an extended temperature range, this may not be true for the activation energy. Once again in good agreement with prediction that $\beta_{G(unfold)(T)}$ must decrease with temperature for $T > T_S$, Tokmakoff and coworkers find that $\beta_{G(unfold)(T)}$ for ubiquitin decreases with temperature (0.77 at 53 °C and 0.67 at 67 °C).[85] Note that although raw data of the said groups and their conclusion that the position of the TSE shifts closer to the NSE as the temperature is raised for $T > T_S$ is in agreement with the predictions of the equations derived here, their Hammond-postulate-based inference of the structure of the TSE is flawed from the perspective of the parabolic approximation.

Now, at the midpoint of thermal ($T_m$) or cold denaturation ($T_c$), $\Delta G_{D\text{-}N(T)} = 0$; therefore, Eqs. (1) and (2) become

$$m_{\text{TS-D}(T)}\Big|_{T=T_c,T_m} = \frac{m_{\text{D-N}}\left(\omega - \sqrt{\alpha\omega}\right)}{(\omega - \alpha)} \Rightarrow \beta_{\text{T(fold)}(T)}\Big|_{T=T_c,T_m} = \frac{\omega - \sqrt{\alpha\omega}}{\omega - \alpha} \quad (36)$$

$$m_{\text{TS-N}(T)}\Big|_{T=T_c,T_m} = \frac{m_{\text{D-N}}\left(\sqrt{\alpha\omega} - \alpha\right)}{(\omega - \alpha)} \Rightarrow \beta_{\text{T(unfold)}(T)}\Big|_{T=T_c,T_m} = \frac{\sqrt{\alpha\omega} - \alpha}{\omega - \alpha} \quad (37)$$

Substituting Eqs. (36) and (37), and $\sqrt{\varphi_{(T)}}\Big|_{T=T_c,T_m} = \sqrt{\lambda\omega} = m_{\text{D-N}}\sqrt{\alpha\omega}$ in (33) and (35), respectively, and simplifying gives

$$\beta_{\text{G(fold)}(T)}\Big|_{T=T_c,T_m} = \frac{\sqrt{\alpha\omega} - \alpha}{\omega - \alpha} \equiv \beta_{\text{T(unfold)}(T)}\Big|_{T=T_c,T_m} \quad (38)$$

$$\beta_{\text{G(unfold)}(T)}\Big|_{T=T_c,T_m} = \frac{\omega - \sqrt{\alpha\omega}}{\omega - \alpha} \equiv \beta_{\text{T(fold)}(T)}\Big|_{T=T_c,T_m} \quad (39)$$

Simply put, at the midpoint of cold or heat denaturation, the position of the TSE relative to the DSE along the entropic RC is identical to the position of the TSE relative to the NSE along the SASA-RC (**Figure 19A**). Similarly, the position of the TSE relative to the NSE along the entropic RC is identical to the position of the TSE relative to the DSE along the SASA-RC (**Figure 19B**). Dividing Eq. (38) by (39) gives

$$\frac{\beta_{\text{G(fold)}(T)}}{\beta_{\text{G(unfold)}(T)}}\Bigg|_{T=T_c,T_m} = \frac{\beta_{\text{T(unfold)}(T)}}{\beta_{\text{T(fold)}(T)}}\Bigg|_{T=T_c,T_m} \Rightarrow \frac{-\Delta S_{\text{TS-D}(T)}}{\Delta S_{\text{TS-N}(T)}}\Bigg|_{T=T_c,T_m} = \frac{m_{\text{TS-N}(T)}}{m_{\text{TS-D}(T)}}\Bigg|_{T=T_c,T_m} \quad (40)$$

This seemingly obvious relationship has far deeper physical meaning. Simplifying further and recasting gives

$$\frac{\Delta S_{\text{N-TS}(T)}}{\Delta S_{\text{TS-D}(T)}}\Bigg|_{T=T_c,T_m} = \frac{m_{\text{TS-D}(T)}}{m_{\text{TS-N}(T)}}\Bigg|_{T=T_c,T_m} = \sqrt{\frac{\omega}{\alpha}}\Bigg|_{T=T_c,T_m} = \sqrt{\frac{\sigma^2_{\text{DSE}(T)}}{\sigma^2_{\text{NSE}(T)}}}\Bigg|_{T=T_c,T_m} = \frac{\sigma_{\text{DSE}(T)}}{\sigma_{\text{NSE}(T)}}\Bigg|_{T=T_c,T_m} \quad (41)$$

Thus, at the temperatures $T_c$ and $T_m$ where the concentration of the DSE and the NSE are identical, the ratio of the slopes of the folding and unfolding arms of the chevron determined at the said temperatures are a measure of the ratio of the change in entropies for the partial folding reactions $[TS] \rightleftharpoons N$ and $D \rightleftharpoons [TS]$, or the square root of the ratio of the Gaussian variances of the DSE ($\sigma^2_{DSE(T)}$) and the NSE ($\sigma^2_{NSE(T)}$) along the SASA-RC, or equivalently, the ratio of the standard deviations of the DSE ($\sigma_{DSE(T)}$) and the NSE ($\sigma_{NSE(T)}$) Gaussians (**Figure 19−figure supplement 1**; see Paper-I for the relationship between force constants, Gaussian variances and equilibrium stability). A corollary is that irrespective of the primary sequence, or the topology of the native state, or the residual structure in the DSE, if for a spontaneously folding two-state system at constant pressure and solvent conditions it is found that at a certain temperature the ratio of the distances by which the denatured and the native conformers must travel from the mean of their ensemble to reach the TSE along the SASA RC is identical to the ratio of the standard deviations of the Gaussian distribution of the SASA of the conformers in the DSE and the NSE, then at this temperature the Gibbs energy of unfolding or folding must be zero.

As an aside, the reader will note that $\beta_{G(fold)(T)}$ and $\beta_{G(unfold)(T)}$ are equivalent to the Brønsted exponents alpha and beta, respectively, in physical organic chemistry; and their classical interpretation is that they are a measure of the structural similarity of the transition state to either the reactants or the products.[81] If the introduction of a systematic perturbation (often a change in structure *via* addition or removal of a substituent, pH, solvent etc.) generates a reaction-series, and if for this reaction-series it is found that alpha is close to zero (or beta close to unity), then it implies that the energetics of the transition state is perturbed to the same extent as that of the reactant, and hence inferred that the structure of the transition state is very similar to that of the reactant. Conversely, if alpha is close to unity (or beta is almost zero), it implies that the energetics of the transition state is perturbed to the same extent as the product, and hence inferred that the transition state is structurally similar to the product. Although the Brønsted exponents in many cases can be invariant with the degree of perturbation (i.e., a constant slope leading to linear free energy relationships),[70,86] this is not necessarily true, especially if the degree of perturbation is substantial (Fig. 3 in Cohen and Marcus, 1968; Fig. 1 in Kresge, 1975).[14,72,81] Further, this seemingly straightforward and logical Hammond-postulate-based conversion of Brønsted exponents to similarity or dissimilarity of the structure of the transition states to either of the ground states nevertheless fails for those systems with Brønsted exponents greater than unity and less than zero (see page 1897 in Kresge, 1974).[24,81,87-91]

To summarise, a comparison of the position of the TSE along the solvent ($\beta_{T(T)}$), heat capacity ($\beta_{H(T)}$), and entropic ($\beta_{G(T)}$) RCs leads to three important general conclusions (**Figure 20**): (*i*) as long as $\Delta SASA_{D-N}$ is large, and by extension $\Delta C_{pD-N}$ is large and positive, the position of the TSE relative to the ground states along the various RCs is neither constant nor a simple linear function of temperature when investigated over a large temperature range; (*ii*) for a given temperature, the position of the TSE along the RC depends on the choice of the RC; and (*iii*) although the algebraic sum of $\beta_{T(fold)(T)}$ and $\beta_{T(unfold)(T)}$, $\beta_{H(fold)(T)}$ and $\beta_{H(unfold)(T)}$, and $\beta_{G(fold)(T)}$ and $\beta_{G(unfold)(T)}$ must be unity for a two-state system for any particular temperature, individually they can be positive, negative, or zero. Consequently, the notion that the atomic structure of the transiently populated reaction-states in protein folding can be inferred from their position along the said RCs is flawed.[78]

## Temperature-dependence of Φ-values

Φ-value analysis is a variation of the Brønsted procedure introduced by Fersht and co-workers which when properly implemented claims to provide a near-atomic-level description of the transiently populated reaction-states in protein folding.[79,80] In this procedure, the primary sequence of the target protein is modified using protein engineering, and the effect of these perturbations are quantified through a parameter Φ ($0 \leq \Phi \leq 1$) which by definition is the ratio of mutation-induced change in the Gibbs activation energy of folding/unfolding to the corresponding change in equilibrium stability. According to the canonical formulation, when $\Phi_{F(T)} = 0$ (Φ-value for folding), it implies that the energetics of the TSE is perturbed to the same extent as that of the DSE upon mutation, and hence *inferred* that the said reaction-states are structurally identical with respect to the site of mutation. In contrast, when $\Phi_{F(T)} = 1$, it implies that the energetics of the TSE is perturbed to the same extent as that of the NSE, and hence *inferred* that the structure at the site of mutation is identical in both the TSE and the NSE. Partial Φ-values are difficult to interpret and are thought to be due to partially developed interactions in the TSE, or multiple routes to the TSE. Thus, while Φ *per se* is the slope a two-point Brønsted plot, the conversion of this value to relative-structure is based on the Hammond postulate and the canonical range: The Hammond postulate provides the licence to infer structure from energetics, and the canonical scale enables one to infer how similar or dissimilar the TSE is to either the DSE or the NSE. Assuming that the prefactor is identical for the wild type and the mutant proteins, we may write for the partial folding ($D \rightleftharpoons [TS]$) and unfolding ($N \rightleftharpoons [TS]$) reactions

$$\Phi_{F(T)} = \frac{RT \ln\left(k_{f(\text{wt})(T)}/k_{f(\text{mut})(T)}\right)}{\Delta G_{\text{N-D(mut)}(T)} - \Delta G_{\text{N-D(wt)}(T)}} = \frac{\Delta G_{\text{TS-D(mut)}(T)} - \Delta G_{\text{TS-D(wt)}(T)}}{\Delta G_{\text{N-D(mut)}(T)} - \Delta G_{\text{N-D(wt)}(T)}} \tag{42}$$

$$\Phi_{U(T)} = \frac{RT \ln\left(k_{u(\text{mut})(T)}/k_{u(\text{wt})(T)}\right)}{\Delta G_{\text{D-N(wt)}(T)} - \Delta G_{\text{D-N(mut)}(T)}} = \frac{\Delta G_{\text{TS-N(wt)}(T)} - \Delta G_{\text{TS-N(mut)}(T)}}{\Delta G_{\text{D-N(wt)}(T)} - \Delta G_{\text{D-N(mut)}(T)}} \tag{43}$$

where the subscripts "wt" and "mut" denote the reference or the wild type, and the structurally perturbed protein, respectively, and $\Phi_{U(T)}$ is the Φ-value for unfolding. Inspection of Eqs. (42) and (43) shows that for a two-state system, $\Phi_{F(T)} + \Phi_{U(T)} = 1$. Now, although the primary sequence is intact in thermal denaturation experiments, we can readily calculate the temperature-dependence of Φ values for folding and unfolding using the protein at one unique temperature as the internal reference or the wild type, and protein at all the rest of the temperatures as the mutants. Thus, if the protein at $T_S$ is defined as the internal reference or the wild type, Eqs. (42) and (43) become

$$\Phi_{F(\text{internal})(T)} = \frac{RT \ln k_{f(T)}\big|_{T=T_S} - RT \ln k_{f(T)}}{\Delta G_{\text{N-D}(T)} - \Delta G_{\text{N-D}(T_S)}} = \frac{\Delta G_{\text{TS-D}(T)} - \Delta G_{\text{TS-D}(T_S)}}{\Delta G_{\text{N-D}(T)} - \Delta G_{\text{N-D}(T_S)}} \tag{44}$$

$$\Phi_{U(\text{internal})(T)} = \frac{RT \ln k_{u(T)} - RT \ln k_{u(T)}\big|_{T=T_S}}{\Delta G_{\text{D-N}(T_S)} - \Delta G_{\text{D-N}(T)}} = \frac{\Delta G_{\text{TS-N}(T_S)} - \Delta G_{\text{TS-N}(T)}}{\Delta G_{\text{D-N}(T_S)} - \Delta G_{\text{D-N}(T)}} \tag{45}$$

Similarly, if the protein at $T_m$ is defined as the internal reference or the wild type, Eqs. (42) and (43) become

$$\Phi_{F(\text{internal})(T)} = \frac{\Delta G_{\text{TS-D}(T)} - \Delta G_{\text{TS-D}(T_m)}}{\Delta G_{\text{N-D}(T)} - \Delta G_{\text{N-D}(T_m)}} = \frac{\Delta G_{\text{TS-D}(T)} - \Delta G_{\text{TS-D}(T_m)}}{\Delta G_{\text{N-D}(T)}} = \frac{\Delta G_{\text{TS-D}(T)} - x}{y} \tag{46}$$

$$\Phi_{U(\text{internal})(T)} = \frac{\Delta G_{\text{TS-N}(T_m)} - \Delta G_{\text{TS-N}(T)}}{\Delta G_{\text{D-N}(T_m)} - \Delta G_{\text{D-N}(T)}} = \frac{\Delta G_{\text{TS-N}(T_m)} - \Delta G_{\text{TS-N}(T)}}{-\Delta G_{\text{D-N}(T)}} = \frac{x - \Delta G_{\text{TS-N}(T)}}{y} \tag{47}$$

where $x = \Delta G_{\text{TS-D}(T_m)} = \Delta G_{\text{TS-N}(T_m)}$ and $y = \Delta G_{\text{N-D}(T)} \equiv -\Delta G_{\text{D-N}(T)}$ (the denominator reduces to a single quantity since $\Delta G_{\text{D-N}(T_m)} \equiv -\Delta G_{\text{N-D}(T_m)} = 0$). The parameters $\Phi_{\text{F(internal)}(T)}$ and $\Phi_{\text{U(internal)}(T)}$ (which are obviously undefined for the reference temperatures) when interpreted according to the canonical Φ-value framework (i.e., the notion that $0 \leq \Phi \leq 1$) are a measure of the *global* similarity or dissimilarity of the structure of the TSE to either the DSE or the NSE. Thus, if $\Phi_{\text{F(internal)}(T)} = 0$, it implies that the energetics of the TSE is perturbed to the same extent as that of the DSE upon a perturbation in temperature, and hence inferred that the global structure of the TSE is identical to that of the DSE. Conversely, if $\Phi_{\text{F(internal)}(T)} = 1$, it implies that the energetics of the TSE is perturbed to the same extent as the NSE upon a perturbation in temperature, and hence inferred that the global structure of the TSE is identical to that of the NSE.

Inspection of **Figures 21** and **Figure 21−figure supplements 1, 2, 3** and **4** immediately demonstrates that: (*i*) irrespective of which temperature is defined as the internal reference (i.e., the wild type), $\Phi_{\text{F(internal)}(T)}$ must be a minimum and $\Phi_{\text{U(internal)}(T)}$ must be a maximum at $T_S$ (see **Appendix**); (*ii*) the magnitude of $\Phi_{\text{F(internal)}(T)}$ is always the least, and the magnitude of $\Phi_{\text{U(internal)}(T)}$ is always the greatest when the protein at $T_S$ is defined as the reference or the wild type protein, and any deviation in the definition of the reference temperature from $T_S$ must lead to a uniform increase in $\Phi_{\text{F(internal)}(T)}$ and a uniform decrease in $\Phi_{\text{U(internal)}(T)}$ for all temperatures; (*iii*) although the algebraic sum of $\Phi_{\text{F(internal)}(T)}$ and $\Phi_{\text{U(internal)}(T)}$ is unity for all temperatures, the notion that they must independently be restricted to $0 \leq \Phi \leq 1$ is flawed; and (*iv*) although both Leffler $\beta_{G(T)}$ and Fersht Φ values are derived from changes in Gibbs activation energies for folding and unfolding relative to changes in equilibrium stability upon a perturbation in temperature, their response is not the same since the equations that govern their behaviour are not the same. While the magnitude of the Leffler $\beta_{G(T)}$ is independent of the reference owing to it being the ratio of the derivatives of the change in Gibbs energies with respect to temperature, the magnitude of $\Phi_{\text{(internal)}(T)}$ is dependent on the definition of the reference state. For example, if the protein at $T_S$ is defined as the wild type, then $\beta_{G(\text{fold})(T)} \approx \Phi_{\text{F(internal)}(T)}$ and $\beta_{G(\text{unfold})(T)} \approx \Phi_{\text{U(internal)}(T)}$ around the temperature of maximum stability; but as the temperature deviates from $T_S$, $\beta_{G(\text{fold})(T)}$ increases far more steeply than $\Phi_{\text{F(internal)}(T)}$, and $\beta_{G(unfold)(T)}$ decreases far more steeply than $\Phi_{U(internal)(T)}$ such that for $T \neq T_S$ we have $\beta_{G(fold)(T)} > \Phi_{F(internal)(T)}$ and $\beta_{G(unfold)(T)} < \Phi_{U(internal)(T)}$ (**Figure 21−figure supplement 3**). In contrast, if the protein at $T_m$ is defined as the wild type, then we have: (*i*) $\beta_{G(fold)(T)} < \Phi_{F(internal)(T)}$ for $T_c < T < T_m$ and $\beta_{G(fold)(T)} > \Phi_{F(internal)(T)}$ for $T < T_c$ and $T > T_m$; and (*ii*) $\beta_{G(unfold)(T)} > \Phi_{U(internal)(T)}$ for $T_c < T < T_m$ and $\beta_{G(unfold)(T)} < \Phi_{U(internal)(T)}$ for $T < T_c$ and $T > T_m$ (**Figure 21−figure supplement 4**). The point we are trying to make is that a comparison of the position of the TSE along Leffler $\beta_{G(T)}$ and $\Phi_{(internal)(T)}$ RCs is not straightforward since both $\beta_{G(T)}$ and $\Phi_{(internal)(T)}$ are temperature-dependent, and importantly respond differently to temperature-perturbation; and even if we restrict the comparison to one particular temperature, the answer we get is still subjective since the magnitude of $\Phi_{(internal)(T)}$ is dependent on how we define the wild type.[92]

Although the mathematical formalism for why the extrema of $\Phi_{F(internal)(T)}$ (which is a minimum) and $\Phi_{U(internal)(T)}$ (which is a maximum) must always occur precisely at $T_S$ has been shown in the appendix, it is instructive to examine the same graphically. Inspection of **Figure 21−figure supplements 5, 6** and **7** demonstrates that this is a consequence of $\Delta G_{TS-D(T)}$ and $\Delta G_{N-D(T)}$ being a minimum, and $\Delta G_{TS-N(T)}$ and $\Delta G_{D-N(T)}$ being a maximum at $T_S$. Subtracting the reference Gibbs energies from the numerator and the denominator (Eq. (44)) has the effect of lowering the $\Delta G_{TS-D(T)}$ curve and raising the $\Delta G_{N-D(T)}$, such that the value of the said curves are zero at the reference temperature, but the shapes of the curves are not altered in any way (**Figure 21−figure supplement 5**). On the other hand, for $\Delta G_{TS-N(T)}$ and $\Delta G_{D-N(T)}$ curves (Eq. (45)), apart from the value of the curves becoming zero at the reference, it causes them to flip vertically (**Figure 21−figure supplement 6**). Consequently, if we divide the transformed Gibbs activation energies by the transformed equilibrium Gibbs energies, we end up with $\Phi_{F(internal)(T)}$ and $\Phi_{U(internal)(T)}$ which are a minimum and a maximum, respectively, at $T_S$ (**Figure 21−figure supplement 7**).

Now that the process that leads to the temperature-dependence of $\Phi$ has been addressed, the question is "Can we infer the structure of the TSE as being similar to either the DSE or the NSE from these data?" The answer is "no" for several reasons. First, as argued earlier, the Hammond postulate cannot be valid for protein folding; and because the structural interpretation of $\Phi$ values is based on the Hammond postulate, it too must be deemed fallacious. Second, even if we accept the premise that Hammond postulate is applicable to protein folding, the inference that the global structure of the TSE as being denatured-like for $\Phi_{F(internal)(T)} = 0$, and native-like for $\Phi_{F(internal)(T)} = 1$ is flawed since $\Phi$ values need not necessarily be restricted to $0 \leq \Phi \leq 1$ (**Figure 21−figure supplement 2**). Third, even if we summarily exclude those wild types that lead to anomalous $\Phi$ values as being unsuitable for $\Phi$ analysis, we still have a problem since even within the restricted set of wild types that yield $0 \leq \Phi \leq 1$, their magnitude depends on the definition of the wild type; consequently, for the same temperature, the degree of structure in the TSE relative to that in the DSE appears to increase as the definition of the wild type deviates from $T_S$ (**Figure 21−figure supplement 1**). If we try to circumvent this interpretational problem by arguing that the "inference of the structure of the TSE" is always relative to the residual structure in the DSE, and that changing the definition of what constitutes the wild type will invariably affect $\Phi$ values, then we can't really say much about the structure of the TSE without first solving the structure of the DSE. Fourth, even if through a judicious combination of various structural and biophysical methods (residual dipolar couplings, paramagnetic relaxation enhancement, small angle X-ray scattering, single molecule spectroscopy etc.), and computer simulation, we are able to determine the residual structure in the DSE,[93-96] the structural interpretation of $\Phi$ values leads to physically unrealistic scenarios. For example, inspection of **Figure 21A** shows that around room temperature (298 K) $\Phi_{F(internal)(T)} \approx 0.18$. A canonical interpretation of this number implies that the global structure of the TSE is very similar to that of the DSE. However, inspection of **Figure 2−figure supplement 1A** shows that the denatured conformer has buried ~70% of the total SASA to reach the TSE (i.e., advanced by about 70% along the SASA-RC). Similarly, inspection of **Figure 5A** shows that $\Delta G_{TS-D(T)} = 2.6$ kcal.mol$^{-1}$ at 298 K (note that this is not a small number that can be ignored since $\Delta G_{D-N(T)} = 2.1$ kcal.mol$^{-1}$ at 298 K). Further, we have shown earlier in the section on the "Inapplicability of the Hammond postulate to protein folding," that even when two reaction-states have identical SASA, Gibbs energies, enthalpies, and entropies, they need not necessarily have identical structure. Thus, the question is: How can we conclude with any measure of certainty that the global structure of the TSE is very similar to that of the DSE at 298 K when they have such a large difference in SASA, and a substantial difference in Gibbs energy? To illustrate why it is difficult to rationalize the theoretical basis of $\Phi$ analysis, it is instructive to directly examine the ratio of the Gibbs activation energies and the difference in Gibbs energy between the ground states (**Figure 21−figure supplement 8**). It is immediately apparent that the ratios are a complex function of temperature; and although we can readily provide an explanation for the particular features of these complex dependences, it is difficult to see how subtracting reference energies from the numerator and denominator of the ratios $\Delta G_{\text{TS-D}(T)}/\Delta G_{\text{N-D}(T)}$ and $\Delta G_{\text{TS-N}(T)}/\Delta G_{\text{D-N}(T)}$ allows us to divine the structure of the TSE to a near-atomic resolution. This is once again readily apparent from the complex non-linear relationship between equilibrium stability and the rate constants (**Figure 21−figure supplement 9**).

To further illuminate the difficulty in rationalizing the Φ-value procedure, it is instructive to apply Eqs. (44) and (45) to treat enthalpies. Thus, for the partial folding ($D \rightleftharpoons [TS]$) and unfolding ($N \rightleftharpoons [TS]$) reactions we have

$$\Phi_{H_F(\text{internal})(T)} = \frac{\Delta H_{\text{TS-D}(T)} - \Delta H_{\text{TS-D}(T_S)}}{\Delta H_{\text{N-D}(T)} - \Delta H_{\text{N-D}(T_S)}} \tag{48}$$

$$\Phi_{H_U(\text{internal})(T)} = \frac{\Delta H_{\text{TS-N}(T_S)} - \Delta H_{\text{TS-N}(T)}}{\Delta H_{\text{D-N}(T_S)} - \Delta H_{\text{D-N}(T)}} \tag{49}$$

where the parameters $\Phi_{H_F(\text{internal})(T)}$ and $\Phi_{H_U(\text{internal})(T)}$ are the "*enthalpic analogues*" of $\Phi_{F(\text{internal})(T)}$ and $\Phi_{U(\text{internal})(T)}$, respectively (the subscript "H" indicates we are using enthalpy instead of Gibbs energy), when the protein at the temperature $T_S$ is defined as the wild type. Now, if we apply an analogous version of the canonical interpretation given by Fersht and coworkers, it implies that when $\Phi_{H_F(\text{internal})(T)} = 0$, the enthalpy of the TSE is perturbed to the same extent as that of the DSE upon a perturbation in temperature; and when $\Phi_{H_F(\text{internal})(T)} = 1$, it implies that the enthalpy of the TSE is perturbed to the same extent as that of the NSE. It is easy to see that just as $\Phi_{F(\text{internal})(T)}$ and $\Phi_{U(\text{internal})(T)}$ are the *Fersht-analogues* of the Leffler $\beta_{G(\text{fold})(T)}$ and $\beta_{G(\text{unfold})(T)}$, respectively (see entropic RC), the parameters $\Phi_{H_F(\text{internal})(T)}$ and $\Phi_{H_U(\text{internal})(T)}$ are similarly the Fersht-analogues of the Leffler $\beta_{H(\text{fold})(T)}$ and $\beta_{H(\text{unfold})(T)}$, respectively (see heat capacity RC).

Inspection of **Figure 22** and its supplements immediately demonstrates that the same anomalies that prevent a straightforward structural interpretation of $\Phi_{F(\text{internal})(T)}$ and $\Phi_{U(\text{internal})(T)}$ also emerge if we try to assign structure to their enthalpic analogues, $\Phi_{H_F(\text{internal})(T)}$ and $\Phi_{H_U(\text{internal})(T)}$. First, although the algebraic sum of $\Phi_{H_F(\text{internal})(T)}$ and $\Phi_{H_U(\text{internal})(T)}$ is unity for all temperatures, they need not independently be restricted to a canonical range of $0 \leq \Phi \leq 1$ (**Figure 22**). Second, the magnitude of $\Phi_{H_F(\text{internal})(T)}$ and $\Phi_{H_U(\text{internal})(T)}$ are dependent on the definition of the wild type (**Figure 22−figure supplement 1**). Third, changing the definition of the wild type has a dramatic effect on the relationship between the Leffler $\beta_{H(T)}$ and its analogue, the Fersht $\Phi_{H(\text{internal})(T)}$. Consequently, the question of whether Leffler $\beta_{H(T)}$ underestimates or overestimates structure is dependent on how we analyse the system (**Figure 22−figure supplements 2** and **3**). Fourth, just as the temperature-dependent position of the TSE relative to the ground states depends on the choice of the RC (**Figure 20**), we see that $\Phi_{(\text{internal})(T)}$ and its enthalpic analogue, $\Phi_{H(\text{internal})(T)}$, change at different rates upon a perturbation in temperature (**Figure 22−figure supplement 4**). The difficulty in rationalizing how subtracting reference values from the numerator and the denominator of Eqs. (48) and (49) can yield residue-level information is once again apparent from the complex dependence of the ratios $\partial \ln k_{f(T)} / \partial \ln K_{\text{N-D}(T)} = \Delta H_{\text{TS-D}(T)} / \Delta H_{\text{N-D}(T)}$ and $\partial \ln k_{u(T)} / \partial \ln K_{\text{D-N}(T)} = \Delta H_{\text{TS-N}(T)} / \Delta H_{\text{D-N}(T)}$ on temperature (**Figure 22−figure supplement 5**).

## Comparison of theoretical and experimental Φ-values obtained from structural perturbation across 31 two-state systems

Given that the framework of Φ-value analysis was primarily developed to be used in conjunction with structural rather than temperature perturbation, and despite its anomalies has been used extensively for more than twenty years to divine the structures of the TSEs of not just globular but also membrane proteins, it is imperative to demonstrate that the notion that the structure of the TSE cannot be inferred from Φ-values is also valid for structural perturbation.[97-101] Although a detailed reappraisal is beyond the scope of this article and will be presented elsewhere, because we have questioned the validity of Φ analysis, one is compelled to provide some justification in this article.

Consider the wild type of a hypothetical two-state folder whose equilibrium stability and the mean length of the RC at constant temperature, pressure and solvent conditions are given by $\Delta G_{\text{D-N}(T)}$ = 6 kcal.mol$^{-1}$ and $m_{\text{D-N}}$ = 2 kcal.mol$^{-1}$.M$^{-1}$, respectively. Although not necessarily true and addressed elsewhere, to limit the number of hypothetical scenarios to a manageable number, we will assume that the force constants of the DSE and the NSE-parabolas of the wild type and all its mutants are given by α = 1 M$^2$.mol.kcal$^{-1}$ and ω = 30 M$^2$.mol.kcal$^{-1}$. The effect of single point mutations on the wild type may be classified into a total of five unique scenarios (**Figure 23A**).

**Case I** (**Quadrant *x*2**): The introduced mutation causes a concomitant decrease in both the stability and the mean length of the RC (i.e., $\Delta G_{\text{D-N}(T)(\text{wt})} > \Delta G_{\text{D-N}(T)(\text{mut})}$ and $m_{\text{D-N}(\text{wt})} > m_{\text{D-N}(\text{mut})}$). This is equivalent to the introduced mutation causing the separation between the vertices of the DSE and the NSE-parabolas along the abscissa and ordinate to decrease (**Figure 23−figure supplement 1A**).

**Case II** (**Quadrant *y*1**): The introduced mutation causes a decrease in stability but concomitantly causes an increase in the mean length of the RC (i.e., $\Delta G_{\text{D-N}(T)(\text{wt})} > \Delta G_{\text{D-N}(T)(\text{mut})}$ and $m_{\text{D-N}(\text{wt})} < m_{\text{D-N}(\text{mut})}$). This is equivalent to the mutation causing a decrease in the separation between the vertices of the DSE and the NSE-parabolas along the ordinate, but an increase along the abscissa (**Figure 23−figure supplement 1B**).

**Case III** (**Quadrant *x*1**): The introduced mutation leads to an increase in stability but concomitantly causes a decrease in the mean length of the RC (i.e., $\Delta G_{\text{D-N}(T)(\text{wt})} < \Delta G_{\text{D-N}(T)(\text{mut})}$ and $m_{\text{D-N}(\text{wt})} > m_{\text{D-N}(\text{mut})}$). This is equivalent to the mutation causing an increase in the separation between the vertices of the DSE and the NSE-parabolas along the ordinate, but a decrease along the abscissa (**Figure 23−figure supplement 1C**).

**Case IV** (**Quadrant *y*2**): The introduced mutation leads to a concomitant increase in both the stability and the mean length of the RC (i.e., $\Delta G_{\text{D-N}(T)(\text{wt})} < \Delta G_{\text{D-N}(T)(\text{mut})}$ and $m_{\text{D-N}(\text{wt})} < m_{\text{D-N}(\text{mut})}$). This is equivalent to the mutation causing an increase in the separation between the vertices of the DSE and the NSE-parabolas along the ordinate and the abscissa (**Figure 23−figure supplement 1D**).

**Case V**: The introduced mutation leads to a change in stability but has no effect on the mean length of the RC ($m_{\text{D-N}(\text{wt})} = m_{\text{D-N}(\text{mut})}$). This is equivalent to the mutation causing an increase or a decrease in the separation between the vertices of the DSE and the NSE-parabolas along the ordinate, but the separation along the abscissa is invariant (**Figure 23−figure supplement 2**).

In summary, what we done is taken a pair of intersecting parabolas of differing curvature ($\omega > \alpha$), and systematically varied the separation between their vertices along the abscissa ($m_{\text{D-N}}$) and ordinate ($\Delta G_{\text{D-N}(T)}$) without changing the curvature of the parabolas. Once this is done, we can calculate *a priori* the position of the *curve-crossings* relative to the vertex of the DSE-parabola along the abscissa (i.e., $m_{\text{TS-D}(T)}$; Eq. (1)) and ordinate (i.e., $\Delta G_{\text{TS-D}(T)}$; Eq. (3)). Once the $\Delta G_{\text{TS-D}(T)}$ values for all combinations of $\Delta G_{\text{D-N}(T)}$ and $m_{\text{D-N}}$ are obtained (each combination is equivalent to a point mutation), $\Phi_{F(T)}$ values can be readily calculated using Eq. (50) by arbitrarily choosing one particular combination of $\Delta G_{\text{D-N}(T)}$ (= 6 kcal.mol$^{-1}$) and $m_{\text{D-N}}$ (= 2 kcal.mol$^{-1}$.M$^{-1}$) as the wild type.

$$\Phi_{F(\text{theory})(T)} = \frac{\Delta G_{\text{TS-D(mut)}(T)} - \Delta G_{\text{TS-D(wt)}(T)}}{\Delta G_{\text{N-D(mut)}(T)} - \Delta G_{\text{N-D(wt)}(T)}} = \frac{\alpha\left[\left(m_{\text{TS-D(mut)}(T)}\right)^2 - \left(m_{\text{TS-D(wt)}(T)}\right)^2\right]}{\Delta\Delta G_{\text{D-N(wt-mut)}(T)}} \quad (50)$$

**Figure 23A** which has been generated by plotting the theoretical $\Phi_{F(T)}$ values as a function of $\Delta\Delta G_{\text{D-N(wt-mut)}(T)}$ leads to two important conclusions: (*i*) $\Phi_{F(T)}$ values are not restricted to $0 \leq \Phi \leq 1$, and that the perceived unusualness of anomalous or non-classical $\Phi$ values is a consequence of flawed canonical limits; and (*ii*) the magnitude of $\Phi_{F(T)}$ values decrease as the difference in stability between the wild type and the mutant proteins increase, and at once debunks the idea that one must use an arbitrary $\Delta\Delta G_{\text{D-N(wt-mut)}(T)}$ cut-off ($\pm$ 0.6 kcal.mol$^{-1}$ according to the Fersht lab, and $\pm$ 1.7 kcal.mol$^{-1}$ according to Sanchez and Kiefhaber) for $\Phi_{F(T)}$ values to be interpretable.[98,102] While it is true that $\Phi$ values would be error prone when $|\Delta\Delta G_{\text{D-N(wt-mut)}(T)}|$ is less than the error with which one can determine $\Delta G_{\text{D-N}(T)}$ of both the wild type and the mutant proteins (typically about $\pm$ 5-10% of $\Delta G_{\text{D-N}(T)}$),[103] the increase in the magnitude of $\Phi_{F(T)}$ values when $\Delta\Delta G_{\text{D-N(wt-mut)}(T)}$ approaches zero (the vertical asymptotes) is a mathematical certainty and not because of error as is commonly argued. Nevertheless, because these conclusions are based on the results of a model that is purely hypothetical, they would naturally be meaningless without experimental validation. Thus, as a test of the hypothesis, experimental $\Phi_{F(T)}$ values in water were calculated according to Eq. (51) using published kinetic data of a total of 1064 proteins (1035 mutants + 29 wild types) from 31 two-state systems (details of the systems analysed will be provided elsewhere).

$$\Phi_{\text{F(experimental)}(T)} = \frac{RT\ln\left(k_{f(\text{wt})(T)}/k_{f(\text{mut})(T)}\right)}{RT\left[\ln\left(\frac{k_{f(\text{wt})(T)}}{k_{u(\text{wt})(T)}}\right) - \ln\left(\frac{k_{f(\text{mut})(T)}}{k_{u(\text{mut})(T)}}\right)\right]} = \frac{\Delta\Delta G_{\text{TS-D(mut-wt)}(T)}}{\Delta\Delta G_{\text{D-N(wt-mut)}(T)}} \tag{51}$$

The remarkable agreement between theoretical prediction and experimental $\Phi_{F(T)}$ values is immediately apparent from an overlay of the said datasets (**Figure 23B**), and serves as arguably one of the most rigorous tests of the hypothesis for the following reasons: (1) The space enclosed by the curves in **Figure 23A** is complex and restricted. Therefore, if the experimental $\Phi_{F(T)}$ values fall within this restricted theoretical space it would be highly unlikely for it to be purely due to some dramatic coincidence. (2) The sample size of experimental dataset is sufficiently large (1035 mutations), and the two-state systems investigated include α, β, and α/β proteins (note that α and β refer to secondary structure in this context and not to the force constant of the DSE or the Tanford beta value, respectively), with size ranging from 37 to 107 residues. (3) The published kinetic data used to calculate experimental $\Phi_{F(T)}$ values were acquired by various labs under varying solvent conditions (buffers, co-solvents and pH; denaturant is either guanidine hydrochloride or urea) and temperature (as low as 278 K to as high as 301.16 K), over a period of about two decades using a variety of experimental methods, including infrared laser-induced and electrical discharge temperature-jump relaxation measurements, stopped flow and manual mixing experiments, and lineshape analysis of exchange-broadened NMR resonances. These results, including those on the temperature-dependence of $\Phi_{F(T)}$ values lead to an important conclusion: Because the canonical scale itself has no basis, Φ-value-based interpretation of the structure of the transiently populated protein reaction-states is dubious. Further, because we are able to fit 1035 experimental Φ-values from 31 two-state systems to a set of theoretical curves despite having zero structural information, a corollary is that one cannot draw any conclusion regarding the structure of the TSE purely from mutation-induced change in the rate constants.

## CONCLUDING REMARKS

Although the temperature-dependent behaviour of FBP28 WW was analysed in great detail using the theory developed in the Papers I and II, and novel conclusions have been drawn, this is by no means sufficient since we have barely addressed the physical chemistry underlying the effect of temperature on the Gibbs energies, the enthalpies, the entropies, and the heat capacities of activation for folding and unfolding. These aspects will be dealt with in the accompanying articles. Further, there is a good reason why we have given little importance to the actual values of the reference temperatures and instead focussed on what they actually mean, and how they relate to each other. Although the remarks in Table 1 are valid for all reference temperatures, except for the values of the equilibrium reference temperatures ($T_c$, $T_H$, $T_S$, and $T_m$), the values for the rest of them can change depending on the values of the force constants. However, what will not change is the inter-relationship between them. The nature of this limitation will be addressed when the mechanism of action of denaturants is investigated.

## METHODS

The temperature-dependence of $\Delta G_{D-N(T)}$ of FBP28 WW wild type (**Figure 1**) was simulated according to Eq. (A1) using $T_m$ = 337.2 K, $\Delta H_{D-N(Tm)}$ = 26.9 kcal.mol$^{-1}$ and $\Delta C_{pD-N}$ = 417 cal.mol$^{-1}$.K$^{-1}$ (Table 1 in Petrovich et al., 2006).[4] The values of $k^0$ = 2180965 s$^{-1}$, $\alpha$ = 7.594 M$^2$.mol.kcal$^{-1}$, $\omega$ = 85.595 M$^2$.mol.kcal$^{-1}$, and $m_{D-N}$ = 0.82 kcal.mol$^{-1}$.M$^{-1}$ were extracted from the chevron of FBP28 WW (acquired at 283.16 K in 20 mM 3-[morpholino] propanesulfonic acid, ionic strength adjusted to150 mM with Na$_2$SO$_4$, pH 6.5) by fitting it to a modified chevron-equation using non-linear regression as described in Paper-I. The data required to simulate the chevron ($k_{f(H_2O)(T)}$, $k_{u(H_2O)(T)}$, $m_{TS-D(T)}$ and $m_{TS-N(T)}$) were taken from Table 4 in Petrovich et al., 2006.[4] Once the parameters $\Delta H_{D-N(Tm)}$, $T_m$, $\Delta C_{pD-N}$, $m_{D-N}$, the force constants $\alpha$ and $\omega$, and $k^0$ are known, the left-hand side of all the equations in this article may be readily calculated for any temperature. Note that the spring constants, $k^0$, $m_{D-N}$, and $\Delta C_{pD-N}$ are temperature-invariant.

## COMPETING FINANCIAL INTERESTS

The author declares no competing financial interests.

## COPYRIGHT INFORMATION

# APPENDIX

## The temperature-dependence of $\Delta G_{\text{D-N}(T)}$, $\Delta H_{\text{D-N}(T)}$, and $\Delta S_{\text{D-N}(T)}$ functions

The temperature-dependence of the change in Gibbs energy, enthalpy and entropy of two-state systems upon unfolding at equilibrium are given by[6]

$$\Delta H_{\text{D-N}(T)} = \Delta H_{\text{D-N}(T_m)} + \int_{T_m}^{T} \Delta C_{p\text{D-N}(T)}\, dT = \Delta H_{\text{D-N}(T_m)} + \Delta C_{p\text{D-N}}(T-T_m) \tag{A1}$$

$$\begin{aligned}\Delta S_{\text{D-N}(T)} &= \Delta S_{\text{D-N}(T_m)} + \int_{T_m}^{T} \frac{\Delta C_{p\text{D-N}(T)}}{T}\, dT = \Delta S_{\text{D-N}(T_m)} + \Delta C_{p\text{D-N}} \ln\left(\frac{T}{T_m}\right) \\ &= \left(\frac{\Delta H_{\text{D-N}(T_m)}}{T_m}\right) + \Delta C_{p\text{D-N}} \ln\left(\frac{T}{T_m}\right)\end{aligned} \tag{A2}$$

$$\Delta G_{\text{D-N}(T)} = \Delta H_{\text{D-N}(T_m)}\left(1-\frac{T}{T_m}\right) + \Delta C_{p\text{D-N}}(T-T_m) + T\Delta C_{p\text{D-N}} \ln\left(\frac{T_m}{T}\right) \tag{A3}$$

where $\Delta H_{\text{D-N}(T)}$, $\Delta H_{\text{D-N}(T_m)}$ and $\Delta S_{\text{D-N}(T)}$, $\Delta S_{\text{D-N}(T_m)}$ denote the equilibrium enthalpies and entropies of unfolding, respectively, at any given temperature, and at the midpoint of thermal denaturation ($T_m$), respectively, for a given two-state folder under defined solvent conditions. The temperature-invariant and the temperature-dependent difference in heat capacity between the DSE and NSE are denoted by $\Delta C_{p\text{D-N}}$ and $\Delta C_{p\text{D-N}(T)}$, respectively.

## The first derivatives of $m_{\text{TS-D}(T)}$, $m_{\text{TS-N}(T)}$, $\beta_{\text{T(fold)}(T)}$ and $\beta_{\text{T(unfold)}(T)}$ with respect to temperature

The first derivative of $m_{\text{TS-D}(T)}$ is given by

$$\frac{\partial m_{\text{TS-D}(T)}}{\partial T} = \frac{\Delta S_{\text{D-N}(T)}}{2\sqrt{\varphi}} = \frac{\Delta C_{p\text{D-N}}}{2\sqrt{\varphi}} \ln\left(\frac{T}{T_S}\right) \tag{A4}$$

Because $\beta_{\text{T(fold)}(T)} = m_{\text{TS-D}(T)}/m_{\text{D-N}}$, we also have

$$\frac{\partial \beta_{\text{T(fold)}(T)}}{\partial T} = \frac{1}{m_{\text{D-N}}} \frac{\partial m_{\text{TS-D}(T)}}{\partial T} = \frac{\Delta C_{p\text{D-N}}}{2m_{\text{D-N}}\sqrt{\varphi}} \ln\left(\frac{T}{T_S}\right) \tag{A5}$$

Since $\partial m_{\text{TS-D}(T)}/\partial T$ and $\partial \beta_{\text{T(fold)}(T)}/\partial T$ are physically undefined for $\varphi < 0$, their algebraic sign at any given temperature is determined by the $\ln(T/T_S)$ term. This leads to three scenarios: (*i*) for $T < T_S$ we have $\partial m_{\text{TS-D}(T)}/\partial T < 0$ and $\partial \beta_{\text{T(fold)}(T)}/\partial T < 0$; (*ii*) for $T > T_S$ we have $\partial m_{\text{TS-D}(T)}/\partial T > 0$ and $\partial \beta_{\text{T(fold)}(T)}/\partial T > 0$; and (*iii*) for $T = T_S$ we have $\partial m_{\text{TS-D}(T)}/\partial T = 0$ and $\partial \beta_{\text{T(fold)}(T)}/\partial T = 0$.

Because $m_{\text{TS-N}(T)} = (m_{\text{D-N}} - m_{\text{TS-D}(T)})$ for a two-state system, and $\beta_{\text{T(unfold)}(T)} = m_{\text{TS-N}(T)}/m_{\text{D-N}}$, we have

$$\frac{\partial m_{\text{TS-N}(T)}}{\partial T} = -\frac{\partial m_{\text{TS-D}(T)}}{\partial T} = -\frac{\Delta S_{\text{D-N}(T)}}{2\sqrt{\varphi}} = \frac{\Delta C_{p\text{D-N}}}{2\sqrt{\varphi}}\ln\left(\frac{T_S}{T}\right) \tag{A6}$$

$$\frac{\partial \beta_{\text{T(unfold)}(T)}}{\partial T} = \frac{1}{m_{\text{D-N}}}\frac{\partial m_{\text{TS-N}(T)}}{\partial T} = \frac{\Delta C_{p\text{D-N}}}{2m_{\text{D-N}}\sqrt{\varphi}}\ln\left(\frac{T_S}{T}\right) \tag{A7}$$

Eqs. (A6) and (A7) once again lead to three scenarios: (*i*) for $T < T_S$ we have $\partial m_{\text{TS-N}(T)}/\partial T > 0$ and $\partial \beta_{\text{T(unfold)}(T)}/\partial T > 0$; (*ii*) for $T > T_S$ we have $\partial m_{\text{TS-N}(T)}/\partial T < 0$ and $\partial \beta_{\text{T(unfold)}(T)}/\partial T < 0$; and (*iii*) for $T = T_S$ we have $\partial m_{\text{TS-N}(T)}/\partial T = 0$ and $\partial \beta_{\text{T(unfold)}(T)}/\partial T = 0$.

**The second derivatives of $m_{\text{TS-D}(T)}$ and $m_{\text{TS-N}(T)}$ with respect to temperature**

Differentiating Eq. (A4) with respect to temperature gives

$$\frac{\partial^2 m_{\text{TS-D}(T)}}{\partial T^2} = \frac{\partial}{\partial T}\left(\frac{\Delta S_{\text{D-N}(T)}}{2\sqrt{\varphi}}\right) = \frac{1}{2}\frac{\partial}{\partial T}\left(\frac{\Delta S_{\text{D-N}(T)}}{\sqrt{\varphi}}\right) \tag{A8}$$

Simplifying Eq. (A8) yields

$$\frac{\partial^2 m_{\text{TS-D}(T)}}{\partial T^2} = \frac{1}{4T\varphi\sqrt{\varphi}}\left[2\varphi\Delta C_{p\text{D-N}} + T\left(\Delta S_{\text{D-N}(T)}\right)^2(\omega-\alpha)\right] \tag{A9}$$

Similarly, we may show that

$$\frac{\partial^2 m_{\text{TS-N}(T)}}{\partial T^2} = -\frac{\partial^2 m_{\text{TS-D}(T)}}{\partial T^2} = -\frac{1}{4T\varphi\sqrt{\varphi}}\left[2\varphi\Delta C_{p\text{D-N}} + T\left(\Delta S_{\text{D-N}(T)}\right)^2(\omega-\alpha)\right] \tag{A10}$$

## Expression for the temperature-dependence of the observed rate constant

The observed rate constant $k_{obs(T)}$ for a two-state system is the sum of $k_{f(T)}$ and $k_{u(T)}$.[104] Therefore, we can write

$$k_{obs(T)} = k_{f(T)} + k_{u(T)} \Rightarrow \ln k_{obs(T)} = \ln\left(k_{f(T)} + k_{u(T)}\right) \tag{A11}$$

Substituting Eqs. (5) and (6) in (A11) gives

$$\ln k_{obs(T)} = \ln\left[k^0 \exp\left(-\frac{\alpha\left(\omega m_{D\text{-}N} - \sqrt{\varphi}\right)^2}{RT(\omega-\alpha)^2}\right) + k^0 \exp\left(-\frac{\omega\left(\sqrt{\varphi} - \alpha m_{D\text{-}N}\right)^2}{RT(\omega-\alpha)^2}\right)\right] \tag{A12}$$

## Expressions to demonstrate why the extrema of $\Phi_{F(internal)(T)}$ and $\Phi_{U(internal)(T)}$ must occur at $T_S$

Differentiating Eq. (44) with respect to temperature gives

$$\frac{\partial \Phi_{F(internal)(T)}}{\partial T} = \frac{\partial}{\partial T}\left(\frac{\Delta\Delta G_{TS\text{-}D(T\text{-}T_{Ref})}}{\Delta\Delta G_{N\text{-}D(T\text{-}T_{Ref})}}\right)$$

$$= \left(\frac{\Delta\Delta G_{N\text{-}D(T\text{-}T_{Ref})}\frac{\partial}{\partial T}\left(\Delta\Delta G_{TS\text{-}D(T\text{-}T_{Ref})}\right) - \Delta\Delta G_{TS\text{-}D(T\text{-}T_{Ref})}\frac{\partial}{\partial T}\left(\Delta\Delta G_{N\text{-}D(T\text{-}T_{Ref})}\right)}{\left(\Delta\Delta G_{N\text{-}D(T\text{-}T_{Ref})}\right)^2}\right) \tag{A13}$$

$$\Rightarrow \frac{\partial \Phi_{F(internal)(T)}}{\partial T} = \frac{\Delta\Delta G_{TS\text{-}D(T\text{-}T_{Ref})}\Delta S_{N\text{-}D(T)} - \Delta\Delta G_{N\text{-}D(T\text{-}T_{Ref})}\Delta S_{TS\text{-}D(T)}}{\left(\Delta\Delta G_{N\text{-}D(T\text{-}T_{Ref})}\right)^2} \tag{A14}$$

where the protein at the temperature $T_{Ref}$ is by definition the wild type protein. Because $\Delta S_{N\text{-}D(T)}$ and $\Delta S_{TS\text{-}D(T)}$ are both zero at $T_S$, irrespective of $T_{Ref}$, the derivative of $\Phi_{F(internal)(T)}$ will be zero at $T_S$. Similarly, we can show by differentiating Eq. (45) that

$$\frac{\partial \Phi_{U(internal)(T)}}{\partial T} = \frac{\Delta\Delta G_{D\text{-}N(T_{Ref}\text{-}T)}\Delta S_{TS\text{-}N(T)} - \Delta\Delta G_{TS\text{-}N(T_{Ref}\text{-}T)}\Delta S_{D\text{-}N(T)}}{\left(\Delta\Delta G_{D\text{-}N(T_{Ref}\text{-}T)}\right)^2} \tag{A15}$$

Once again, since $\Delta S_{D\text{-}N(T)}$ and $\Delta S_{TS\text{-}N(T)}$ are both zero at $T_S$, irrespective of $T_{Ref}$, the derivative of $\Phi_{U(internal)(T)}$ will be zero at $T_S$.

# Table 1: Reference temperatures

| Temperature | Value | Remark |
|---|---|---|
| $T_\alpha$ | 182 K | A two-state system is physically undefined for $T < T_\alpha$ |
| $T_{S(\alpha)}$ | 184.4 K | $m_{\text{TS-N}(T)} = 0$, $\Delta H_{\text{TS-N}(T)} = \Delta S_{\text{TS-N}(T)} = \Delta G_{\text{TS-N}(T)} = 0$, $k_{u(T)} = k^0$ |
| $T_{C_p\text{TS-N}(\alpha)}$ | 201 K | $\Delta C_{p\text{TS-N}(T)} = 0$ |
| $T_c$ | 223.6 K | Midpoint of cold denaturation, $\Delta G_{\text{D-N}(T)} = 0$, $k_{f(T)} = k_{u(T)}$ |
| $T_{H(\text{TS-N})}$ | 264.3 K | $\Delta H_{\text{TS-N}(T)} = 0$, $k_{u(T)}$ is a minimum |
| $T_H$ | 272.9 K | $\Delta H_{\text{TS-D}(T)} = \Delta H_{\text{TS-N}(T)}$, $\Delta H_{\text{D-N}(T)} = 0$, $\Delta H_{\text{TS-D}(T)} > 0$, $\Delta H_{\text{TS-N}(T)} > 0$, |
| $T_S$ | 278.8 K | $\Delta S_{\text{TS-D}(T)} = \Delta S_{\text{TS-N}(T)} = \Delta S_{\text{D-N}(T)} = 0$, $\Delta G_{\text{D-N}(T)}$ is a maximum |
| $T_{H(\text{TS-D})}$ | 311.4 K | $\Delta H_{\text{TS-D}(T)} = 0$, $k_{f(T)}$ is a maximum |
| $T_m$ | 337.2 K | Midpoint of heat denaturation, $\Delta G_{\text{D-N}(T)} = 0$, $k_{f(T)} = k_{u(T)}$ |
| $T_{C_p\text{TS-N}(\omega)}$ | 361.7 K | $\Delta C_{p\text{TS-N}(T)} = 0$ |
| $T_{S(\omega)}$ | 384.5 K | $m_{\text{TS-N}(T)} = 0$, $\Delta H_{\text{TS-N}(T)} = \Delta S_{\text{TS-N}(T)} = \Delta G_{\text{TS-N}(T)} = 0$, $k_{u(T)} = k^0$ |
| $T_\omega$ | 388 K | A two-state system is physically undefined for $T > T_\omega$ |

**FIGURES**

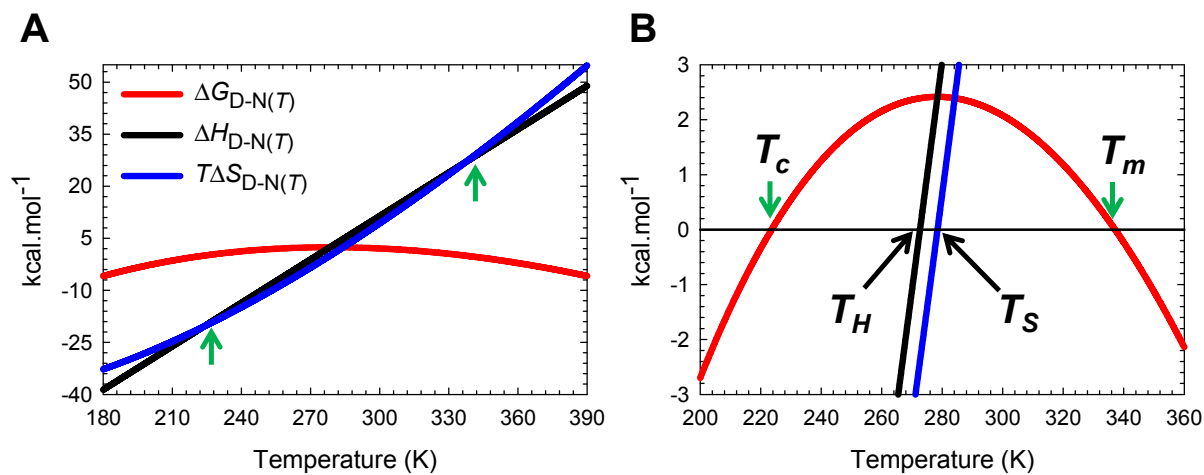

**Figure 1.**

**Stability curve for the unfolding reaction $N \rightleftharpoons D$.**

**(A)** Temperature-dependence of $\Delta H_{D-N(T)}$, $\Delta S_{D-N(T)}$ and $\Delta G_{D-N(T)}$ according to Eqs. (A1), (A2) and (A3), respectively. The green pointers identify the cold ($T_c$) and heat ($T_m$) denaturation temperatures. The slopes of the red and black curves are given by $\partial \Delta G_{D-N(T)}/\partial T = -\Delta S_{D-N(T)}$ and $\partial \Delta H_{D-N(T)}/\partial T = \Delta C_{p\,D-N}$, respectively. **(B)** An appropriately scaled version of the plot on the left. $T_H$ is the temperature at which $\Delta H_{D-N(T)} = 0$, and $T_S$ is the temperature at which $\Delta S_{D-N(T)} = 0$. The values of the reference temperatures are given in **Table 1**.

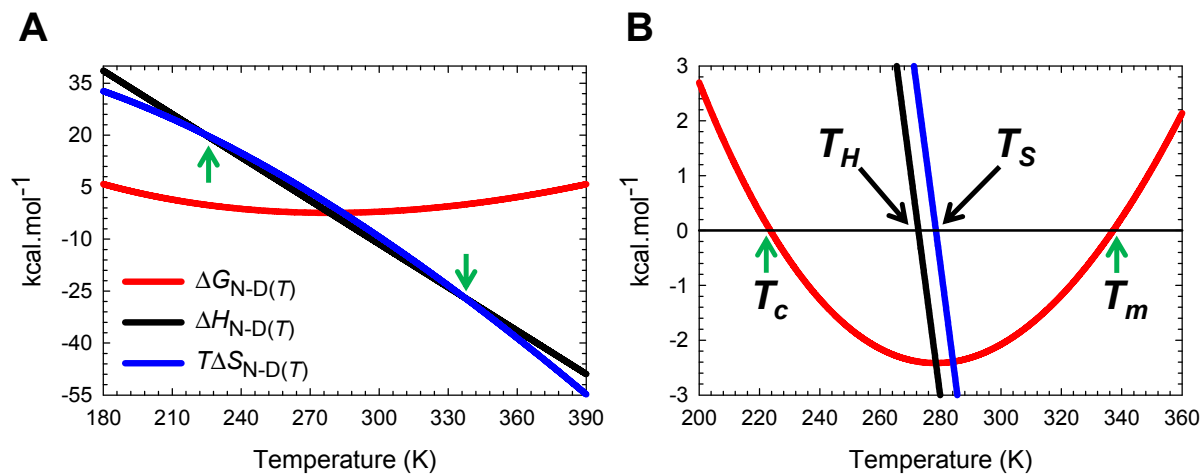

**Figure 1−figure supplement 1.**

**Stability curve for the folding reaction $D \rightleftharpoons N$.**

**(A)** Temperature-dependence of $\Delta G_{\text{N-D}(T)}$, $\Delta H_{\text{N-D}(T)}$, and $T\Delta S_{\text{N-D}(T)}$. The green pointers identify $T_c$ and $T_m$. The slopes of the red and black curves are given by $\partial \Delta G_{\text{N-D}(T)}/\partial T = -\Delta S_{\text{N-D}(T)}$ and $\partial \Delta H_{\text{N-D}(T)}/\partial T = \Delta C_{p\text{N-D}}$, respectively. **(B)** An appropriately scaled version of plot on the left. The reference temperatures are as described in the parent figure.

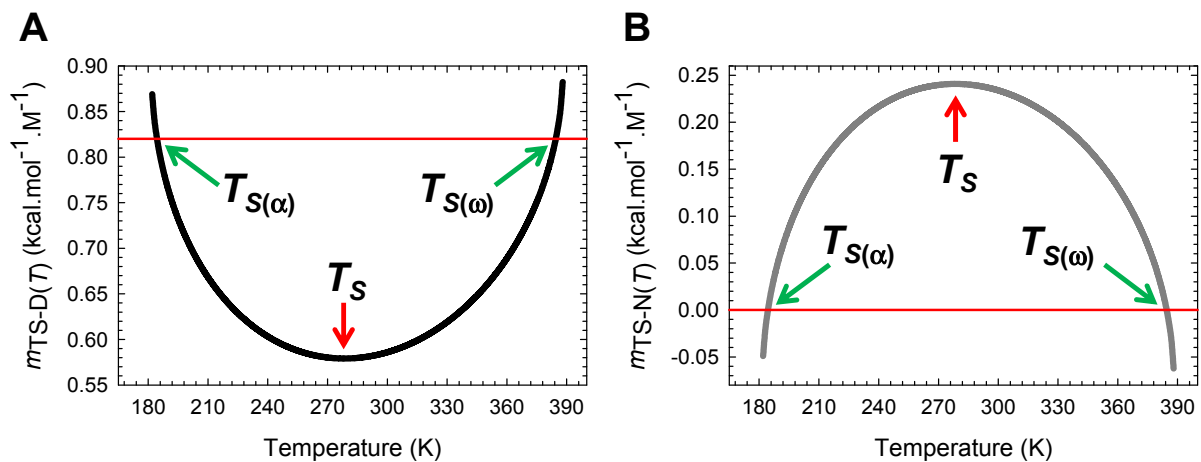

**Figure 2.**

**Temperature-dependence of $m_{TS-D(T)}$ and $m_{TS-N(T)}$.**

**(A)** $m_{TS-D(T)}$ is a minimum at $T_S$, is identical to $m_{D-N}$ at $T_{S(\alpha)}$ and $T_{S(\omega)}$, and is greater than $m_{D-N}$ for $T_\alpha \leq T < T_{S(\alpha)}$ and $T_{S(\omega)} < T \leq T_\omega$. The slope of this curve is given by $\Delta S_{D-N(T)}/2\sqrt{\varphi}$ **(B)** $m_{TS-N(T)}$ is a maximum at $T_S$, zero at $T_{S(\alpha)}$ and $T_{S(\omega)}$, and negative for $T_\alpha \leq T < T_{S(\alpha)}$ and $T_{S(\omega)} < T \leq T_\omega$. The slope of this curve is given by $\Delta S_{N-D(T)}/2\sqrt{\varphi}$. While the slopes of these curves are related to the activation entropies, the second derivatives of these functions with respect to temperature are related to the heat capacities of activation as shown in Paper-II. The values of the reference temperatures are given in **Table 1**.

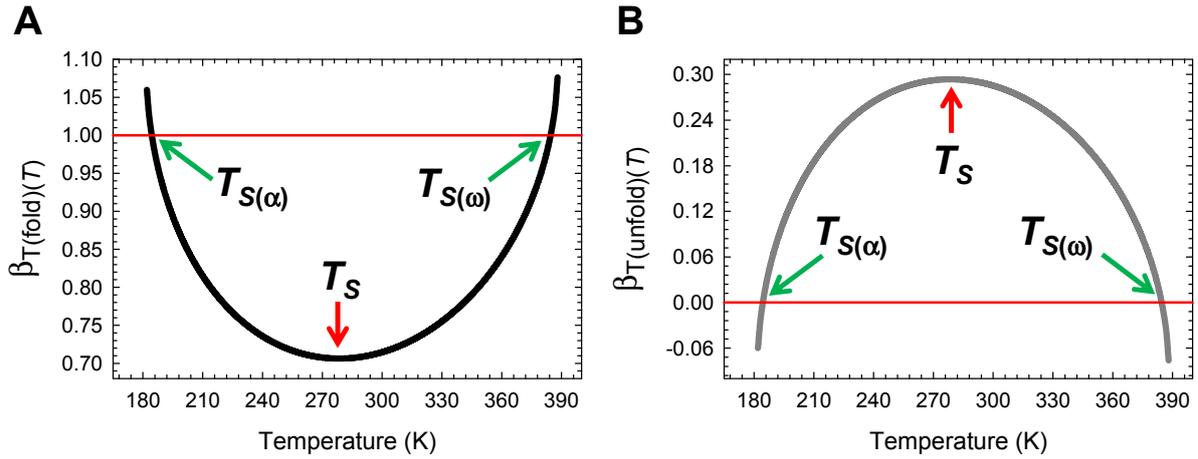

**Figure 2−figure supplement 1.**

**Temperature-dependence of $\beta_{T(fold)(T)}$ and $\beta_{T(unfold)(T)}$.**

**(A)** $\beta_{T(fold)(T)}$ is a minimum at $T_S$, unity at $T_{S(\alpha)}$ and $T_{S(\omega)}$, and greater than unity for $T_\alpha \leq T < T_{S(\alpha)}$ and $T_{S(\omega)} < T \leq T_\omega$. The slope of this curve is given by $\Delta S_{\text{D-N}(T)} / m_{\text{D-N}} 2\sqrt{\varphi}$ **(B)** $\beta_{T(unfold)(T)}$ is a maximum at $T_S$, zero at $T_{S(\alpha)}$ and $T_{S(\omega)}$, and negative for $T_\alpha \leq T < T_{S(\alpha)}$ and $T_{S(\omega)} < T \leq T_\omega$. The slope of this curve is given by $\Delta S_{\text{N-D}(T)} / m_{\text{D-N}} 2\sqrt{\varphi}$. From the perspective of Tanford's framework, the SASA of the TSE is the least native-like at $T_S$ but becomes progressively more native-like as the temperature deviates from the $T_S$, and is identical to the SASA of the NSE at $T_{S(\alpha)}$ and $T_{S(\omega)}$; and for $T_\alpha \leq T < T_{S(\alpha)}$ and $T_{S(\omega)} < T \leq T_\omega$, the TSE is more compact than the NSE.

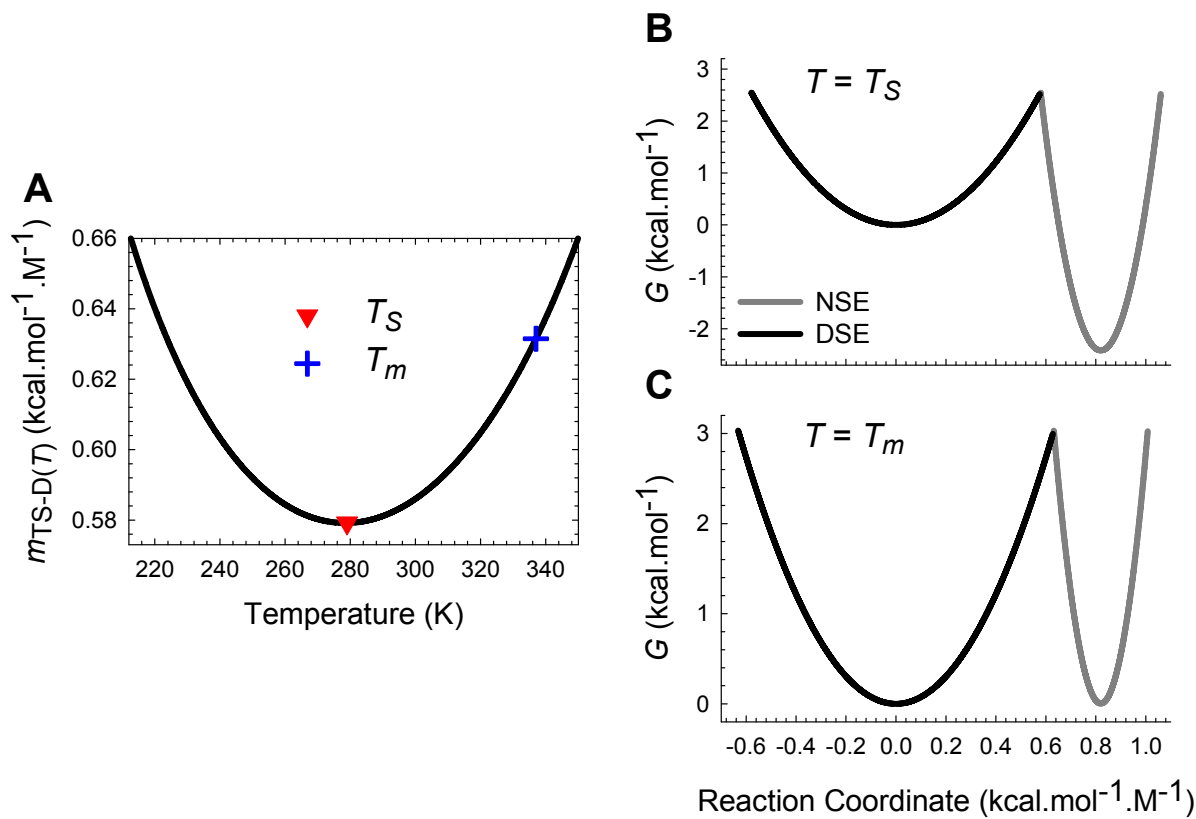

**Figure 3.**

**Marcus *curve-crossings* at $T_S$ and $T_m$.**

**(A) Figure 2A** reproduced for comparison. **(B)** *Curve-crossing* at $T_S$ where $\Delta G_{D-N(T)}$ is a maximum and purely enthalpic (**Figure 1**). The relevant parameters are as follows: $\Delta G_{TS-D(T)}$ = 2.547 kcal.mol$^{-1}$, $\Delta G_{TS-N(T)}$ = 4.964 kcal.mol$^{-1}$, $\Delta G_{D-N(T)}$ = 2.417 kcal.mol$^{-1}$, $k_{f(T)}$ = 22009 s$^{-1}$, $k_{u(T)}$ = 280.8 s$^{-1}$, $m_{TS-D(T)}$ = 0.5792 kcal.mol$^{-1}$.M$^{-1}$ and $m_{TS-N(T)}$ = 0.2408 kcal.mol$^{-1}$.M$^{-1}$. **(C)** *Curve-crossing* at $T_m$ and $T_c$ where $\Delta G_{TS-D(T)}$ = $\Delta G_{TS-N(T)}$ = 3.032 kcal.mol$^{-1}$, $\Delta G_{D-N(T)}$ = 0, $k_{f(T)}$ = $k_{u(T)}$ = 23618 s$^{-1}$, $m_{TS-D(T)}$ = 0.6319 kcal.mol$^{-1}$.M$^{-1}$ and $m_{TS-N(T)}$ = 0.1881 kcal.mol$^{-1}$.M$^{-1}$. The DSE and the NSE-parabolas are given by $G_{DSE(r,T)} = \alpha r^2$ and $G_{NSE(r,T)} = \omega(m_{D-N} - r)^2 - \Delta G_{D-N(T)}$, respectively, $\alpha$ = 7.594 M$^2$.mol.kcal$^{-1}$, $\omega$ = 85.595 M$^2$.mol.kcal$^{-1}$, $m_{D-N}$ (0.82 kcal.mol$^{-1}$.M$^{-1}$) is the separation between the vertices of the DSE and the NSE-parabolas along the abscissa, and $r$ is any point on the abscissa. The abscissae are identical for plots B and C. The values of the reference temperatures are given in **Table 1**.

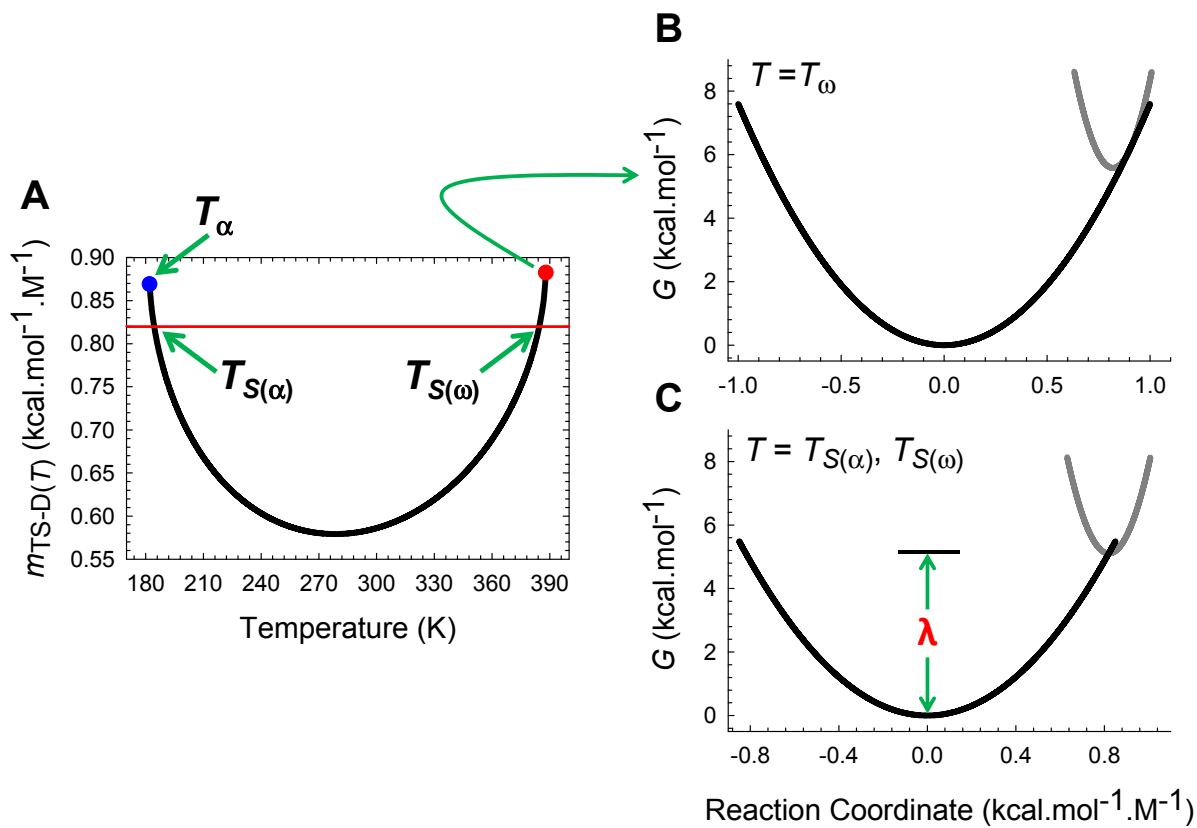

**Figure 4.**

**Marcus *curve-crossings* at $T_{S(\omega)}$ and $T_\omega$.**

**(A) Figure 2A** reproduced for comparison. The blue and red dots represent $T_\alpha$ and $T_\omega$, respectively. The red reference line represents $m_{D-N}$. **(B)** *Curve-crossing* at $T_\omega$ where $m_{TS-D(T)} > m_{D-N}$. The relevant parameters are as follows: $\Delta G_{TS-D(T)} = 5.9136$ kcal.mol$^{-1}$, $\Delta G_{TS-N(T)} = 0.3338$ kcal.mol$^{-1}$, $\Delta G_{D-N(T)} = -5.5798$ kcal.mol$^{-1}$, $k_{f(T)} = 1017$ s$^{-1}$, $k_{u(T)} = 1414594$ s$^{-1}$, $m_{TS-D(T)} = 0.8824$ kcal.mol$^{-1}$.M$^{-1}$ and $m_{TS-N(T)} = -0.0624$ kcal.mol$^{-1}$.M$^{-1}$. **(C)** *Curve-crossing* at $T_{S(\alpha)}$ and $T_{S(\omega)}$ where $m_{TS-D(T)} = m_{D-N} = 0.82$ kcal.mol$^{-1}$.M$^{-1}$, $m_{TS-N(T)} = 0$, $\Delta G_{TS-N(T)} = 0$, $\Delta G_{TS-D(T)} = \lambda = 5.106$ kcal.mol$^{-1}$, $\Delta G_{D-N(T)} = -\lambda$, and $k_{u(T)} = k^0 = 2180965$ s$^{-1}$. The parabolas have been generated as described in the legend for **Figure 3**. The values of the reference temperatures are given in **Table 1**. The rate at which the *curve-crossing* shifts with stability relative to the vertex of the DSE-parabola is given by $\partial m_{TS-D(T)} / \partial \Delta G_{D-N(T)} = -1/2\sqrt{\varphi}$.

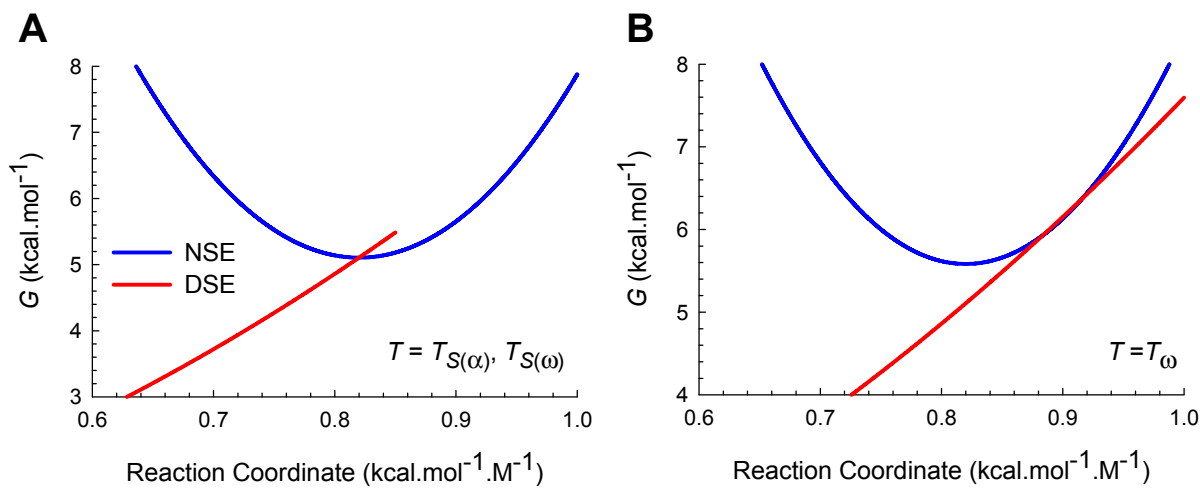

**Figure 4−figure supplement 1.**

**An appropriately scaled view of Marcus *curve-crossings* at $T_{S(\omega)}$ and $T_{\omega}$.**

**(A)** *Curve-crossing* at $T_{S(\alpha)}$ and $T_{S(\omega)}$ where $m_{TS\text{-}D(T)} = m_{D\text{-}N}$, $m_{TS\text{-}N(T)} = 0$, $\Delta G_{TS\text{-}N(T)} = 0$, $\Delta G_{TS\text{-}D(T)} = \alpha\left(m_{D\text{-}N}\right)^2 = \lambda$, $\Delta G_{D\text{-}N(T)} = -\lambda$, and $k_{u(T)} = k^0$. **(B)** *Curve-crossing* at $T_{\omega}$ where $m_{TS\text{-}D(T)} > m_{D\text{-}N}$.

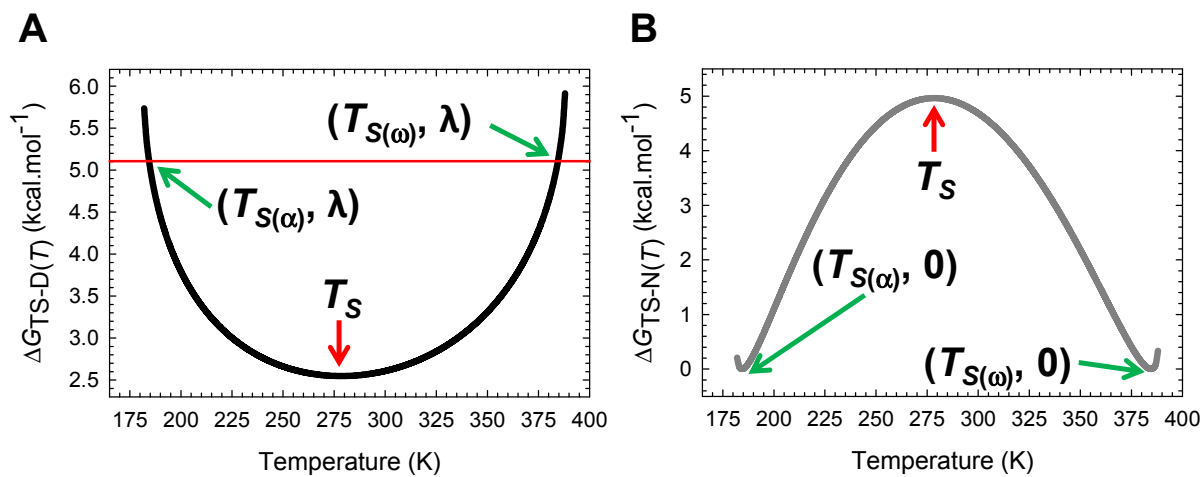

**Figure 5.**

**Temperature-dependence of the Gibbs activation energies for folding and unfolding.**

**(A)** $\Delta G_{\text{TS-D}(T)}$ is a minimum at $T_S$, identical to $\lambda = \alpha\,(m_{\text{D-N}})^2 = 5.106$ kcal.mol$^{-1}$ at $T_{S(\alpha)}$ and $T_{S(\omega)}$, and greater than $\lambda$ for $T_\alpha \leq T < T_{S(\alpha)}$ and $T_{S(\omega)} < T \leq T_\omega$. Note that $\partial \Delta G_{\text{TS-D}(T)}/\partial T = -\Delta S_{\text{TS-D}(T)} = 0$ at $T_S$. **(B)** In contrast to $\Delta G_{\text{TS-D}(T)}$ which has only one extremum, $\Delta G_{\text{TS-N}(T)}$ is a maximum at $T_S$ and a minimum (zero) at $T_{S(\alpha)}$ and $T_{S(\omega)}$; consequently, $\partial \Delta G_{\text{TS-N}(T)}/\partial T = -\Delta S_{\text{TS-N}(T)} = 0$ at $T_{S(\alpha)}$, $T_S$ and $T_{S(\omega)}$. Although unfolding is barrierless at $T_{S(\alpha)}$ and $T_{S(\omega)}$, it is once again barrier-limited for $T_\alpha \leq T < T_{S(\alpha)}$ and $T_{S(\omega)} < T \leq T_\omega$; however, unlike the *conventional barrier-limited* unfolding which is characteristic for $T_{S(\alpha)} < T < T_{S(\omega)}$, these two regimes fall under the *Marcus-inverted-region* and can be rationalized from **Figures 2**, **4**, and their figure supplements. The values of the reference temperatures are given in **Table 1**.

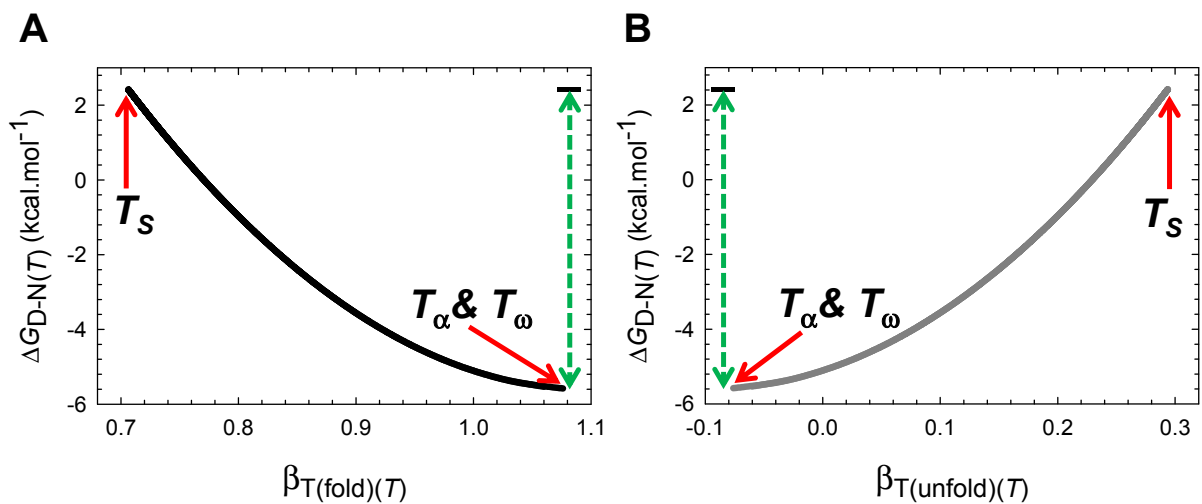

**Figure 5−figure supplement 1.**

**The principle of least displacement.**

**(A)** The stability of a two-state system at constant pressure and solvent conditions is the greatest when the denatured conformers are displaced the least from the mean of their ensemble along the SASA-RC to reach the TSE. The length of the green dotted line is identical to $\Delta G_{\text{D-N}(T_S)} + \left[\lambda\omega/(\omega-\alpha)\right]$, where $\Delta G_{\text{D-N}(T_S)}$ is the stability at $T_S$. The slope of this curve equals $-2m_{\text{D-N}}\sqrt{\varphi}$. **(B)** $\Delta G_{\text{D-N}(T)}$ will be the greatest when the native conformers expose the greatest amount of SASA to reach the TSE. The slope of this curve equals $2m_{\text{D-N}}\sqrt{\varphi}$.

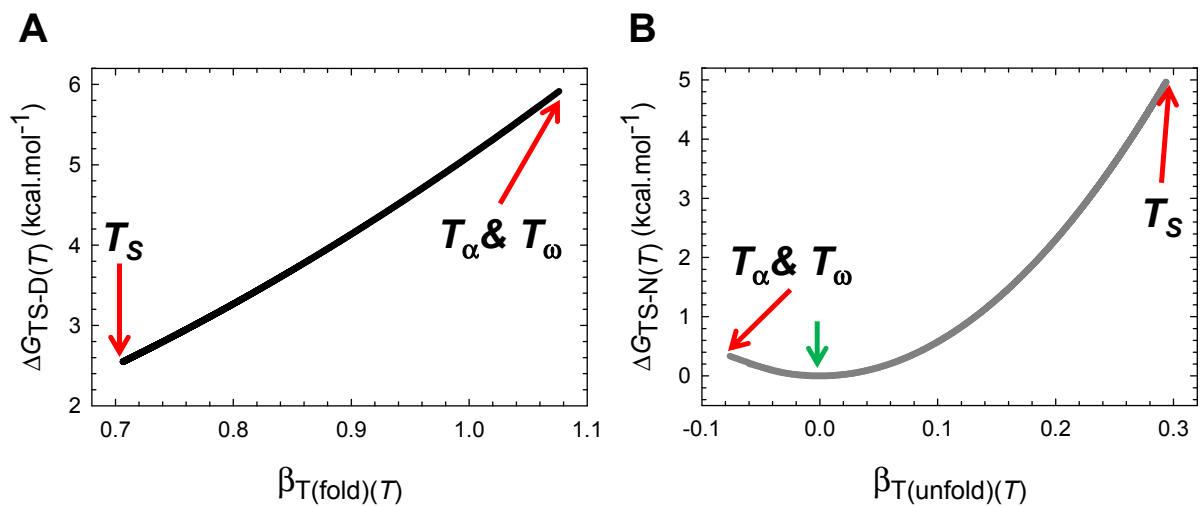

**Figure 5−figure supplement 2.**

**Gibbs activation energies as a function of the position of the TSE along the RC.**

**(A)** $\Delta G_{TS\text{-}D(T)}$ is the least when the denatured conformers bury the least amount of SASA to reach the TSE. The slope of this curve equals $2\lambda\beta_{T(fold)(T)}$. **(B)** $\Delta G_{TS\text{-}N(T)}$ is the greatest when the native conformers expose the greatest amount of SASA to reach the TSE. The green pointer indicates $T_{S(\alpha)}$ and $T_{S(\omega)}$ where $m_{TS\text{-}D(T)} = m_{D\text{-}N}$, $m_{TS\text{-}N(T)} = \beta_{T(unfold)(T)} = 0$, $\Delta G_{TS\text{-}N(T)} = 0$, $\Delta G_{TS\text{-}D(T)} = \alpha(m_{D\text{-}N})^2 = \lambda$, and $\Delta G_{D\text{-}N(T)} = -\lambda$. The slope of this curve equals $2\omega m_{TS\text{-}N(T)} m_{D\text{-}N}$. Because $m_{TS\text{-}N(T)} < 0$ for $T_\alpha \leq T < T_{S(\alpha)}$ and $T_{S(\omega)} < T \leq T_\omega$, the slope is negative for the part that is to the left of the green pointer.

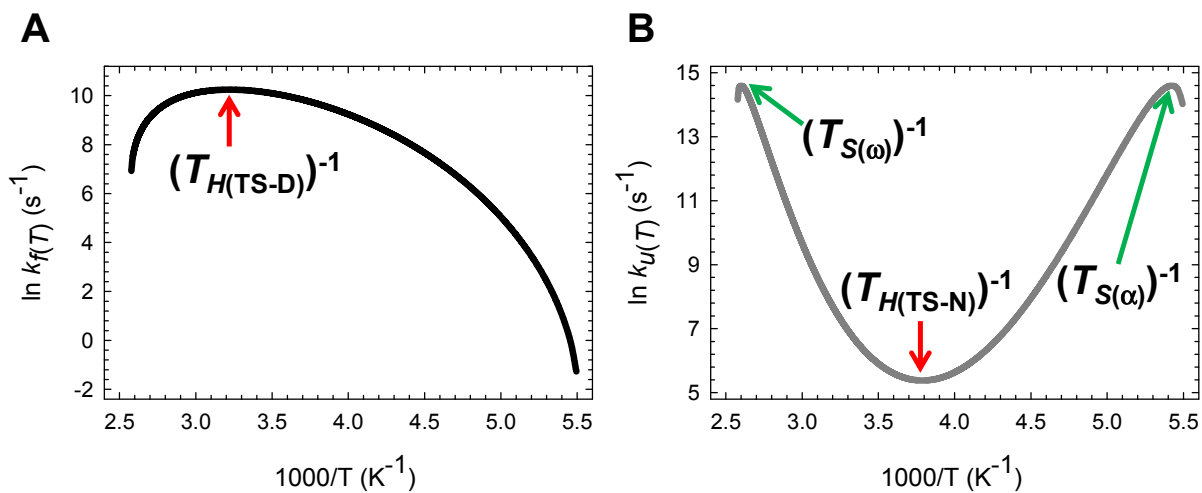

**Figure 6.**

**Arrhenius plots for the temperature-dependence of the rate constants.**

(**A**) $k_{f(T)}$ is a maximum and $\Delta H_{\text{TS-D}(T)} = 0$ at $T_{H(\text{TS-D})}$. The slope of this curve is given by $-\Delta H_{\text{TS-D}(T)}/R$. (**B**) Unlike $k_{f(T)}$ which has only one extremum, $k_{u(T)}$ is a minimum at $T_{H(\text{TS-N})}$ and a maximum at $T_{S(\alpha)}$ and $T_{S(\omega)}$. Consequently, $\Delta H_{\text{TS-N}(T)} = 0$ at $T_{S(\alpha)}$, $T_{H(\text{TS-N})}$ and $T_{S(\omega)}$. The slope of this curve is given by $-\Delta H_{\text{TS-N}(T)}/R$. When $T = T_{S(\alpha)}$ or $T_{S(\omega)}$, we have a unique scenario: $m_{\text{TS-N}(T)} = \Delta G_{\text{TS-N}(T)} = \Delta H_{\text{TS-N}(T)} = 0 \Rightarrow \Delta S_{\text{TS-N}(T)} = 0$, and $k_{u(T)} = k^0$. Although unfolding is barrier-limited for $T_\alpha \leq T < T_{S(\alpha)}$ and $T_{S(\omega)} < T \leq T_\omega$, leading to $k_{u(T)} < k^0$, these ultra-low and high temperature regimes fall under the *Marcus-inverted-regime* as compared to the *conventional barrier-limited* unfolding which is characteristic for $T_{S(\alpha)} < T < T_{S(\omega)}$ (the *curve-crossing* occurs in-between the vertices of the DSE and the NSE Gibbs basins) and can be rationalized comprehensively when considered in conjunction with **Figures 2**, **4**, and **5** (see also their figure supplements if any). The maxima of $k_{f(T)}$ and $k_{u(T)}$, as well as the *inverted-region* can be better appreciated on a linear scale as shown in the figure supplement. The values of the reference temperatures are given in **Table 1**.

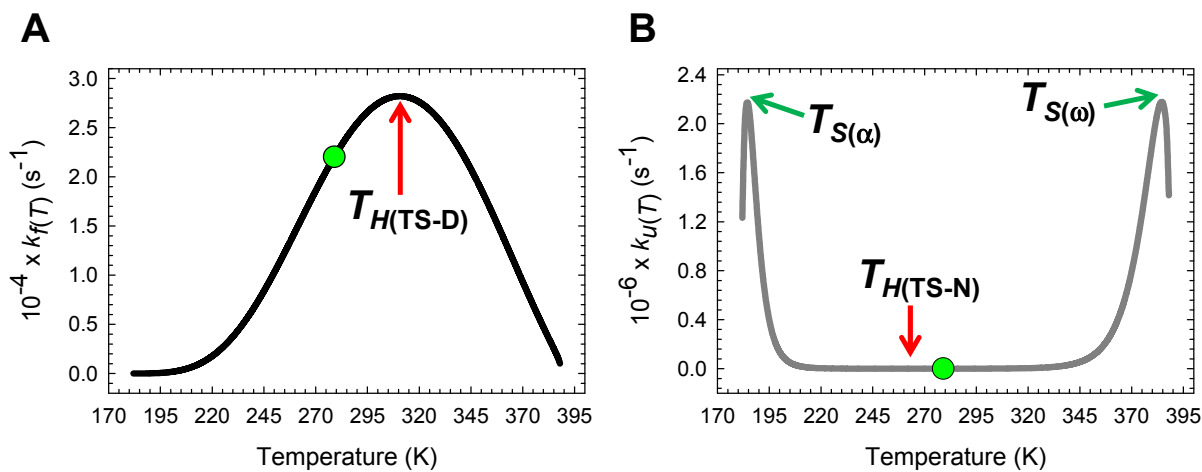

**Figure 6−figure supplement 1.**

**Temperature-dependence of $k_{f(T)}$ and $k_{u(T)}$ on a linear scale.**

**(A)** $k_{f(T)}$ is a maximum and $\Delta H_{TS\text{-}D(T)} = 0$ at $T_{H(TS\text{-}D)}$. The slope of this curve is given by $k_{f(T)} \Delta H_{TS\text{-}D(T)} / RT^2$. **(B)** Unlike $k_{f(T)}$ which has only one extremum, $k_{u(T)}$ is a minimum at $T_{H(TS\text{-}N)}$ and a maximum at $T_{S(\alpha)}$ and $T_{S(\omega)}$. Although the minimum of $k_{u(T)}$ is not apparent on a linear scale, the *barrierless* and *inverted-regimes* for unfolding are readily apparent. The slope of this curve is given by $k_{u(T)} \Delta H_{TS\text{-}N(T)} / RT^2$. The features of these curves arise primarily from the temperature-dependence of the equilibrium constants for the partial folding ($D \rightleftharpoons [TS]$) and unfolding ($N \rightleftharpoons [TS]$) reactions as shown later. The green dots represent $T_S$.

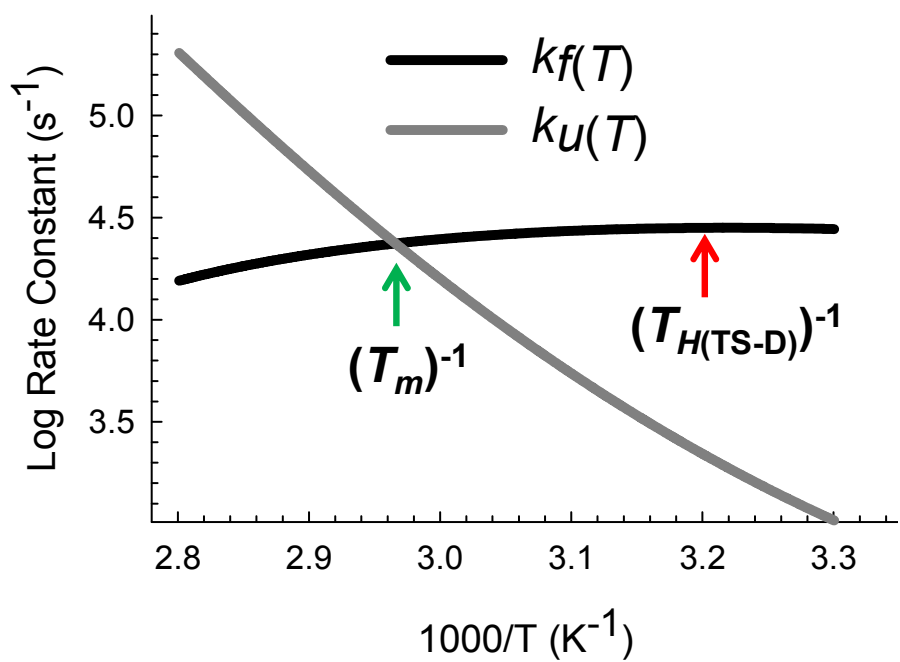

**Figure 6−figure supplement 2.**

**Arrhenius plot for the temperature-dependence of the rate constants with the ordinate on a Log scale (base10).**

A combined and appropriately rescaled version of Figure 6 to enable a ready comparison of the rate constants for FBP28 WW wild type (calculated using parabolic approximation) and the experimental rate constants for ΔNΔC Y11R W30F, a variant of FBP28 WW (reported by Nguyen et al., 2003, Fig. 4A). Note that the intersection of $k_{f(T)}$ and $k_{u(T)}$ is shifted to the left along the abscissa for the wild type FBP28 WW since its $T_m$ is ~ 10 K greater than that of ΔNΔC Y11R W30F (see Table 1 in Nguyen et al., 2003).[25]

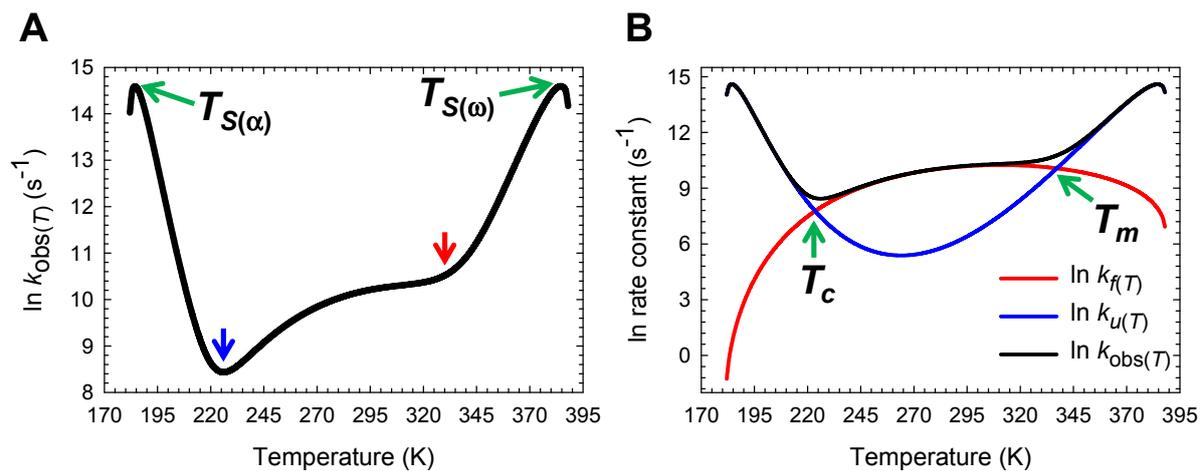

**Figure 7.**

**Temperature-dependence of the observed rate constant.**

**(A)** $k_{obs(T)}$ is a maximum at $T_{S(\alpha)}$ and $T_{S(\omega)}$, and a minimum around $T_c$ (blue pointer). The red pointer indicates $T_m$. The steep increase in $k_{obs(T)}$ at very low and high temperatures is due to $\Delta G_{TS\text{-}N(T)}$ approaching zero as described in previous figures. **(B)** An overlay of $k_{f(T)}$, $k_{u(T)}$ and $k_{obs(T)}$ to illuminate how the features of $k_{obs(T)}$ arise from the sum of $k_{f(T)}$ and $k_{u(T)}$. The slopes of the red and blue curves are given by $\Delta H_{TS\text{-}D(T)}/RT^2$ and $\Delta H_{TS\text{-}N(T)}/RT^2$, respectively.

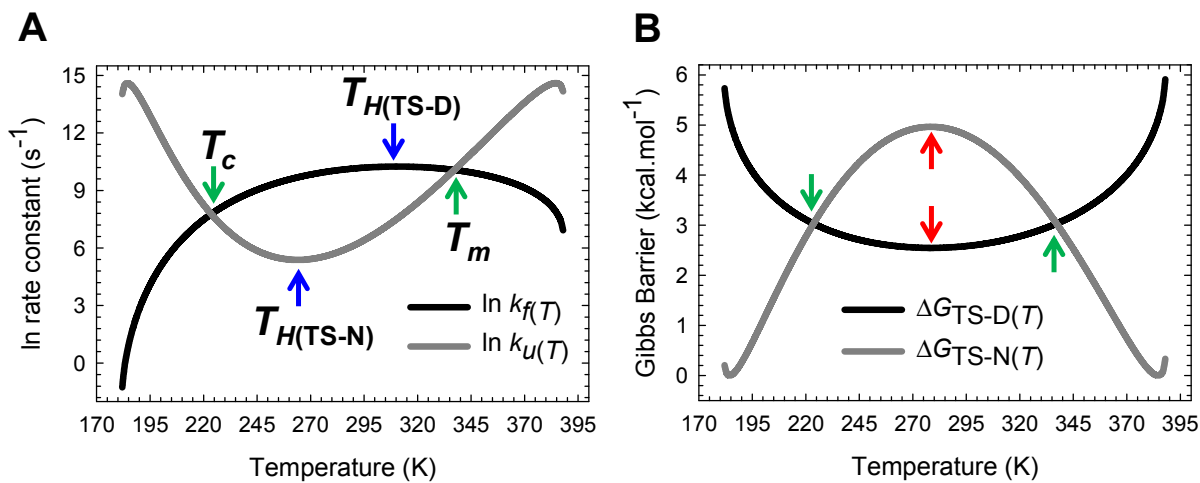

**Figure 7−figure supplement 1.**

**The principle of microscopic reversibility.**

**(A)** $k_{f(T)}$ is a maximum at $T_{H(TS-D)}$ and $k_{u(T)}$ is a minimum at $T_{H(TS-N)}$. The slopes of the black and grey curves are given by $\Delta H_{TS-D(T)}/RT^2$ and $\Delta H_{TS-N(T)}/RT^2$, respectively. **(B)** $\Delta G_{TS-D(T)}$ and $\Delta G_{TS-N(T)}$ are a minimum and a maximum, respectively, at $T_S$ (red pointers) leading to $\Delta G_{D-N(T)}$ being a maximum at $T_S$ (**Figure 1**). Equilibrium stability is thus a consequence or the equilibrium manifestation of the underlying kinetic behaviour. The rate constants are identical at $T_c$ and $T_m$, leading to $\Delta G_{D-N(T)} = RT\ln\left(k_{f(T)}/k_{u(T)}\right) = \Delta G_{TS-N(T)} - \Delta G_{TS-D(T)} = 0$.

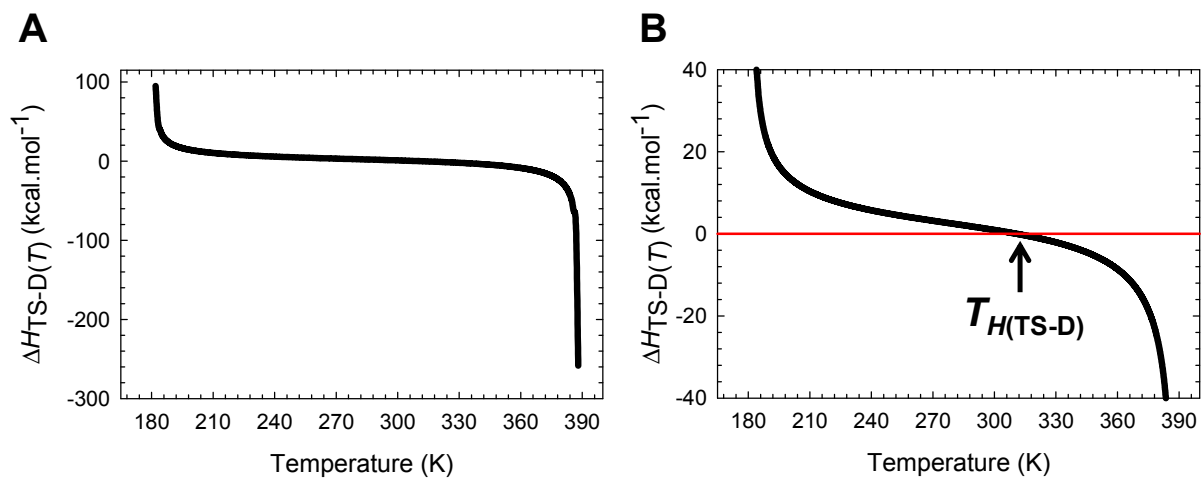

**Figure 8.**

**Temperature-dependence of the activation enthalpy for folding.**

**(A)** The variation in $\Delta H_{TS\text{-}D(T)}$ function with temperature. The slope of this curve varies with temperature, equals $\Delta C_{pTS\text{-}D(T)}$, and is algebraically negative. **(B)** An appropriately scaled version of the plot on the left to illuminate the three important scenarios: (*i*) $\Delta H_{TS\text{-}D(T)} > 0$ for $T_\alpha \leq T < T_{H(TS\text{-}D)}$; (*ii*) $\Delta H_{TS\text{-}D(T)} < 0$ for $T_{H(TS\text{-}D)} < T \leq T_\omega$; and (*iii*) $\Delta H_{TS\text{-}D(T)} = 0$ when $T = T_{H(TS\text{-}D)}$. Note that $k_{f(T)}$ is a maximum at $T_{H(TS\text{-}D)}$.

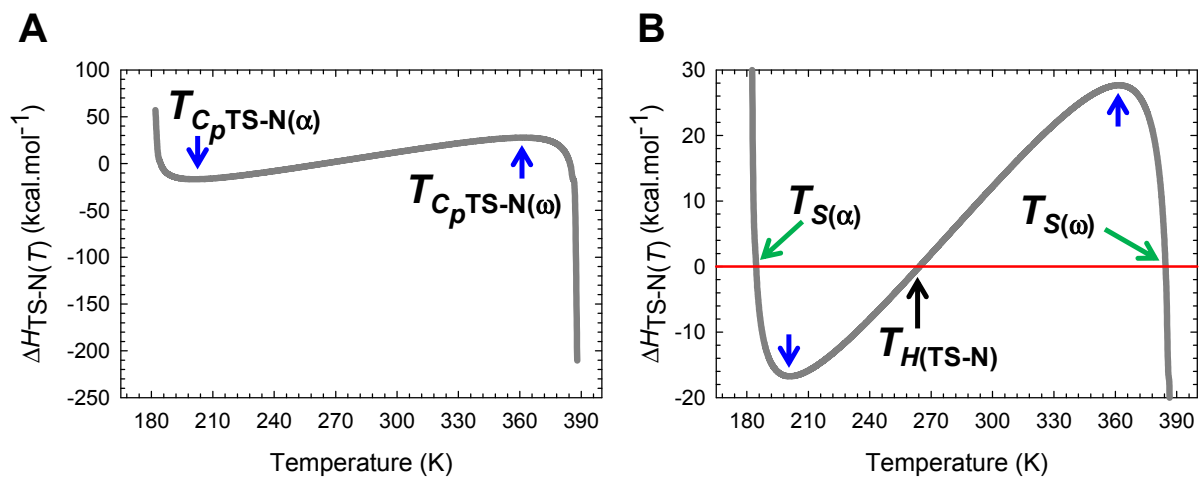

**Figure 9.**

**Temperature-dependence of the activation enthalpy for unfolding.**

**(A)** The variation in $\Delta H_{\text{TS-N}(T)}$ function with temperature. The slope of this curve equals $\Delta C_{p\text{TS-N}(T)}$ and is zero at $T_{C_p\text{TS-N}(\alpha)}$ and $T_{C_p\text{TS-N}(\omega)}$. **(B)** An appropriately scaled version of the figure on the left to illuminate the various temperature-regimes and their implications: (*i*) $\Delta H_{\text{TS-N}(T)} > 0$ for $T_\alpha \leq T < T_{S(\alpha)}$ and $T_{H(\text{TS-N})} < T < T_{S(\omega)}$; (*ii*) $\Delta H_{\text{TS-N}(T)} < 0$ for $T_{S(\alpha)} < T < T_{H(\text{TS-N})}$ and $T_{S(\omega)} < T \leq T_\omega$; and (*iii*) $\Delta H_{\text{TS-N}(T)} = 0$ at $T_{S(\alpha)}$, $T_{H(\text{TS-N})}$, and $T_{S(\omega)}$. Note that at $T_{S(\alpha)}$ and $T_{S(\omega)}$, we have the unique scenario: $m_{\text{TS-N}(T)} = \Delta G_{\text{TS-N}(T)} = \Delta S_{\text{TS-N}(T)} = \Delta H_{\text{TS-N}(T)} = 0$, and $k_{u(T)} = k^0$. The values of the reference temperatures are given in **Table 1**.

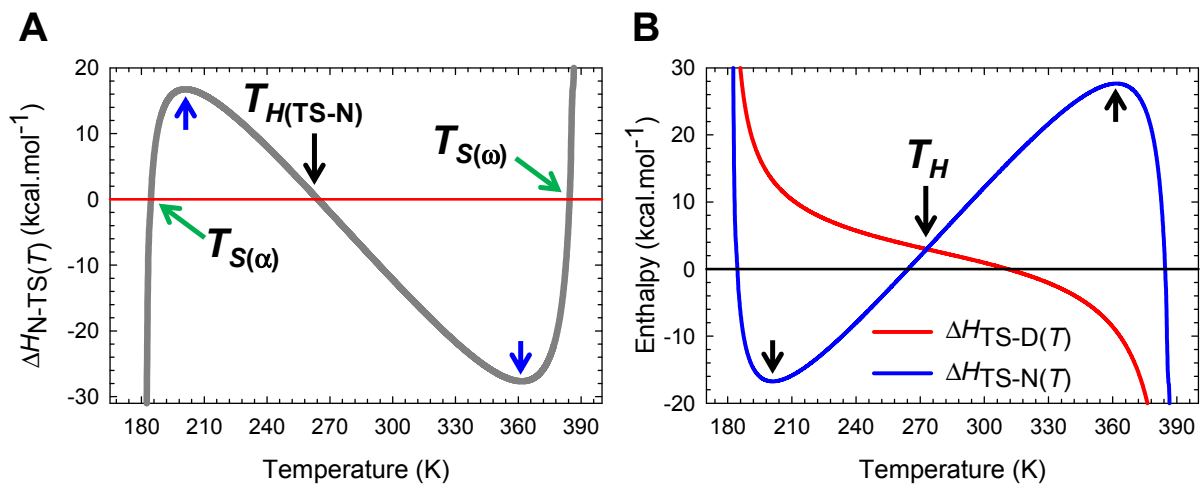

**Figure 9−figure supplement 1.**

**The variation in $\Delta H_{\text{N-TS}(T)}$ with temperature and the intersection of $\Delta H_{\text{TS-D}(T)}$ and $\Delta H_{\text{TS-N}(T)}$ functions.**

**(A)** An appropriately scaled view of the change in enthalpy for the partial folding reaction $[TS] \rightleftharpoons N$. The flux of the conformers from the TSE to the NSE is enthalpically: (*i*) favourable for $T_\alpha \leq T < T_{S(\alpha)}$ and $T_{H(\text{TS-N})} < T < T_{S(\omega)}$ ($\Delta H_{\text{N-TS}(T)} < 0$); (*ii*) unfavourable for $T_{S(\alpha)} < T < T_{H(\text{TS-N})}$ and $T_{S(\omega)} < T \leq T_\omega$ ($\Delta H_{\text{N-TS}(T)} > 0$); and (*iii*) neither favourable nor unfavourable at $T_{S(\alpha)}$, $T_{H(\text{TS-N})}$, and $T_{S(\omega)}$. The blue pointers indicate the temperatures where $\Delta C_{p\text{N-TS}(T)}$ (or $-\Delta C_{p\text{TS-N}(T)}$) is zero. **(B)** The intersection of the $\Delta H_{\text{TS-D}(T)}$ and $\Delta H_{\text{TS-N}(T)}$ functions occurs precisely at $T_H$. The requirement that both $\Delta H_{\text{TS-D}(T)}$ and $\Delta H_{\text{TS-N}(T)}$ be positive at the point of intersection is a consequence of the theoretical relationship: $T_{H(\text{TS-N})} < T_H < T_S < T_{H(\text{TS-D})}$ and must be satisfied by all two-state systems (see Paper II). Note that the net flux of the conformers from the DSE to the NSE at $T_H$ is driven purely by entropy ($\Delta G_{\text{D-N}(T)} = -T\Delta S_{\text{D-N}(T)}$).

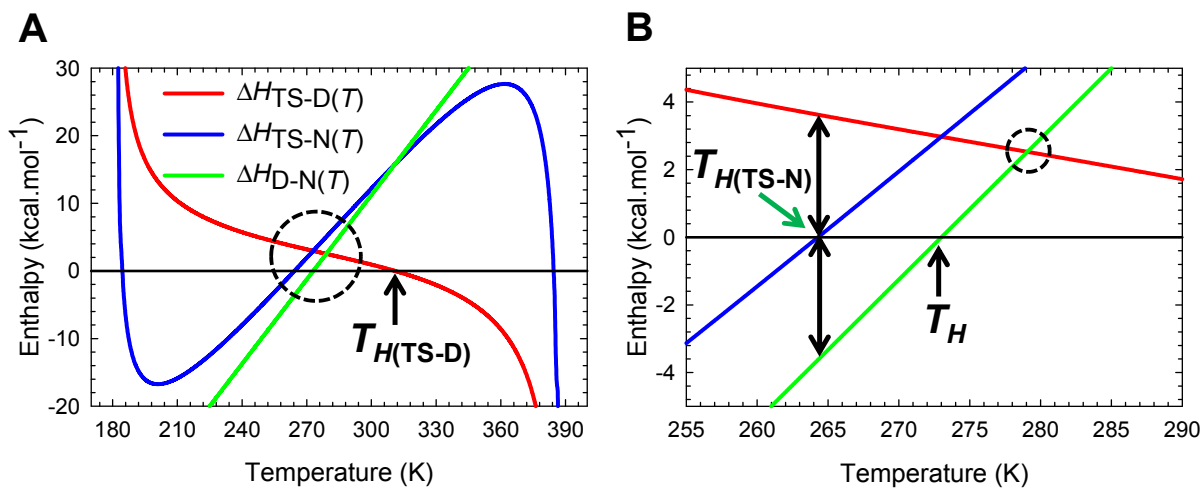

**Figure 9−figure supplement 2.**

**Comparison of equilibrium and activation enthalpies.**

**(A)** $\Delta H_{\text{D-N}(T)}$ for the reaction $N \rightleftharpoons D$ is zero at the temperature where $\Delta H_{\text{TS-D}(T)}$ and $\Delta H_{\text{TS-N}(T)}$ functions intersect (the intersection of green curve and zero reference line must align vertically with the point where the blue and the red curves intersect). The intersection of $\Delta H_{\text{D-N}(T)}$ and $\Delta H_{\text{TS-N}(T)}$ functions (green and blue curves) occurs precisely when $T = T_{H(\text{TS-D})}$. This is expected since $\Delta H_{\text{TS-D}(T)} = 0$ at $T_{H(\text{TS-D})}$. The similarity in the slopes of the $\Delta H_{\text{D-N}(T)}$ and $\Delta H_{\text{TS-N}(T)}$ functions between ~ 240 K and ~ 320 K implies that most of $\Delta C_{p\text{D-N}}$ stems from the first-half of the unfolding reaction $N \rightleftharpoons [TS]$. **(B)** An appropriately scaled view of the encircled area in the figure on the left. When $T = T_{H(\text{TS-N})}$, $\Delta H_{\text{TS-D}(T)}$ is identical to $|\Delta H_{\text{D-N}(T)}|$ or $\Delta H_{\text{N-D}(T)}$. Further, at the temperature where $\Delta H_{\text{TS-D}(T)}$ and $\Delta H_{\text{D-N}(T)}$ functions intersect (i.e., the intersection of the red and the green curves), the absolute enthalpy of the DSE ($H_{\text{D}(T)}$) is exactly half the algebraic sum of the absolute enthalpies of the TSE ($H_{\text{TS}(T)}$) and the NSE ($H_{\text{N}(T)}$), i.e., $H_{\text{D}(T)} = (H_{\text{TS}(T)} + H_{\text{N}(T)})/2$. The various auxiliary relationships that may obtained from the intersection of various state functions are addressed in subsequent publications.

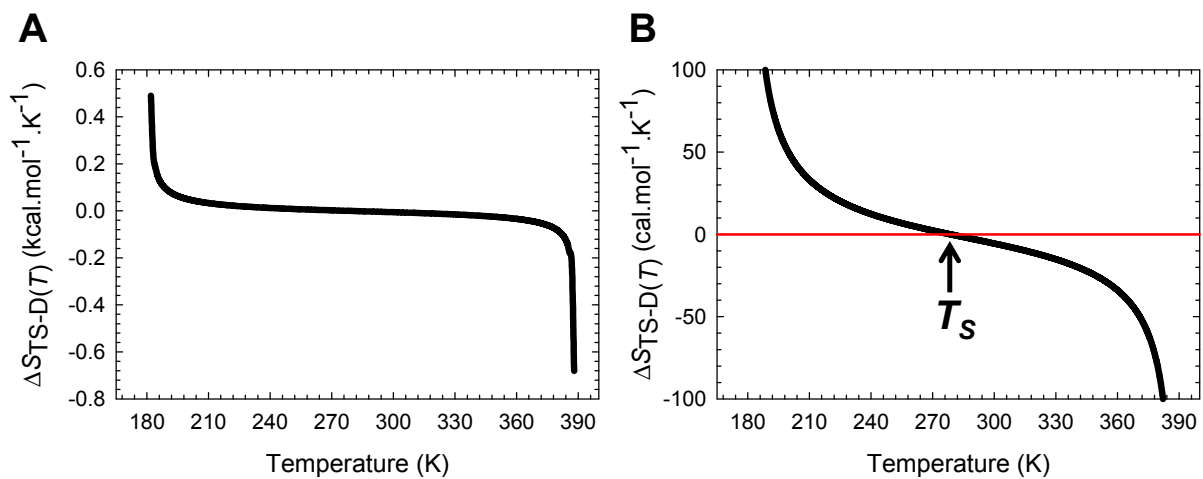

**Figure 10.**

**Temperature-dependence of the activation entropy for folding.**

**(A)** The variation in $\Delta S_{\text{TS-D}(T)}$ function with temperature. The slope of this curve varies with temperature and equals $\Delta C_{p\text{TS-D}(T)}/T$. **(B)** An appropriately scaled version of the figure on the left to illuminate the three temperature regimes and their implications: (*i*) $\Delta S_{\text{TS-D}(T)} > 0$ for $T_\alpha \leq T < T_S$; (*ii*) $\Delta S_{\text{TS-D}(T)} < 0$ for $T_S < T \leq T_\omega$; and (*iii*) $\Delta S_{\text{TS-D}(T)} = 0$ when $T = T_S$.

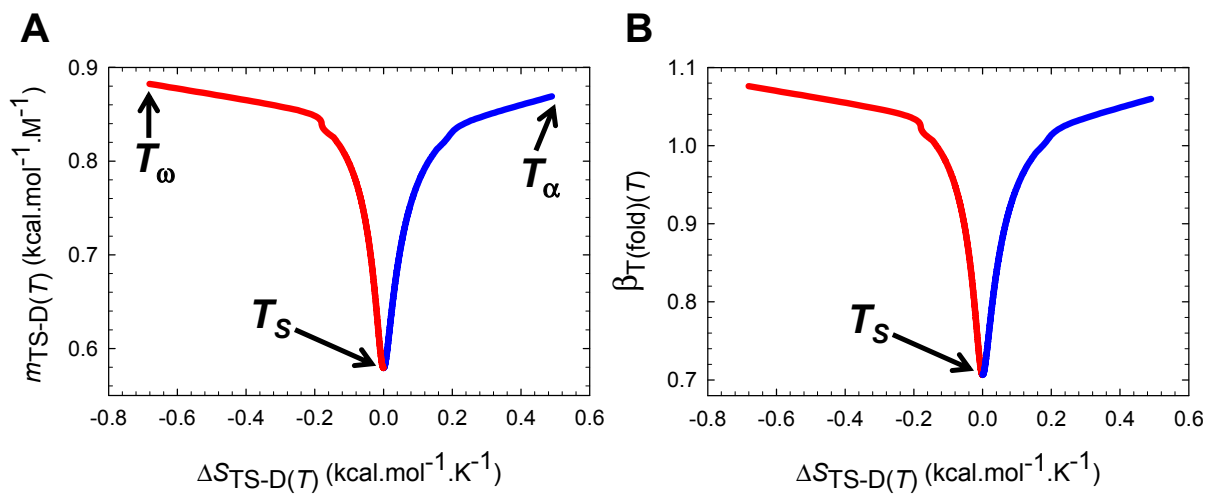

**Figure 10−figure supplement 1.**

**Activation entropy for folding *vs curve-crossing*.**

(A) $\Delta S_{\text{TS-D}(T)}$ is zero when the denatured conformers are displaced the least from the mean of their ensemble to reach the TSE along the SASA-RC. The slope of this curve is given by $T\Delta S_{\text{D-N}(T)} \big/ 2\Delta C_{p\text{TS-D}(T)} \sqrt{\varphi}$ (B) $\Delta S_{\text{TS-D}(T)}$ is zero when the SASA of the TSE is the least native-like. The slope of this curve is given by $T\Delta S_{\text{D-N}(T)} \big/ 2m_{\text{D-N}}\Delta C_{p\text{TS-D}(T)} \sqrt{\varphi}$. The blue and the red sections of the curves represent the temperature regimes $T_\alpha \leq T \leq T_S$ and $T_S \leq T \leq T_\omega$, respectively.

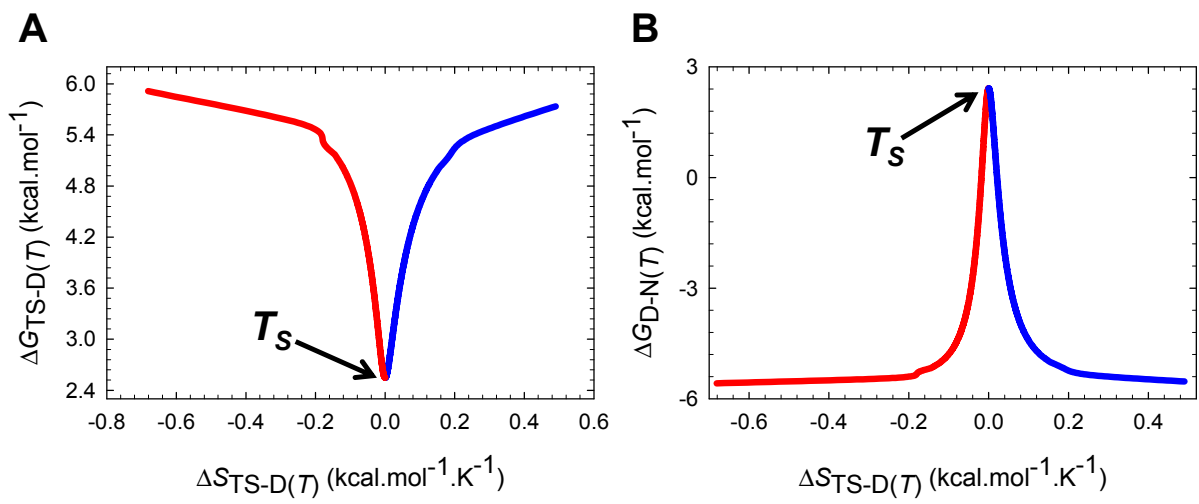

**Figure 10−figure supplement 2.**

**Activation entropy *vs* $\Delta G_{\text{TS-D}(T)}$ and $\Delta G_{\text{D-N}(T)}$.**

**(A)** $\Delta G_{\text{TS-D}(T)}$ is always the least when it is purely enthalpic. The slope of this curve equals $-T\Delta S_{\text{TS-D}(T)}/\Delta C_{p\text{TS-D}(T)}$. **(B)** The stability is always the greatest when the activation entropy for folding is the zero. The slope of this curve equals $-T\Delta S_{\text{D-N}(T)}/\Delta C_{p\text{TS-D}(T)}$. The blue and the red sections of the curves represent the temperature regimes $T_\alpha \leq T \leq T_S$ and $T_S \leq T \leq T_\omega$, respectively.

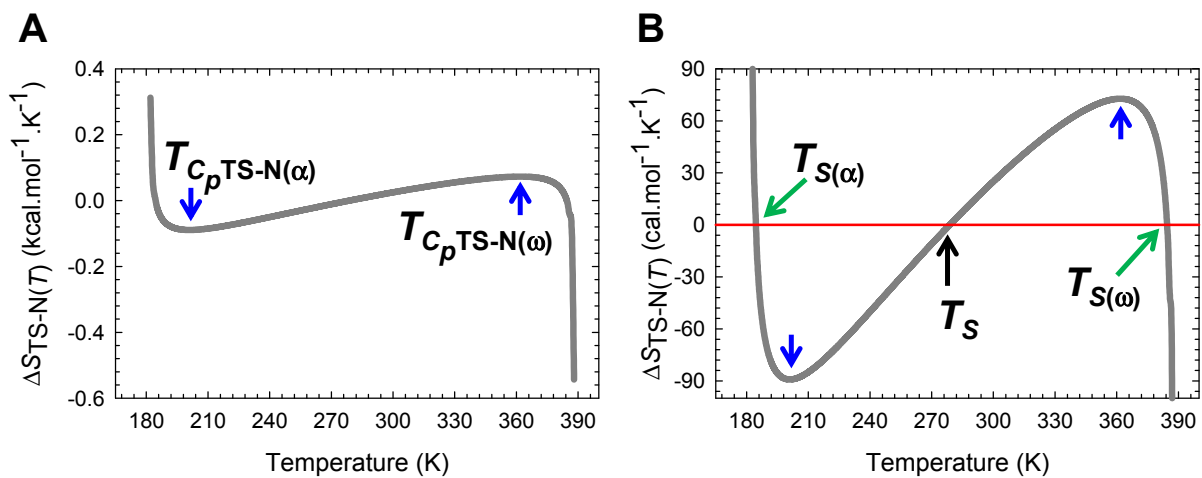

**Figure 11.**

**Temperature-dependence of the activation entropy for unfolding.**

**(A)** The variation in $\Delta S_{\text{TS-N}(T)}$ function with temperature. The slope of this curve, given by $\Delta C_{p\text{TS-N}(T)}/T$, varies with temperature, and is zero at $T_{C_p\text{TS-N}(\alpha)}$ and $T_{C_p\text{TS-N}(\omega)}$. **(B)** An appropriately scaled version of the figure on the left to illuminate the temperature regimes and their implications: (*i*) $\Delta S_{\text{TS-N}(T)} > 0$ for $T_\alpha \leq T < T_{S(\alpha)}$ and $T_S < T < T_{S(\omega)}$; (*ii*) $\Delta S_{\text{TS-N}(T)} < 0$ for $T_{S(\alpha)} < T < T_S$ and $T_{S(\omega)} < T \leq T_\omega$; and (*iii*) $\Delta S_{\text{TS-N}(T)} = 0$ at $T_{S(\alpha)}$, $T_S$, and $T_{S(\omega)}$. Note that at $T_{S(\alpha)}$ and $T_{S(\omega)}$, we have the unique scenario: $m_{\text{TS-N}(T)} = \Delta G_{\text{TS-N}(T)} = \Delta S_{\text{TS-N}(T)} = \Delta H_{\text{TS-N}(T)} = 0$, and $k_{u(T)} = k^0$. The values of the reference temperatures are given in **Table 1**.

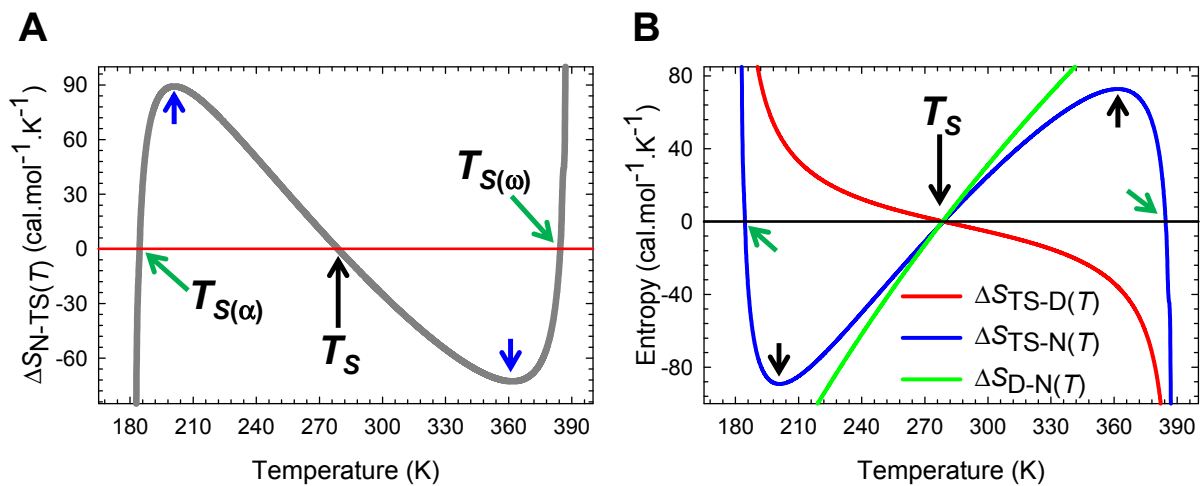

**Figure 11−figure supplement 1.**

**The variation in $\Delta S_{\text{N-TS}(T)}$ with temperature and comparison of equilibrium and activation entropies.**

**(A)** An appropriately scaled view of the change in entropy for the partial folding reaction $[TS] \rightleftharpoons N$. The slope of this curve equals $\Delta C_{p\text{N-TS}(T)}/T$ (or $-\Delta C_{p\text{TS-N}(T)}/T$) and is zero at $T_{C_p\text{TS-N}(\alpha)}$ and $T_{C_p\text{TS-N}(\omega)}$. The flux of the conformers from the TSE to the NSE is entropically: (*i*) unfavourable for $T_\alpha \leq T < T_{S(\alpha)}$ and $T_S < T < T_{S(\omega)}$ ($\Delta S_{\text{N-TS}(T)} < 0$); (*ii*) favourable for $T_{S(\alpha)} < T < T_S$ and $T_{S(\omega)} < T \leq T_\omega$ ($\Delta S_{\text{N-TS}(T)} > 0$); and (*iii*) neutral at $T_{S(\alpha)}$, $T_S$, and $T_{S(\omega)}$. **(B)** An overlay of $\Delta S_{\text{D-N}(T)}$, $\Delta S_{\text{TS-D}(T)}$ and $\Delta S_{\text{TS-N}(T)}$ functions. Unlike the $\Delta H_{\text{TS-D}(T)}$ and $\Delta H_{\text{TS-N}(T)}$ functions which must be positive at the point of intersection (**Figure 9−figure supplement 1B**), theory dictates that both $\Delta S_{\text{TS-D}(T)}$ and $\Delta S_{\text{TS-N}(T)}$ functions must independently be equal to zero at $T_S$, leading to the unique scenario: $S_{D(T)} = S_{TS(T)} = S_{N(T)}$. The similarity in the slopes of the $\Delta S_{\text{D-N}(T)}$ and $\Delta S_{\text{TS-N}(T)}$ functions between ~ 240 K and ~ 320 K implies that most of $\Delta C_{p\text{D-N}}$ stems from the first-half of the unfolding reaction $N \rightleftharpoons [TS]$. Consequently at $T_S$, $\Delta G_{\text{D-N}(T)} = \Delta H_{\text{D-N}(T)}$, i.e., the net flux of the conformers from the DSE to the NSE is driven purely by enthalpy.

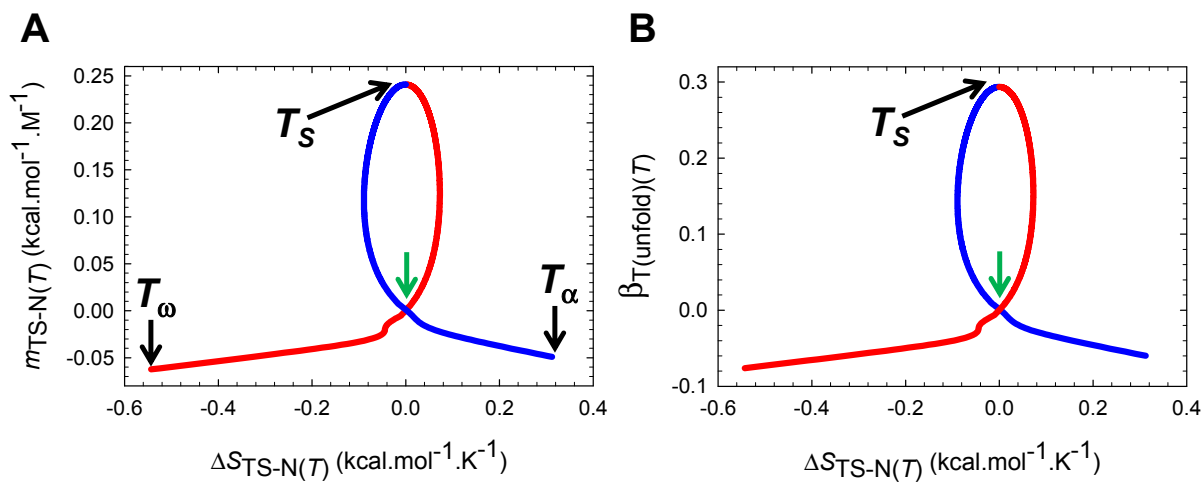

**Figure 11−figure supplement 2.**

**Activation entropy for unfolding *vs curve-crossing*.**

**(A)** Unlike the $\Delta S_{TS\text{-}D(T)}$ function which is zero only once, $\Delta S_{TS\text{-}N(T)}$ is zero once when the native conformers are displaced the greatest to reach the TSE ($T_S$), and twice when this displacement is zero (green pointer; $T_{S(\alpha)}$ and $T_{S(\omega)}$). The slope of this curve is given by $-T\Delta S_{D\text{-}N(T)}\big/2\Delta C_{p\,TS\text{-}N(T)}\sqrt{\varphi}$. **(B)** $\Delta S_{TS\text{-}N(T)}$ is zero once when the difference in SASA between the TSE and the NSE is the greatest ($T_S$), and twice when the SASA of the TSE is identical to that of the NSE (green pointer; $T_{S(\alpha)}$ and $T_{S(\omega)}$). The slope of this curve is given by $-T\Delta S_{D\text{-}N(T)}\big/2m_{D\text{-}N}\Delta C_{p\,TS\text{-}N(T)}\sqrt{\varphi}$. The blue and the red sections of the curves represent the temperature regimes $T_\alpha \leq T \leq T_S$ and $T_S \leq T \leq T_\omega$, respectively.

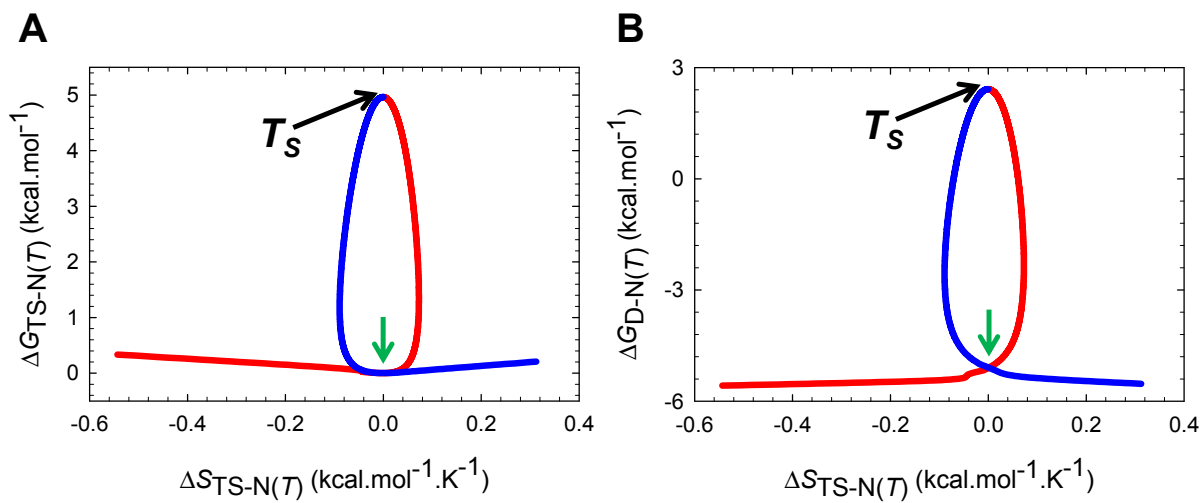

**Figure 11−figure supplement 3.**

**Activation entropy *vs* $\Delta G_{\text{TS-N}(T)}$ and $\Delta G_{\text{D-N}(T)}$.**

**(A)** $\Delta G_{\text{TS-N}(T)}$ is the greatest and also the least (zero) when it is purely enthalpic, with the former occurring at $T_S$, and the latter occurring at $T_{S(\alpha)}$ and $T_{S(\omega)}$ (green pointer). The slope of this curve equals $-T\Delta S_{\text{TS-N}(T)}/\Delta C_{p\text{TS-N}(T)}$. **(B)** The stability is always the greatest at $T_S$ where the Gibbs barrier to unfolding is purely enthalpic; and at $T_{S(\alpha)}$ and $T_{S(\omega)}$ (green pointer), $\Delta G_{\text{D-N}(T)} = -\lambda$. The slope of this curve equals $-T\Delta S_{\text{D-N}(T)}/\Delta C_{p\text{TS-N}(T)}$. The blue and the red sections of the curves represent the temperature regimes $T_\alpha \leq T \leq T_S$ and $T_S \leq T \leq T_\omega$, respectively.

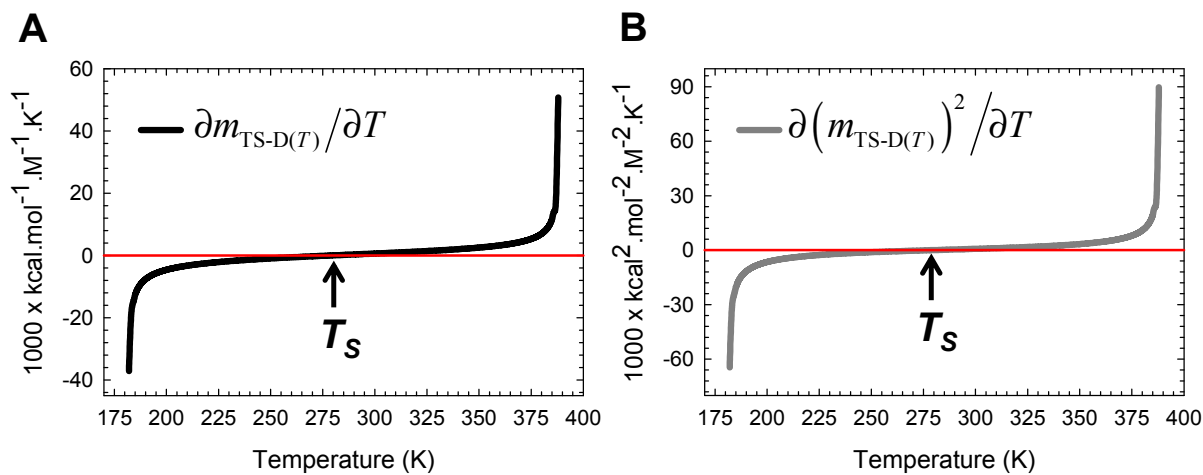

**Figure 11−figure supplement 4.**

**The first derivatives of $m_{\text{TS-D}(T)}$ and the square of $m_{\text{TS-D}(T)}$ with respect to temperature.**

**(A)** $\partial m_{\text{TS-D}(T)}/\partial T$ is negative for $T_\alpha \leq T < T_S$, positive for $T_S < T \leq T_\omega$, and zero at $T_S$ and is dictated by Eq. (A4). **(B)** Because $\partial\left(m_{\text{TS-D}(T)}\right)^2/\partial T = 2m_{\text{TS-D}(T)}\left(\partial m_{\text{TS-D}(T)}/\partial T\right)$ and $m_{\text{TS-D}(T)} > 0$ throughout the temperature regime, the variation of its algebraic sign is identical to that of $\partial m_{\text{TS-D}(T)}/\partial T$. The relationship between $\partial m_{\text{TS-D}(T)}/\partial T$ and $\Delta S_{\text{TS-D}(T)}$ is given by Eq. (8).

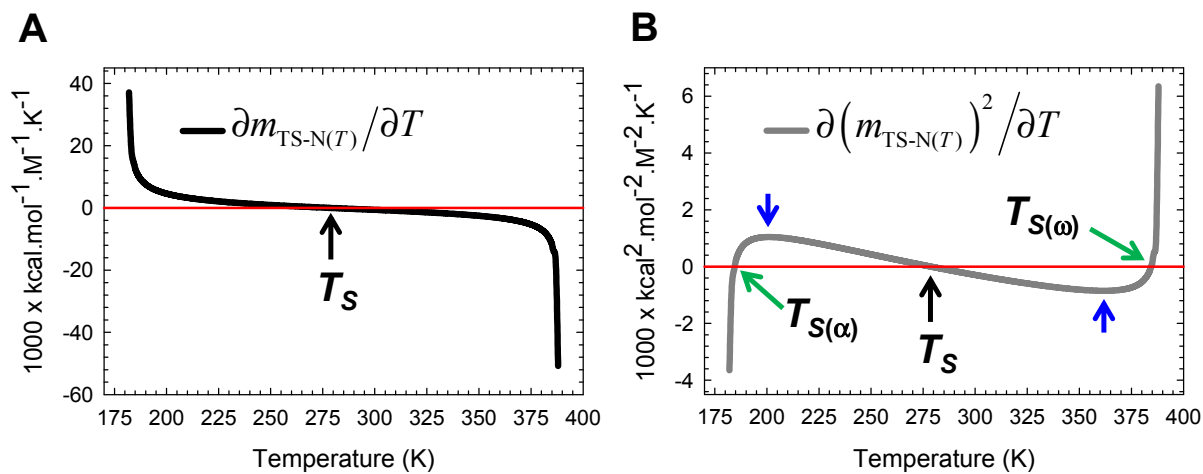

**Figure 11−figure supplement 5.**

**The first derivatives of $m_{\text{TS-N}(T)}$ and the square of $m_{\text{TS-N}(T)}$ with respect to temperature.**

**(A)** $\partial m_{\text{TS-N}(T)}/\partial T$ is positive for $T_\alpha \leq T < T_S$, negative for $T_S < T \leq T_\omega$, and zero at $T_S$ and is governed by Eq. (A6). **(B)** Because $\partial (m_{\text{TS-N}(T)})^2/\partial T = 2 m_{\text{TS-N}(T)} (\partial m_{\text{TS-N}(T)}/\partial T)$ and $m_{\text{TS-N}(T)}$ can be negative, zero or positive depending on the temperature, the variation of its algebraic sign with temperature is far more complex: (*i*) $\partial (m_{\text{TS-N}(T)})^2/\partial T$ is negative for $T_\alpha \leq T < T_{S(\alpha)}$ and $T_S < T < T_{S(\omega)}$; (*ii*) positive for $T_{S(\alpha)} < T < T_S$ and $T_{S(\omega)} < T \leq T_\omega$; and (*iii*) zero at $T_{S(\alpha)}$, $T_S$, and $T_{S(\omega)}$. The relationship between $\partial m_{\text{TS-N}(T)}/\partial T$ and $\Delta S_{\text{TS-N}(T)}$ is given by Eq. (9). The blue pointers indicate the temperatures at which the second derivative of the square of $m_{\text{TS-N}(T)}$ is zero and are identical to the temperatures at which $\Delta C_{p\text{TS-N}(T)}$ is zero.

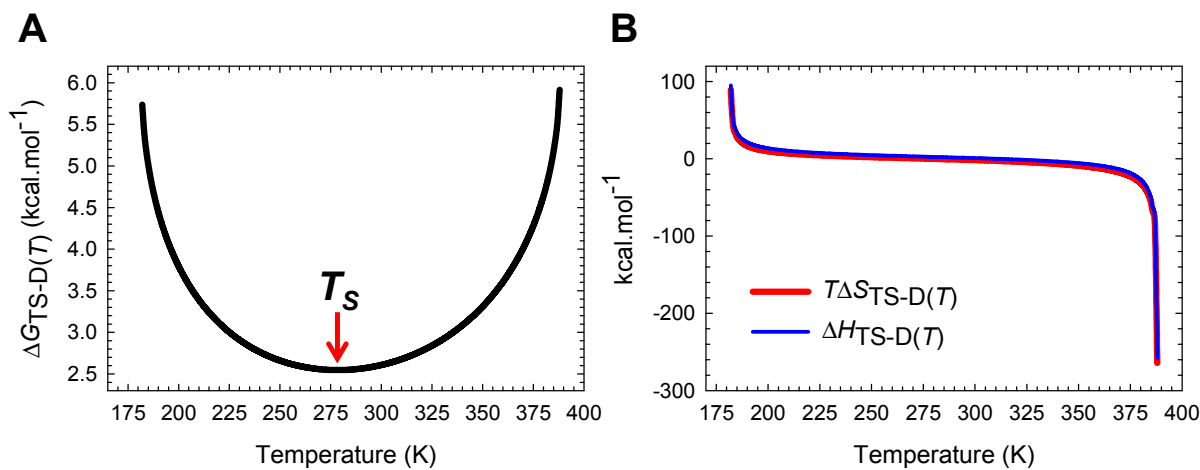

**Figure 12.**

**Entropy-enthalpy compensation for the partial folding reaction $D \rightleftharpoons [TS]$.**

Despite large changes in $\Delta H_{TS\text{-}D(T)}$ (~ 400 kcal.mol$^{-1}$) $\Delta G_{TS\text{-}D(T)}$ which is a minimum at $T_S$, varies only by ~3.4 kcal.mol$^{-1}$ due to compensating changes in $\Delta S_{TS\text{-}D(T)}$. See the appropriately scaled figure supplement for description. The physical basis for entropy-enthalpy compensation is addressed in the accompanying article.

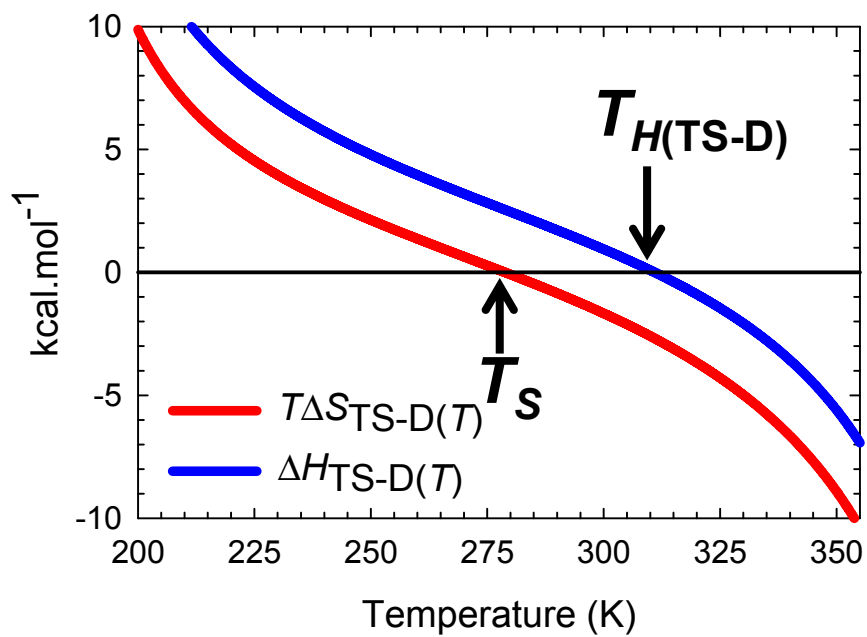

**Figure 12−figure supplement 1.**

**Deconvolution of the Gibbs activation energy for the reaction $D \rightleftharpoons [TS]$.**

This is an appropriately scaled view of **Figure 12B**. For $T_\alpha \leq T < T_S$, $T\Delta S_{\text{TS-D}(T)} > 0$ but is more than offset by unfavourable $\Delta H_{\text{TS-D}(T)}$, leading to incomplete compensation and a positive $\Delta G_{\text{TS-D}(T)}$ ($\Delta H_{\text{TS-D}(T)} - T\Delta S_{\text{TS-D}(T)} > 0$). When $T = T_S$, $\Delta G_{\text{TS-D}(T)}$ is a minimum and purely enthalpic ($\Delta G_{\text{TS-D}(T)} = \Delta H_{\text{TS-D}(T)} > 0$). For $T_S < T < T_{H(\text{TS-D})}$, the activation is enthalpically and entropically disfavoured ($\Delta H_{\text{TS-D}(T)} > 0$ and $T\Delta S_{\text{TS-D}(T)} < 0$) leading to a positive $\Delta G_{\text{TS-D}(T)}$. In contrast, for $T_{H(\text{TS-D})} < T \leq T_\omega$, $\Delta H_{\text{TS-D}(T)} < 0$ but is more than offset by the unfavourable entropy ($T\Delta S_{\text{TS-D}(T)} < 0$), leading once again to a positive $\Delta G_{\text{TS-D}(T)}$. When $T = T_{H(\text{TS-D})}$, $\Delta G_{\text{TS-D}(T)}$ is purely entropic ($\Delta G_{\text{TS-D}(T)} = -T\Delta S_{\text{TS-D}(T)} > 0$) and $k_{f(T)}$ is a maximum.

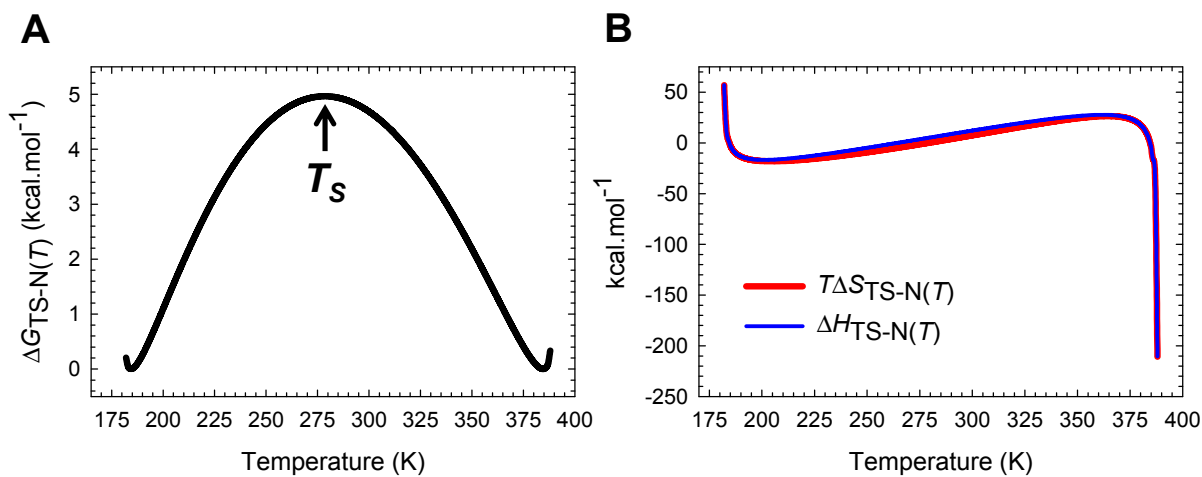

**Figure 13.**

**Entropy-enthalpy compensation for the partial unfolding reaction $N \rightleftharpoons [TS]$.**

Despite large changes in $\Delta H_{TS-N(T)}$, $\Delta G_{TS-N(T)}$ which is a maximum at $T_S$, varies only by ~5 kcal.mol$^{-1}$ due to compensating changes in $\Delta S_{TS-N(T)}$. See the appropriately scaled figure supplement for description.

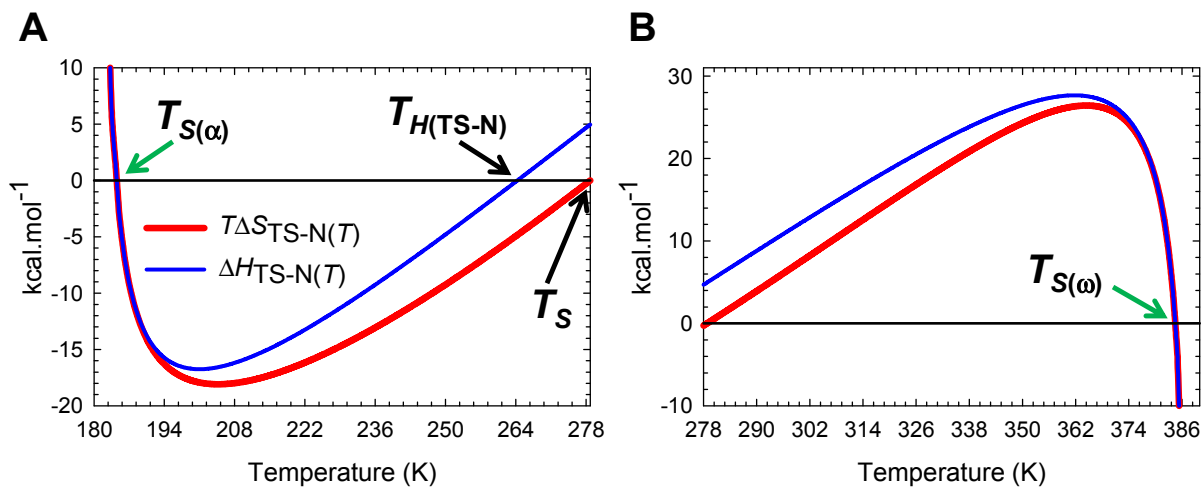

**Figure 13−figure supplement 1.**

**Deconvolution of the Gibbs activation energy for unfolding.**

These are appropriately scaled split views of **Figure 13B**. **(A)** For $T_\alpha \leq T < T_{S(\alpha)}$, $N \rightleftharpoons [TS]$ entropically favoured ($T\Delta S_{\text{TS-N}(T)} > 0$) but is more than offset by endothermic enthalpy ($\Delta H_{\text{TS-N}(T)} > 0$) leading to $\Delta H_{\text{TS-N}(T)} - T\Delta S_{\text{TS-N}(T)} > 0$. When $T = T_{S(\alpha)}$, $\Delta S_{\text{TS-N}(T)} = \Delta H_{\text{TS-N}(T)} = 0 \Rightarrow \Delta G_{\text{TS-N}(T)} = 0$, and $k_{u(T)} = k^0$. For $T_{S(\alpha)} < T < T_{H(\text{TS-N})}$, $N \rightleftharpoons [TS]$ is enthalpically favourable ($\Delta H_{\text{TS-N}(T)} < 0$) but is more than offset by the unfavourable negentropy ($T\Delta S_{\text{TS-N}(T)} < 0$) leading to $\Delta G_{\text{TS-N}(T)} > 0$. When $T = T_{H(\text{TS-N})}$, $\Delta H_{\text{TS-N}(T)} = 0$ for the second time, $\Delta G_{\text{TS-N}(T)}$ is purely due to the negentropy ($\Delta G_{\text{TS-N}(T)} = -T\Delta S_{\text{TS-N}(T)} > 0$), and $k_{u(T)}$ is a minimum. For $T_{H(\text{TS-N})} < T < T_S$, $N \rightleftharpoons [TS]$ is entropically and enthalpically unfavourable ($\Delta H_{\text{TS-N}(T)} > 0$ and $T\Delta S_{\text{TS-N}(T)} < 0$) leading to $\Delta G_{\text{TS-N}(T)} > 0$. When $T = T_S$, $\Delta S_{\text{TS-N}(T)} = 0$ for the second time, and $\Delta G_{\text{TS-N}(T)}$ is a minimum and purely enthalpic ($\Delta G_{\text{TS-N}(T)} = \Delta H_{\text{TS-N}(T)} > 0$). **(B)** For $T_S < T < T_{S(\omega)}$, $N \rightleftharpoons [TS]$ is entropically favourable ($T\Delta S_{\text{TS-N}(T)} > 0$) but is more than offset by the endothermic enthalpy ($\Delta H_{\text{TS-N}(T)} > 0$) leading to a positive $\Delta G_{\text{TS-N}(T)}$. When $T = T_{S(\omega)}$, $\Delta S_{\text{TS-N}(T)} = \Delta H_{\text{TS-N}(T)} = 0$ for the third and the final time, $\Delta G_{\text{TS-N}(T)} = 0$ for the second and final time, and $k_{u(T)} = k^0$. For $T_{S(\omega)} < T \leq T_\omega$, $N \rightleftharpoons [TS]$ is enthalpically favourable ($\Delta H_{\text{TS-N}(T)} < 0$) but is more than offset by the unfavourable negentropy ($T\Delta S_{\text{TS-N}(T)} < 0$), leading to $\Delta G_{\text{TS-N}(T)} > 0$ and $k_{u(T)} < k^0$.

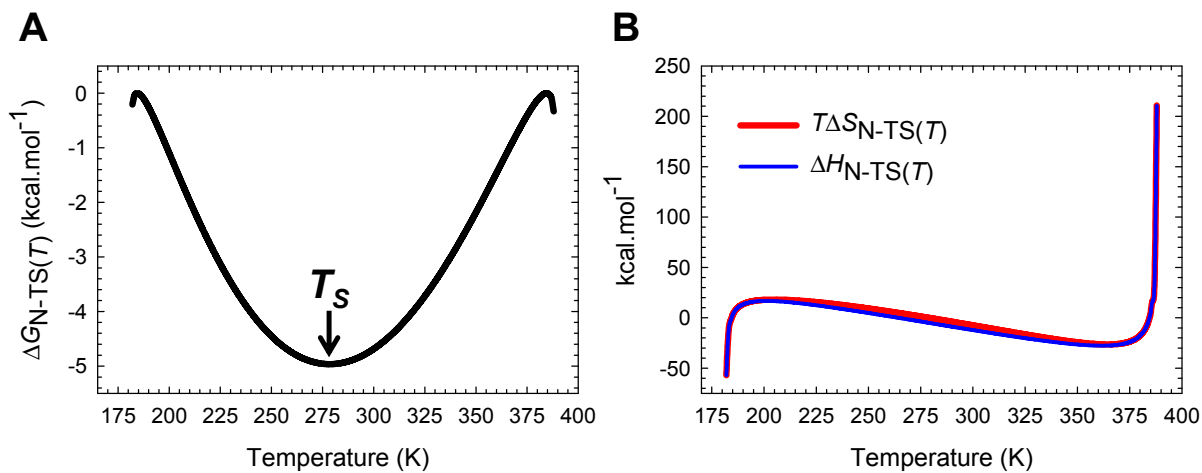

**Figure 13−figure supplement 2.**

**Entropy-enthalpy compensation for the partial folding reaction $[TS] \rightleftharpoons N$**

Despite large changes in $\Delta H_{N\text{-}TS(T)}$, $\Delta G_{N\text{-}TS(T)}$ which is a minimum at $T_S$, varies only by ~5 kcal.mol$^{-1}$ due to compensating changes in $\Delta S_{N\text{-}TS(T)}$. See the appropriately scaled figure supplement for description.

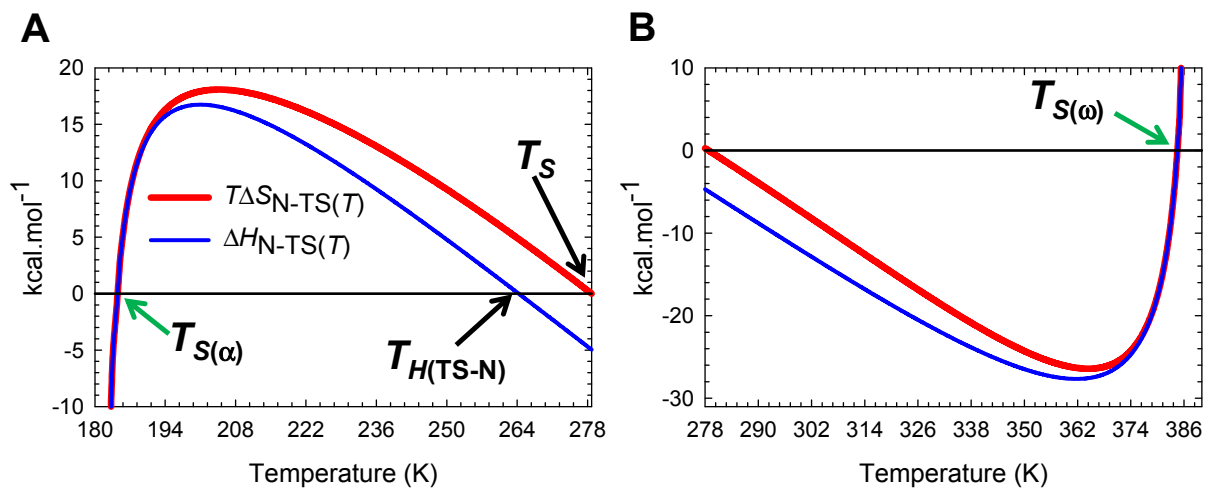

**Figure 13−figure supplement 3.**

**Deconvolution of the change in Gibbs energy for the partial folding reaction $[TS] \rightleftharpoons N$.**

These are appropriately scaled split views of **Figure 13−figure supplement 2B**. **(A)** For $T_\alpha \leq T < T_{S(\alpha)}$, $[TS] \rightleftharpoons N$ is entropically disfavoured ($T\Delta S_{\text{N-TS}(T)} < 0$) but is more than compensated by the exothermic enthalpy ($\Delta H_{\text{N-TS}(T)} < 0$), leading to $\Delta G_{\text{N-TS}(T)} < 0$. When $T = T_{S(\alpha)}$, $\Delta S_{\text{N-TS}(T)} = \Delta H_{\text{N-TS}(T)} = \Delta G_{\text{N-TS}(T)} = 0$, and the net flux of the conformers from the TSE to the NSE is zero. For $T_{S(\alpha)} < T < T_{H(\text{TS-N})}$, $[TS] \rightleftharpoons N$ is enthalpically unfavourable ($\Delta H_{\text{N-TS}(T)} > 0$) but is more than compensated by entropy ($T\Delta S_{\text{N-TS}(T)} > 0$) leading to $\Delta G_{\text{N-TS}(T)} < 0$. When $T = T_{H(\text{TS-N})}$, the net flux from the TSE to the NSE is driven purely by the favourable change in entropy ($\Delta G_{\text{N-TS}(T)} = -T\Delta S_{\text{N-TS}(T)} < 0$). For $T_{H(\text{TS-N})} < T < T_S$, the net flux of the conformers from the TSE to the NSE is entropically and enthalpically favourable ($\Delta H_{\text{N-TS}(T)} < 0$ and $T\Delta S_{\text{N-TS}(T)} > 0$) leading to $\Delta G_{\text{N-TS}(T)} < 0$. When $T = T_S$, the net flux is driven purely by the exothermic change in enthalpy ($\Delta G_{\text{N-TS}(T)} = \Delta H_{\text{N-TS}(T)} < 0$). **(B)** For $T_S < T < T_{S(\omega)}$, $[TS] \rightleftharpoons N$ is entropically unfavourable ($T\Delta S_{\text{N-TS}(T)} < 0$) but is more than compensated by the exothermic enthalpy ($\Delta H_{\text{N-TS}(T)} < 0$) leading to $\Delta G_{\text{N-TS}(T)} < 0$. When $T = T_{S(\omega)}$, $\Delta S_{\text{N-TS}(T)} = \Delta H_{\text{N-TS}(T)} = \Delta G_{\text{N-TS}(T)} = 0$, and the net flux of the conformers from the TSE to the NSE is zero. For $T_{S(\omega)} < T \leq T_\omega$, $[TS] \rightleftharpoons N$ is enthalpically unfavourable ($\Delta H_{\text{N-TS}(T)} > 0$) but is more than compensated by the favourable change in entropy ($T\Delta S_{\text{N-TS}(T)} > 0$), leading to $\Delta G_{\text{N-TS}(T)} < 0$.

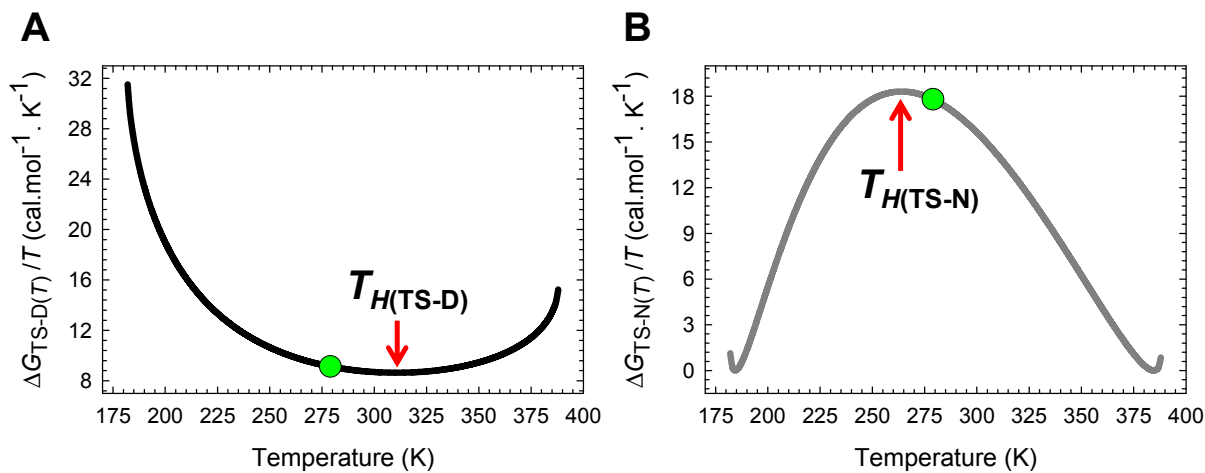

**Figure 14.**

**Temperature-dependence of the Massieu-Planck activation potentials.**

 **(A)** The Massieu-Planck activation potential for folding is a minimum at $T_{H(TS\text{-}D)}$. The slope of this curve is given by $-\Delta H_{TS\text{-}D(T)}/T^2$. **(B)** The Massieu-Planck activation potential for unfolding is a maximum at $T_{H(TS\text{-}N)}$ and a minimum (zero) at $T_{S(\alpha)}$ and $T_{S(\omega)}$. The slope of this curve is given by $-\Delta H_{TS\text{-}N(T)}/T^2$. The temperature $T_S$ at which $\Delta G_{TS\text{-}D(T)}$ and $\Delta G_{TS\text{-}N(T)}$ are a minimum and a maximum, respectively, are indicated by green circles.

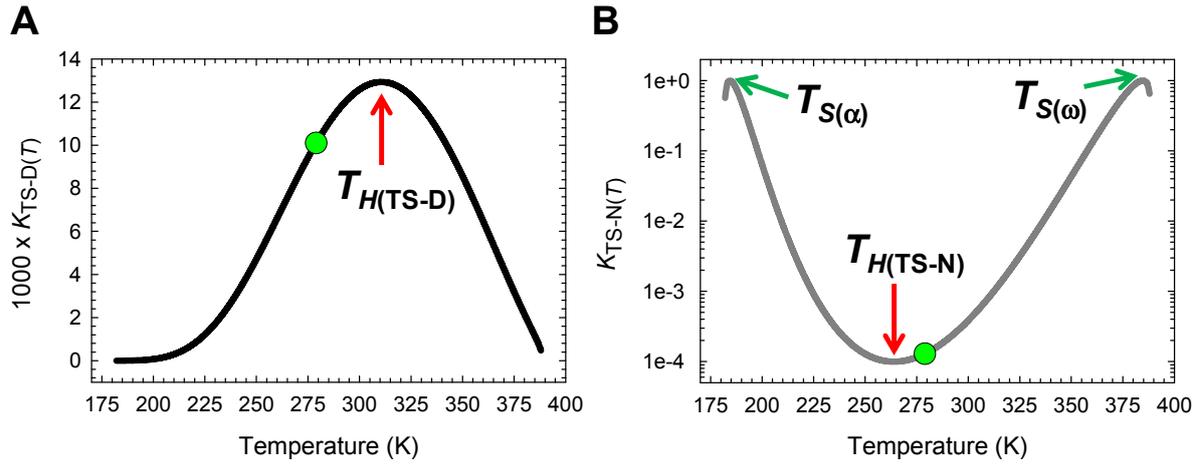

**Figure 14−figure supplement 1.**

**Temperature-dependence of $K_{TS-D(T)}$ and $K_{TS-N(T)}$.**

**(A)** Temperature-dependence of $K_{TS-D(T)} = [TS]/[D]$ for the partial folding reaction $D \rightleftharpoons [TS]$. $K_{TS-D(T)}$ is a maximum not when $\Delta G_{TS-D(T)}$ is a minimum (green circle) but when the Massieu-Planck activation potential for folding, $\Delta G_{TS-D(T)}/T$, is a minimum, and occurs precisely when $T = T_{H(TS-D)}$. The slope of this curve is given by $K_{TS-D(T)} \Delta H_{TS-D(T)}/RT^2$. **(B)** Temperature-dependence of $K_{TS-N(T)} = [TS]/[N]$ for the partial unfolding reaction $N \rightleftharpoons [TS]$. $K_{TS-N(T)}$ is a minimum not when $\Delta G_{TS-N(T)}$ is a maximum (green circle) but when the Massieu-Planck activation potential for unfolding, $\Delta G_{TS-N(T)}/T$, is a maximum, and occurs precisely when $T = T_{H(TS-N)}$. The slope of this curve is given by $K_{TS-N(T)} \Delta H_{TS-N(T)}/RT^2$. Note that $K_{TS-N(T)}$ is unity at $T_{S(\alpha)}$ and $T_{S(\omega)}$. It is not possible to capture the minimum of $K_{TS-N(T)}$ on a linear scale; hence the ordinate is shown on a log scale (base 10). The green circles represent the temperature $T_S$ at which $\Delta G_{D-N(T)}$ and $\Delta G_{TS-N(T)}$ are both a maximum, $\Delta G_{TS-D(T)}$ is a minimum, and the absolute entropies of the DSE, the TSE and the NSE are identical.

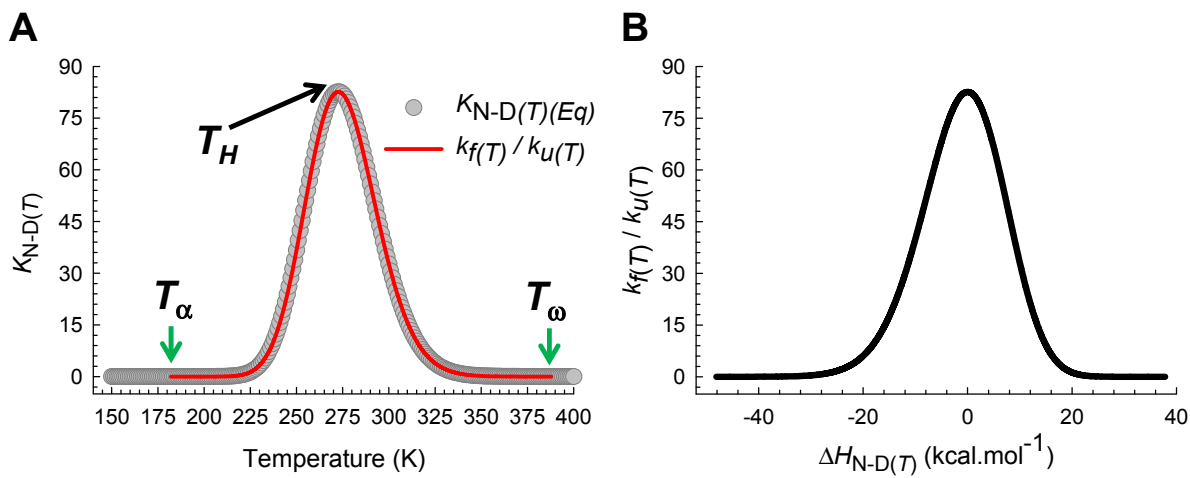

**Figure 14−figure supplement 2.**

**Temperature-dependence of the equilibrium constant for the reaction $D \rightleftharpoons N$.**

**(A)** An overlay of the ratio of the rate constants for folding and unfolding and the equilibrium constant derived from the Gibbs energy of folding at equilibrium. The curve fits to Boltzmann distribution and is a maximum at $T_H$. The slope of this curve is given by $K_{\text{N-D}(T)} \Delta H_{\text{N-D}(T)} / RT^2$. Although the value of $\Delta H_{\text{D-N}(T)}$ can be calculated for any temperature above absolute zero using Eq. (A1), it has physical meaning only for $T_\alpha \leq T \leq T_\omega$. This applies to $\Delta S_{\text{D-N}(T)}$ and $\Delta G_{\text{D-N}(T)}$ as well (Eqs. (A2) and (A3)). **(B)** The solubility of the NSE as compared to the DSE is the greatest when the net flux of the conformers from the DSE to the NSE is driven purely by the difference in entropy between these two reaction-states. The slope of this curve is given by $K_{\text{N-D}(T)} \Delta H_{\text{N-D}(T)} / \Delta C_{p\text{N-D}} RT^2$.

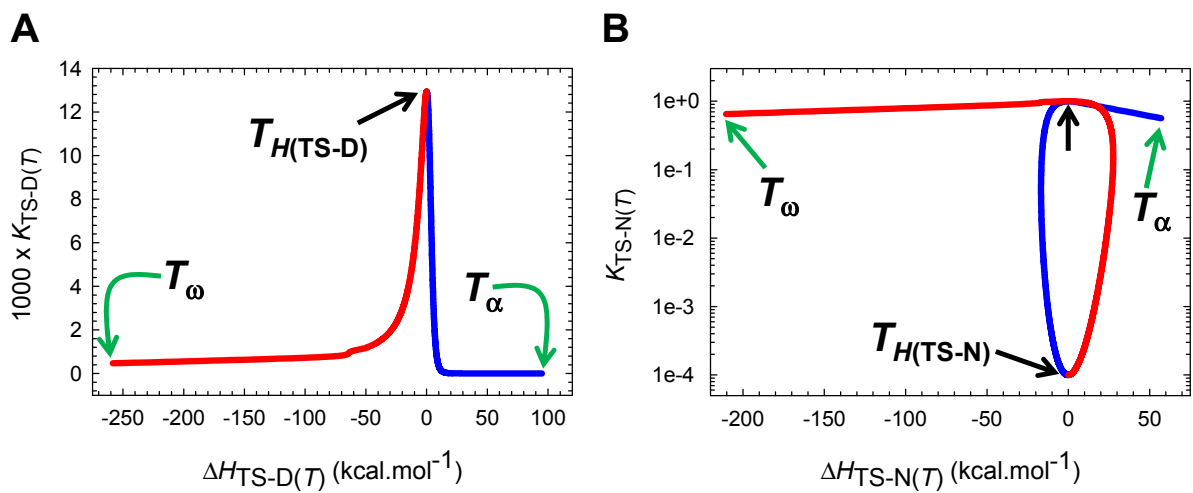

**Figure 14−figure supplement 3.**

**The solubility of the TSE relative to the DSE and the NSE across a broad temperature regime.**

(A) The solubility of the TSE as compared to the DSE is the greatest when $\Delta H_{\text{TS-D}(T)} = 0$, or equivalently, when the Gibbs barrier to folding is purely entropic. The slope of this curve is given by $K_{\text{TS-D}(T)} \Delta H_{\text{TS-D}(T)} / \Delta C_{p\text{TS-D}(T)} RT^2$. The blue and red sections of the curve represent the temperature regimes $T_\alpha \leq T \leq T_{H(\text{TS-D})}$ and $T_{H(\text{TS-D})} \leq T \leq T_\omega$, respectively. (B) The solubility of the TSE as compared to the NSE is the least when $\Delta H_{\text{TS-N}(T)} = 0$ and when the Gibbs barrier to unfolding is purely entropic. The slope of this curve is given by $K_{\text{TS-N}(T)} \Delta H_{\text{TS-N}(T)} / \Delta C_{p\text{TS-N}(T)} RT^2$. The point where the solubility of the TSE is identical to that of the NSE is indicated by the unlabelled black pointer, and described earlier, occurs precisely at $T_{S(\alpha)}$ and $T_{S(\omega)}$. The blue and red sections of the curve represent the temperature regimes $T_\alpha \leq T \leq T_{H(\text{TS-N})}$ and $T_{H(\text{TS-N})} \leq T \leq T_\omega$, respectively. Note that the ordinate is on a log scale (base 10).

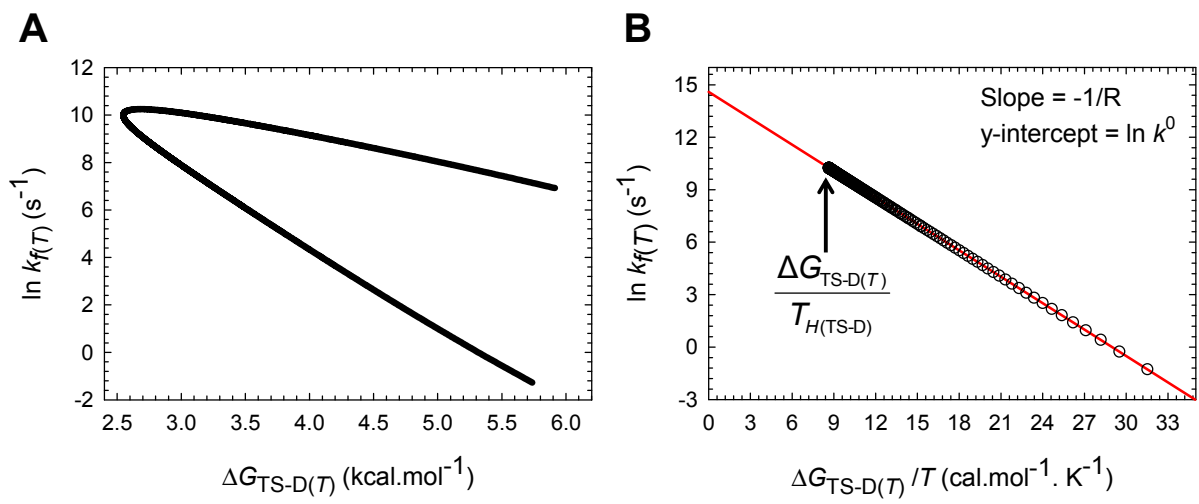

**Figure 14−figure supplement 4.**

**The natural logarithm of $k_{f(T)}$ is linearly dependent on the Massieu-Planck activation potential for folding.**

**(A)** The natural logarithm of $k_{f(T)}$ has a complex dependence on the Gibbs barrier to folding when explored over a large temperature range. The slope of this curve is given by $-\Delta H_{TS-D(T)}/\Delta S_{TS-D(T)}RT^2$. **(B)** The natural logarithm of $k_{f(T)}$ decreases linearly with an increase in the Massieu-Planck activation potential for folding, with the magnitude of the negative slope being given by the reciprocal of the gas constant. The *y*-intercept at zero Massieu-Planck potential yields the value of the prefactor. Naturally, $k_{f(T)}$ is a maximum when the magnitude of the Massieu-Planck function for folding is a minimum, and this occurs precisely at $T_{H(TS-D)}$.

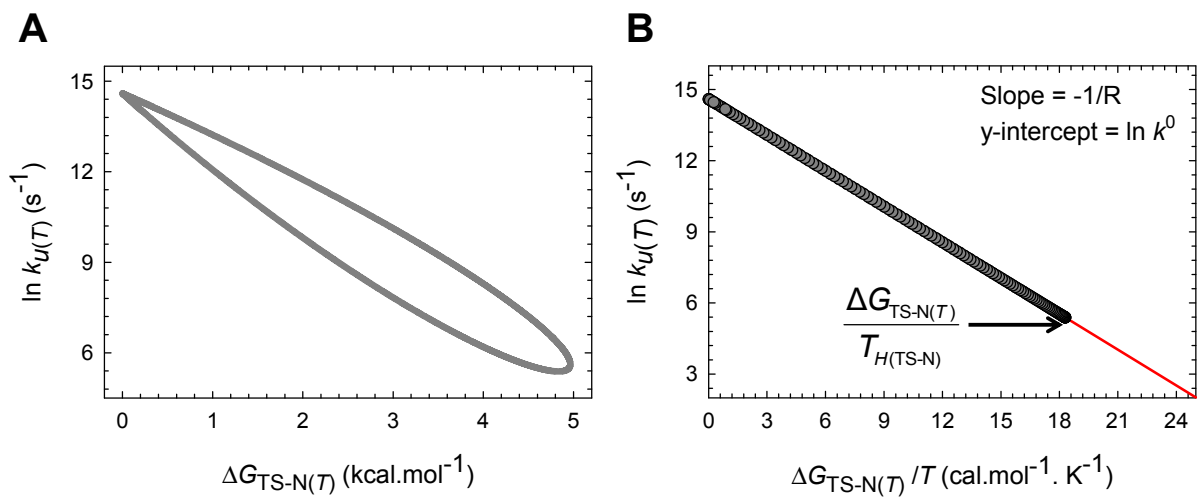

**Figure 14−figure supplement 5.**

**The natural logarithm of $k_{u(T)}$ is linearly dependent on the Massieu-Planck activation potential for unfolding.**

**(A)** The natural logarithm of $k_{u(T)}$ has a complex dependence on the Gibbs barrier to unfolding when explored over a large temperature range. The slope of this curve is given by $-\Delta H_{TS-N(T)}/\Delta S_{TS-N(T)}RT^2$. **(B)** The natural logarithm of $k_{u(T)}$ decreases linearly with an increase in the Massieu-Planck activation potential for unfolding, with the magnitude of the negative slope being given by the reciprocal of the gas constant. The *y*-intercept at zero Massieu-Planck potential yields the value of the prefactor. Naturally, $k_{u(T)}$ is a minimum when the magnitude of the Massieu-Planck function for unfolding is a maximum, and this occurs precisely at $T_{H(TS-N)}$. The reason why the data points for the unfolding rate constants extend all the way to the intercept is because the Gibbs barrier to unfolding becomes zero at $T_{S(\alpha)}$ and $T_{S(\omega)}$.

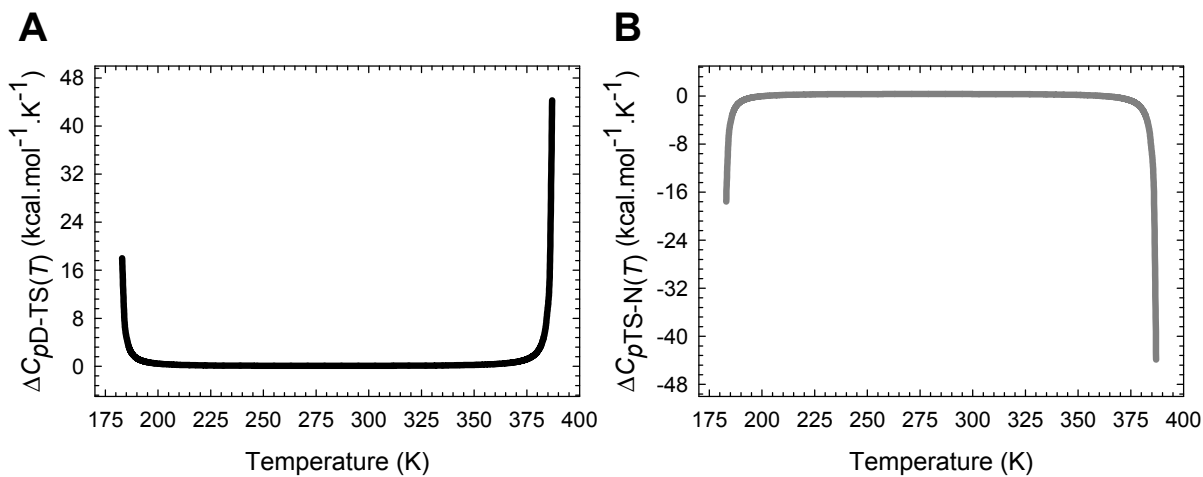

**Figure 15.**

**Temperature-dependence of $\Delta C_{p\text{D-TS}(T)}$ and $\Delta C_{p\text{TS-N}(T)}$ functions.**

**(A)** $\Delta C_{p\text{D-TS}(T)}$ is positive throughout the temperature range and is a minimum at $T_S$ (or $\Delta C_{p\text{TS-D}(T)}$ is a maximum or the least negative at $T_S$). **(B)** $\Delta C_{p\text{TS-N}(T)}$ is a maximum at $T_S$, positive for $T_{C_p\text{TS-N}(\alpha)} < T < T_{C_p\text{TS-N}(\omega)}$, negative for $T_\alpha \leq T < T_{C_p\text{TS-N}(\alpha)}$ and $T_{C_p\text{TS-N}(\omega)} < T \leq T_\omega$, and as described earlier, zero at $T_{C_p\text{TS-N}(\alpha)}$ and $T_{C_p\text{TS-N}(\omega)}$. These aspects can be better appreciated from the appropriately scaled views shown in the figure supplement. Note that the algebraic sum of $\Delta C_{p\text{D-TS}(T)}$ and $\Delta C_{p\text{TS-N}(T)}$ must equal $\Delta C_{p\text{D-N}}$ throughout the temperature-regime.

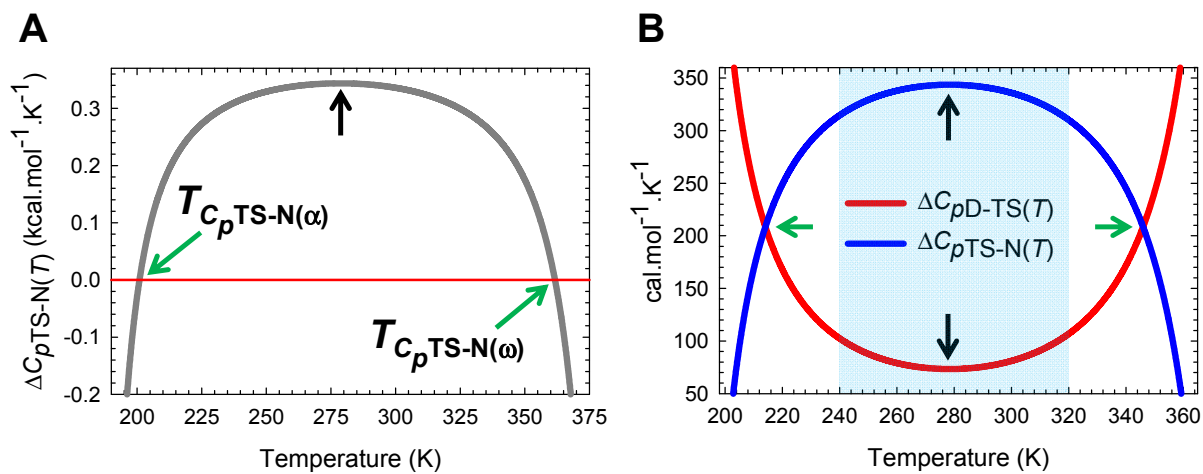

**Figure 15−figure supplement 1.**

**Appropriately scaled $\Delta C_{pD\text{-}TS(T)}$ and $\Delta C_{pTS\text{-}N(T)}$ functions to showcase their features.**

**(A)** $\Delta C_{pTS\text{-}N(T)}$ which is a maximum and positive at $T_S$, decreases with any deviation in temperature from $T_S$, is zero at $T_{C_p TS\text{-}N(\alpha)}$ and $T_{C_p TS\text{-}N(\omega)}$, and negative for $T_\alpha \leq T < T_{C_p TS\text{-}N(\alpha)}$ and $T_{C_p TS\text{-}N(\omega)} < T \leq T_\omega$. **(B)** At the temperatures where the $\Delta C_{pD\text{-}TS(T)}$ and $\Delta C_{pTS\text{-}N(T)}$ functions intersect (214.1K and 345.9 K), the absolute heat capacity of the TSE is exactly half the sum of the absolute heat capacities of the DSE and the NSE. The black pointers indicate that the extrema of $\Delta C_{pD\text{-}TS(T)}$ and $\Delta C_{pTS\text{-}N(T)}$ functions, while the green pointers indicate their intersection. Inspection shows that $\Delta C_{pTS\text{-}N(T)} > \Delta C_{pD\text{-}TS(T)}$ for 240 K < $T$ < 320 K (shaded region), and is approximately five fold greater than $\Delta C_{pTS\text{-}N(T)}$ at $T_S$ (343.7/73.3 = ~ 4.7) despite ~30% and ~70% of the total change in SASA for the unfolding reaction $N \rightleftharpoons D$ occurring in the partial unfolding reactions $N \rightleftharpoons [TS]$ and $[TS] \rightleftharpoons D$, respectively (**Figure 2−figure supplement 1**).

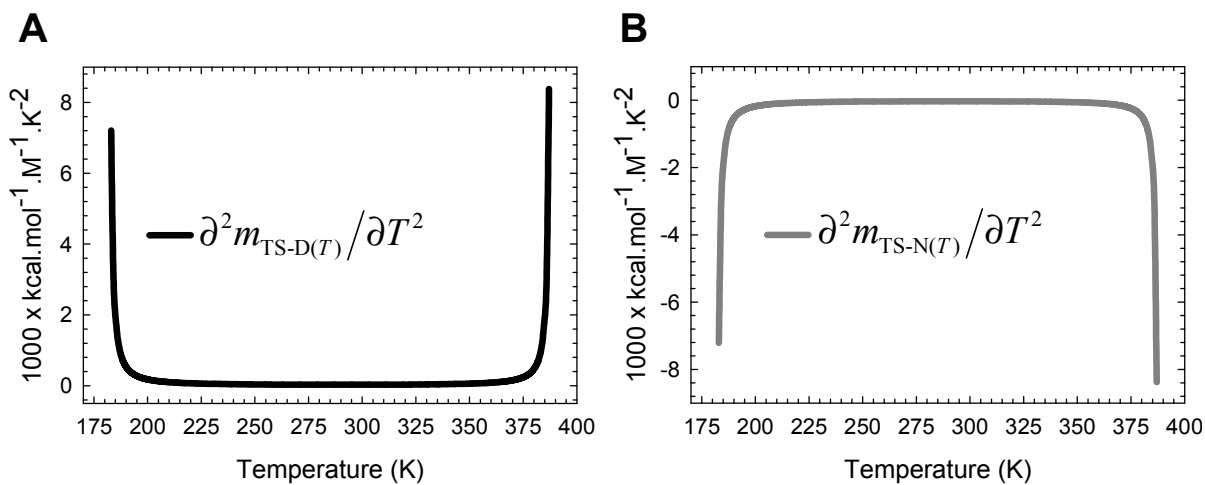

**Figure 15−figure supplement 2.**

**The second derivatives of $m_{TS\text{-}D(T)}$ and $m_{TS\text{-}N(T)}$ with respect to temperature.**

**(A)** The second derivative of $m_{TS\text{-}D(T)}$ according to Eq. (A9). **(B)** The second derivative of $m_{TS\text{-}N(T)}$ according to Eq. (A10). The sole intent of these figures is to demonstrate that the gross features of the temperature-dependence of the heat capacity functions arise primarily from the second derivatives of the temperature-dependent shift in the position of the TSE relative to the vertices of the DSE or the NSE Gibbs parabolas along the RC. See **Figure 15−figure supplement 3** for the location of the extrema of these two functions.

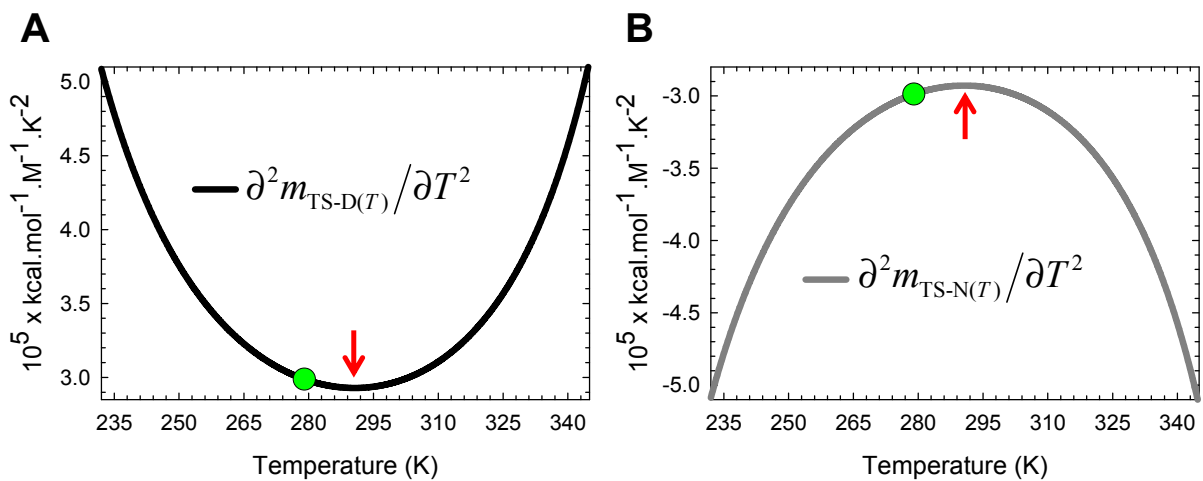

**Figure 15−figure supplement 3.**

**The extrema of the second derivatives of $m_{\text{TS-D}(T)}$ and $m_{\text{TS-N}(T)}$ with respect to temperature are not at $T_S$.**

The sole intent of these appropriately scaled figures is to demonstrate that although the gross features of the temperature-dependence of the heat capacity functions arise predominantly from the second derivatives of the temperature-dependent shift in the position of the TSE along the RC, the minimum of $\partial^2 m_{\text{TS-D}(T)}/\partial T^2$ and the maximum of $\partial^2 m_{\text{TS-N}(T)}/\partial T^2$ do not occur at $T_S$ (green circles), and is apparent from comparison of Eqs. (12), (13), (A9) and (A10).

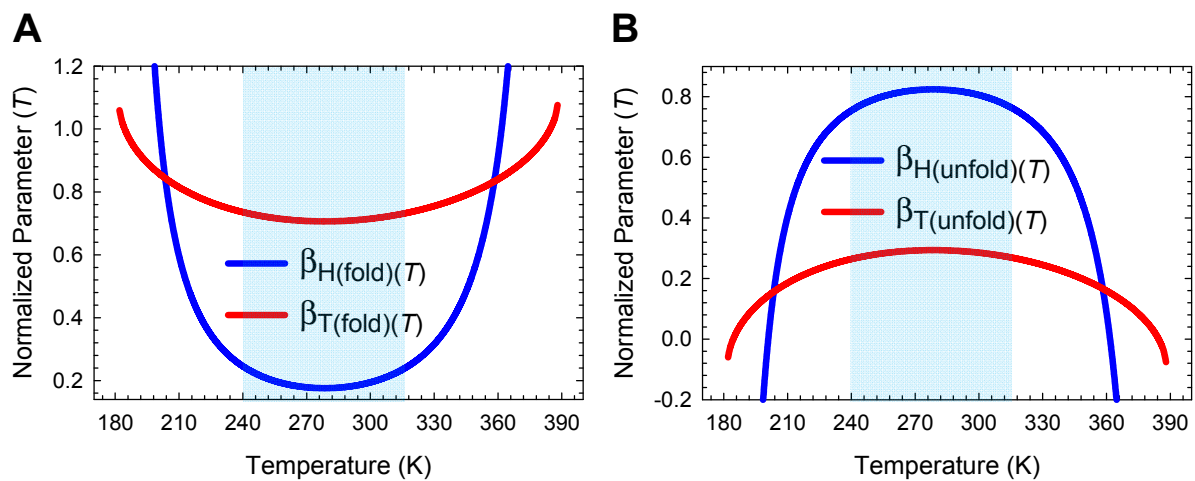

**Figure 16.**

**Comparison of the position of the TSE along the heat capacity RC and the SASA-RC.**

**(A)** Although the temperature-dependences of $\beta_{H(fold)(T)}$ is consistent with $\beta_{T(fold)(T)}$, and both are a minimum at $T_S$, their magnitudes are not even remotely similar across a large temperature regime (240 K < $T$ < 320 K, shaded area); and when $T = T_S$, $\beta_{T(fold)(T)}$ is four fold greater than $\beta_{H(fold)(T)}$ (0.7063/0.1759 = 4.0). Note that the position of the TSE relative to the DSE along the heat capacity and SASA-RCs are identical at the points of intersection (203.6 K and 358.3 K). **(B)** Although the temperature-dependence of $\beta_{H(unfold)(T)}$ is consistent with $\beta_{T(unfold)(T)}$, and both are a maximum at $T_S$, $\beta_{H(unfold)(T)} > \beta_{T(unfold)(T)}$ for 240 K < $T$ < 320 K; and when $T = T_S$, $\beta_{H(unfold)(T)}$ is ~2.8 fold greater than $\beta_{T(unfold)(T)}$ (0.8241/0.2937 = 2.81). The position of the TSE relative to the NSE along the heat capacity and SASA-RCs are identical at the points of intersection (203.6 K and 358.3 K). See **Figure 16−figure supplement 1** for unscaled plots of $\beta_{H(fold)(T)}$ and $\beta_{H(unfold)(T)}$.

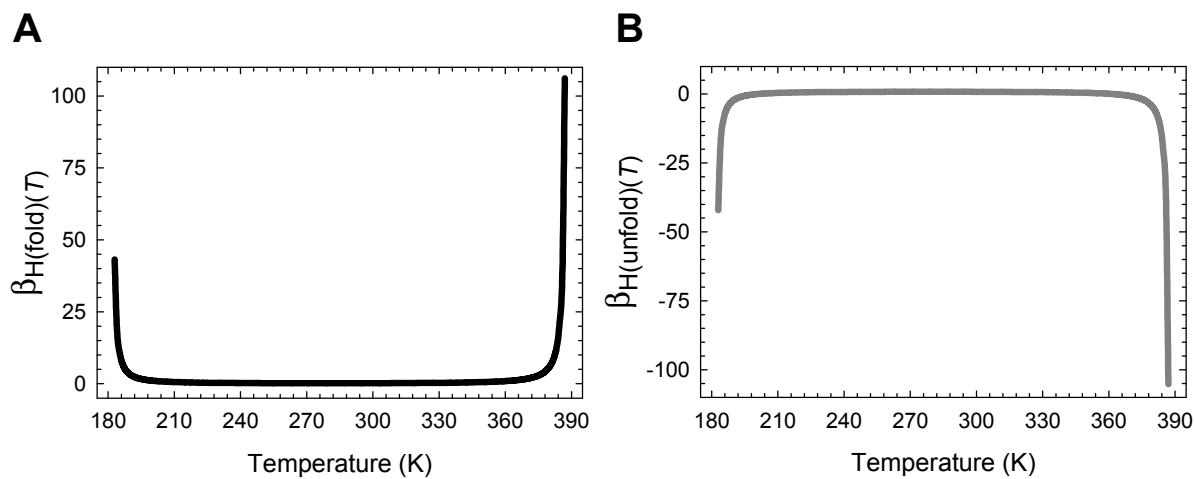

**Figure 16−figure supplement 1.**

**Temperature-dependence of $\beta_{H(fold)(T)}$ and $\beta_{H(unfold)(T)}$.**

**(A)** Variation in $\beta_{H(fold)(T)}$ with temperature according to Eq. (26). **(B)** Variation in $\beta_{H(unfold)(T)}$ with temperature according to Eq. (27). The location of the extrema is not apparent in these figures. Note that although the algebraic sum of $\beta_{H(fold)(T)}$ and $\beta_{H(unfold)(T)}$ must always be unity for a two-state system, they need not be individually restricted to a canonical range of 0 to 1.

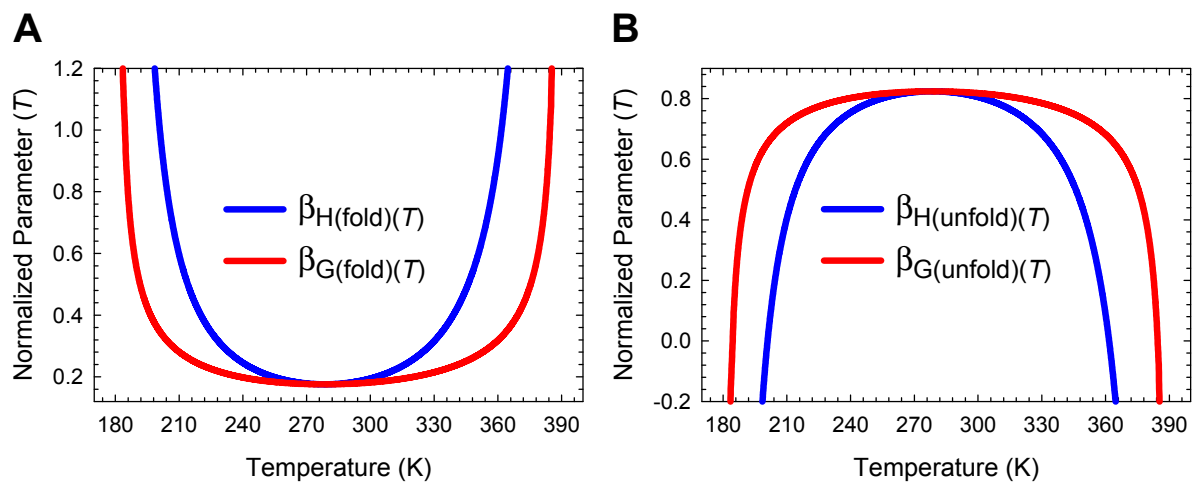

**Figure 17.**

**Comparison of the position of the TSE along the heat capacity and the entropic RCs.**

**(A)** The position of the TSE with respect to the DSE along the heat capacity and the entropic RCs are identical at $T_S$, and non-identical for $T \neq T_S$. **(B)** The position of the TSE with respect to the NSE along the heat capacity and the entropic RCs are identical at $T_S$, and non-identical for $T \neq T_S$. See **Figure 17−figure supplement 1** for unscaled plots of $\beta_{G(fold)(T)}$ and $\beta_{G(unfold)(T)}$.

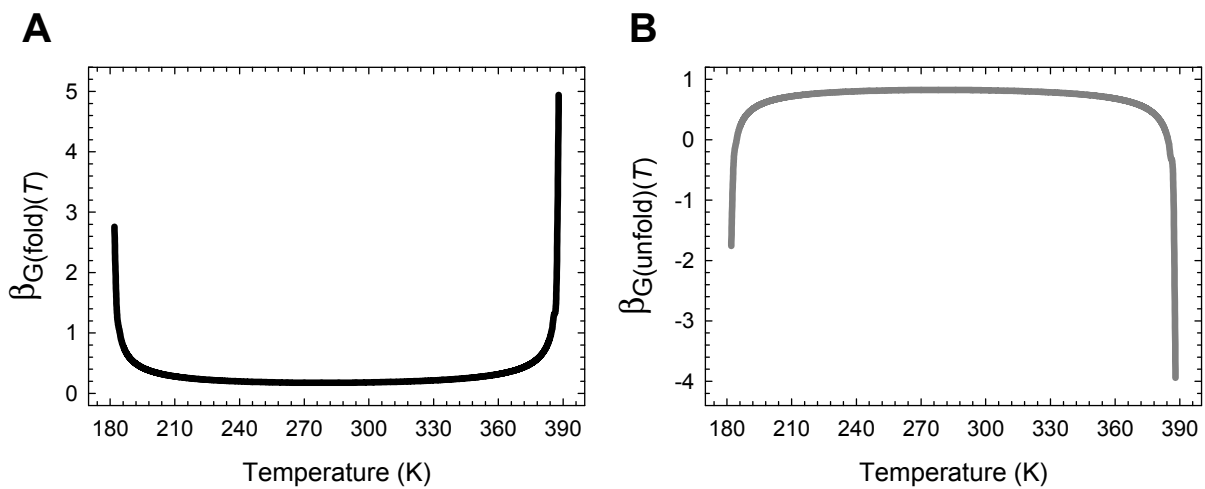

**Figure 17−figure supplement 1.**

**Temperature-dependence of $\beta_{G(fold)(T)}$ and $\beta_{G(unfold)(T)}$.**

**(A)** Variation in $\beta_{G(fold)(T)}$ with temperature according to Eq. (33). **(B)** Variation in $\beta_{G(unfold)(T)}$ with temperature according to Eq. (35). The location of the extrema is not apparent in these figures. Note that although the algebraic sum of $\beta_{G(fold)(T)}$ and $\beta_{G(unfold)(T)}$ must always be unity for a two-state system, they need not be individually restricted to a canonical range of 0 to 1.

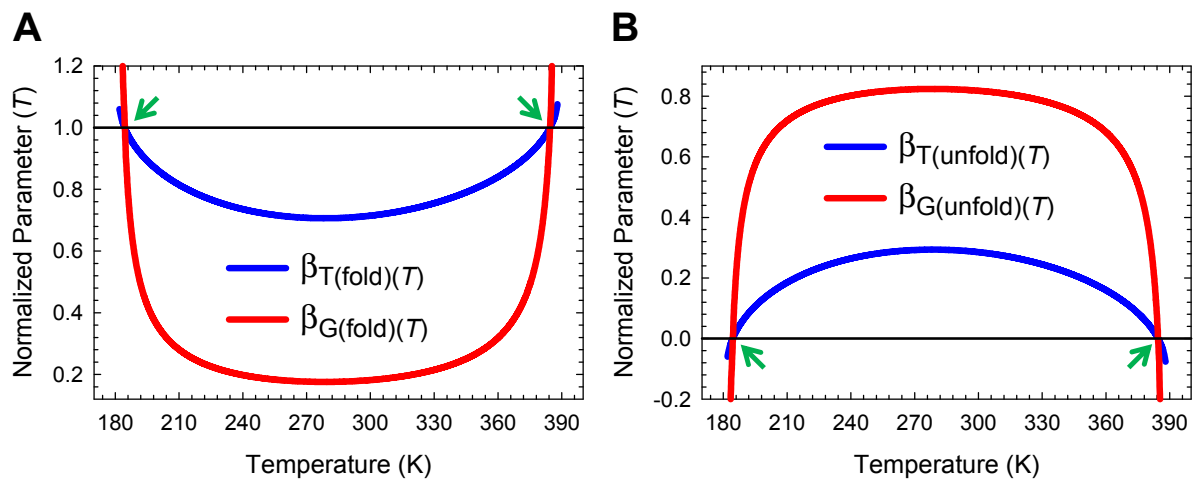

**Figure 18.**

**Comparison of the position of the TSE along the heat capacity and the entropic RCs.**

**(A)** The position of the TSE with respect to the DSE along the SASA and the entropic RCs are identical at $T_{S(\alpha)}$ and $T_{S(\omega)}$, and dissimilar for $T \neq T_{S(\alpha)}$ and $T_{S(\omega)}$. **(B)** The position of the TSE with respect to the NSE along the SASA and the entropic RCs are identical at $T_{S(\alpha)}$ and $T_{S(\omega)}$, and dissimilar for $T \neq T_{S(\alpha)}$ and $T_{S(\omega)}$. The green pointers indicate $T_{S(\alpha)}$ and $T_{S(\omega)}$.

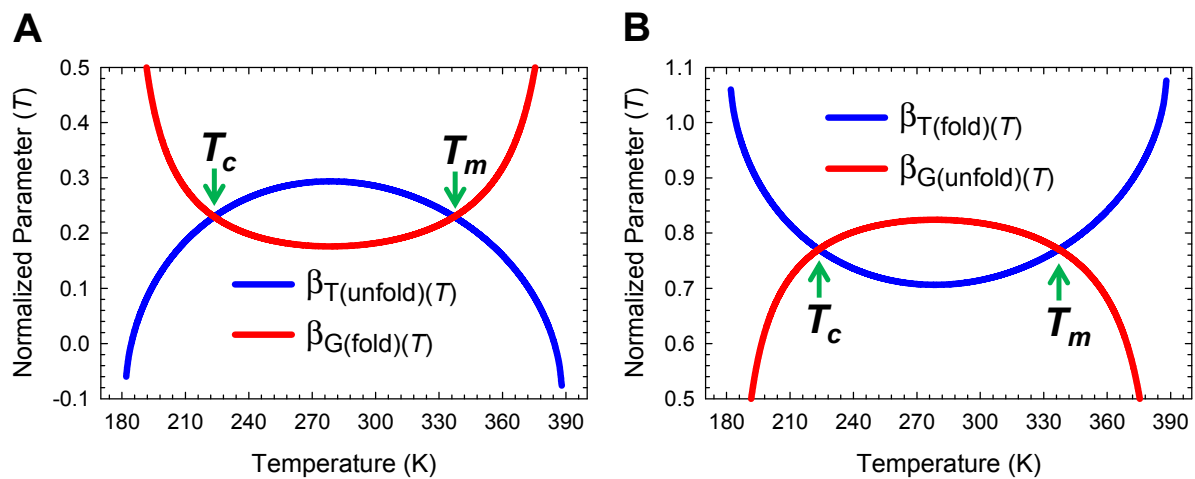

**Figure 19.**

**The intersection of $\beta_{G(T)}$ and $\beta_{T(T)}$ functions.**

**(A)** At the midpoint of cold or heat denaturation, the position of the TSE relative to the DSE along the entropic RC is identical to the position of the TSE relative to the NSE along the SASA-RC. **(B)** The position of the TSE relative to the NSE along the entropic RC is identical to the position of the TSE relative to the DSE along the SASA-RC at the midpoint of cold or heat denaturation.

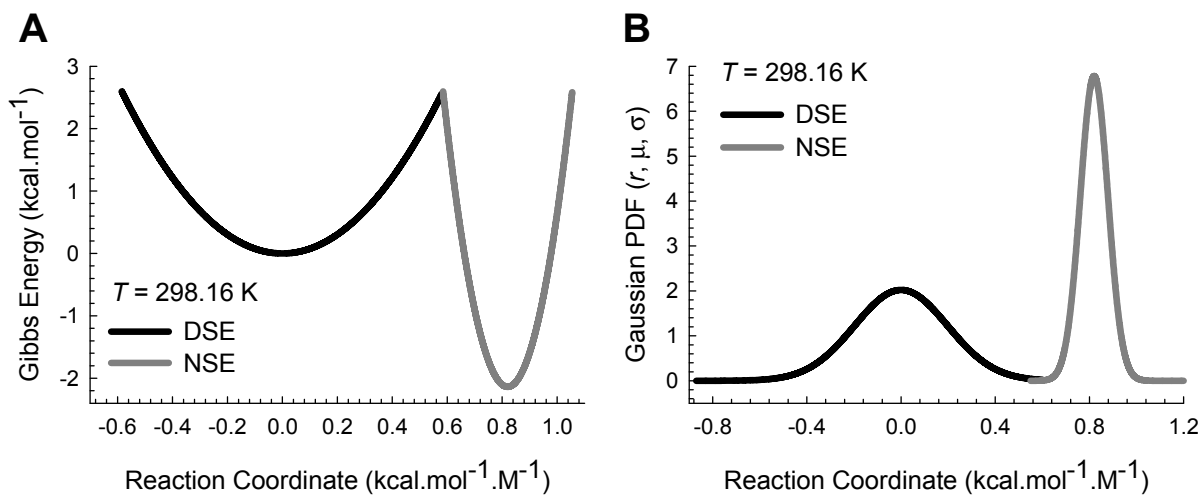

**Figure 19−figure supplement 1.**

**The correspondence between Gibbs parabolas and Gaussian PDFs.**

**(A)** Parabolic Gibbs energy curves with $\alpha$ = 7.594 $M^2.mol.kcal^{-1}$ and $\omega$ = 85.595 $M^2.mol.kcal^{-1}$, $m_{D-N}$ = 0.82 $kcal.mol^{-1}.M^{-1}$ and $\Delta G_{D-N(T)}$ = 2.138 $kcal.mol^{-1}$. The separation between *curve-crossing* and the vertices of the DSE and the NSE-parabolas along the abscissa are 0.5848 $kcal.mol^{-1}.M^{-1}$ and 0.2352 $kcal.mol^{-1}.M^{-1}$, respectively. The absolute values of $\Delta G_{TS-D(T)}$ and $\Delta G_{TS-N(T)}$ are 2.597 $kcal.mol^{-1}$ and 4.735 $kcal.mol^{-1}$, respectively. The parabolas have been generated as described in the legend for **Figure 3**. **(B)** Gaussian PDFs for the DSE and NSE generated using $p(r) = \dfrac{1}{\sqrt{2\pi\sigma^2}} \exp\left[-(r-\mu)^2/2\sigma^2\right]$, where $r$ is any point on the abscissa, $\mu$ = 0 $kcal.mol^{-1}.M^{-1}$ and $\sigma^2 = RT/2\alpha$ for the DSE-Gaussian, and $\mu$ = 0.82 $kcal.mol^{-1}.M^{-1}$ and $\sigma^2 = RT/2\omega$ for the NSE-Gaussian. The units for the Gaussian variances are in $kcal^2.mol^{-2}.M^{-2}$. The relationship between equilibrium stability and the areas enclosed by the DSE and the NSE Gaussians has been addressed in Paper -I.

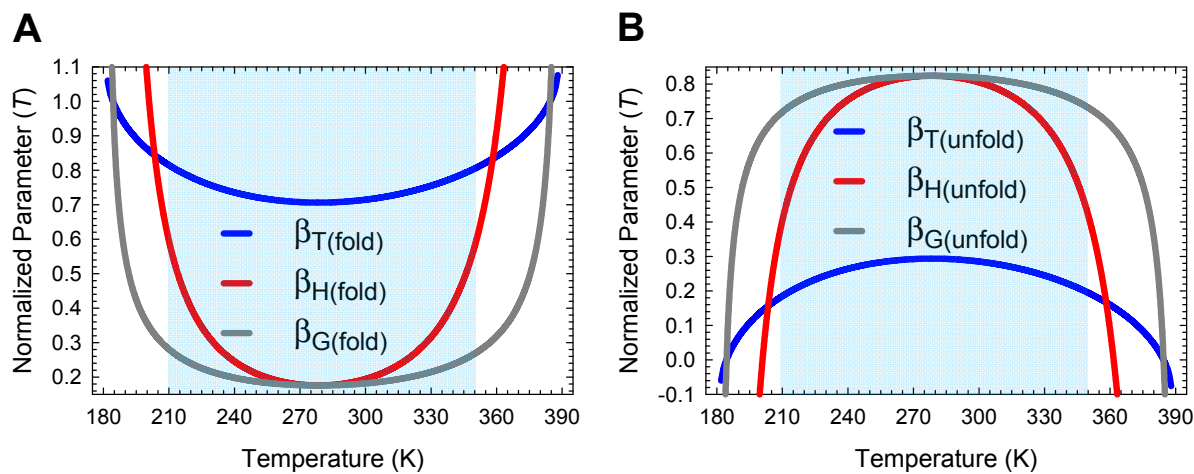

**Figure 20.**

**Comparison of the position of the TSE along the various reaction coordinates.**

The position of the TSE relative to the ground states depends on the choice of the RC and changes in a complex manner with temperature. **(A)** For 210 K~ < $T$ < ~350 K (shaded region), the position of the TSE relative to the DSE is the most advanced along the solvent RC as compared to the heat capacity and entropic RCs; and for $T \neq T_S$, is the most advanced along the heat capacity RC as compared to the entropic RC **(B)** In contrast, for 210 K~ < $T$ < ~350 K and $T \neq T_S$, the position of the TSE relative to the NSE is the most advanced along the entropic RC as compared to the heat capacity RC, and is the least advanced along the SASA-RC..

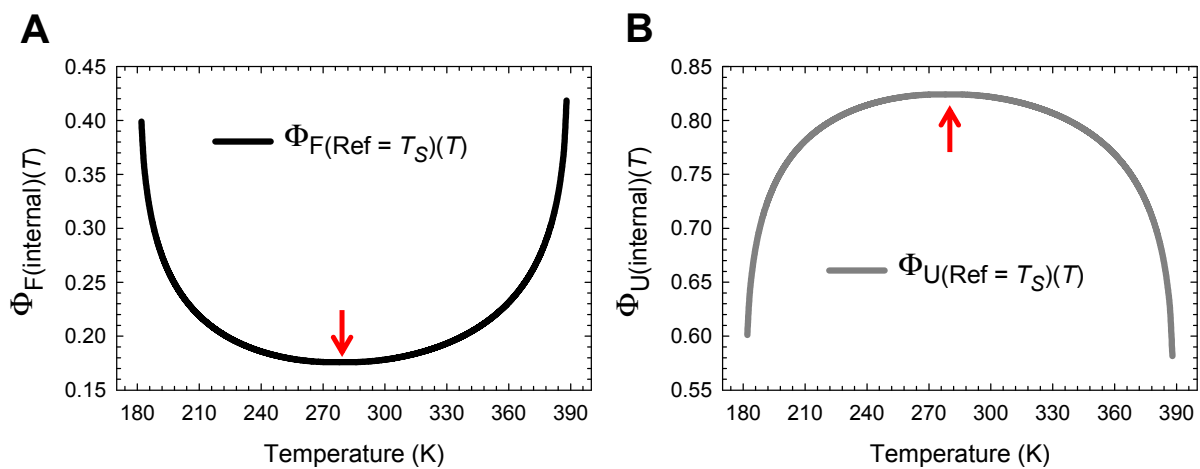

**Figure 21.**

**Temperature-dependence of Φ values when the protein at $T_S$ is defined as the wild type.**

**(A)** Temperature-dependence of $\Phi_{F(internal)(T)}$. **(B)** Temperature-dependence of $\Phi_{U(internal)(T)}$. The red pointers indicate the extrema of the functions. The discontinuities in the curves which must occur at $T_S$ have been removed by mathematically manipulating Eqs. (44) and (45) (manipulated equations not shown). Nevertheless, $\Phi_{F(internal)(T)}$ and $\Phi_{U(internal)(T)}$ are undefined at $T_S$ (i.e., the curves have holes at $T_S$ which is not obvious). Note that the mathematical stipulation that $\Phi_{F(internal)(T)} + \Phi_{U(internal)(T)} = 1$ for a two-state system is satisfied for all temperatures.

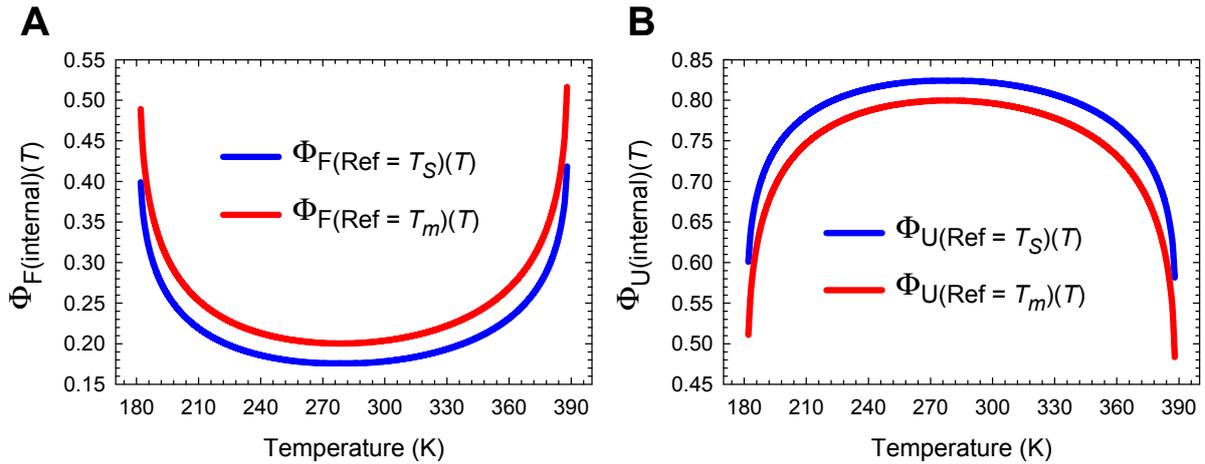

**Figure 21−figure supplement 1.**

**The magnitude of $\Phi_{(internal)(T)}$ is dependent on the definition of the wild type.**

**(A)** $\Phi_{F(internal)(T)}$ calculated using the protein at $T_S$ as the wild type must always be lower than $\Phi_{F(internal)(T)}$ calculated using protein at $T \neq T_S$ as the wild type. **(B)** $\Phi_{U(internal)(T)}$ calculated using the protein at $T_S$ as the wild type must always be greater than $\Phi_{U(internal)(T)}$ calculated using protein at $T \neq T_S$ as the wild type. For the blue curves we have $\Delta G_{(wt)} \equiv \Delta G_{(T_S)}$, and for the red curves, we have $\Delta G_{(wt)} \equiv \Delta G_{(T_m)}$. This notation applies to both equilibrium and activation energies. The blue curves are undefined (0/0) at $T_S$, and the red curves are undefined at $T_c$ and $T_m$. Note that the mathematical stipulation that $\Phi_{F(internal)(T)} + \Phi_{U(internal)(T)} = 1$ for a two-state system is satisfied for both the blue and the red curves for all temperatures.

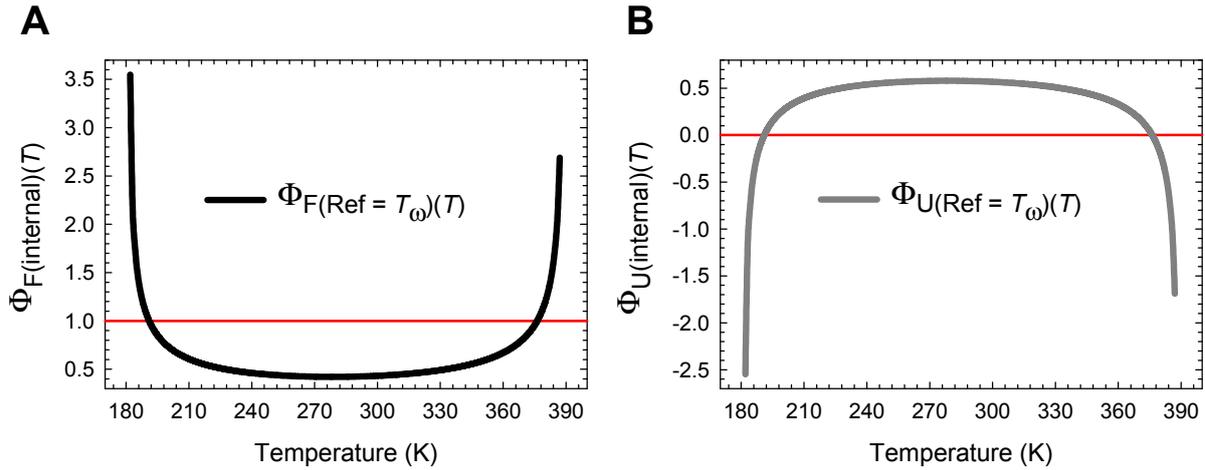

**Figure 21−figure supplement 2.**

**Φ values can be greater than unity, negative or zero depending on the definition of the reference state.**

**(A)** $\Phi_{F(internal)(T)}$ calculated by defining the protein at $T_\omega$ as the wild type. **(B)** $\Phi_{U(internal)(T)}$ calculated by defining the protein at $T_\omega$ as the wild type. Although the mathematical stipulation that $\Phi_{F(internal)(T)} + \Phi_{U(internal)(T)} = 1$ for a two-state system is satisfied for all temperatures, Φ values for folding and unfolding are not restricted to the canonical range of 0 ≤ Φ ≤ 1 when the protein at $T_\omega$ is defined as the reference or the wild type. Note that the curves are undefined at $T_\omega$.

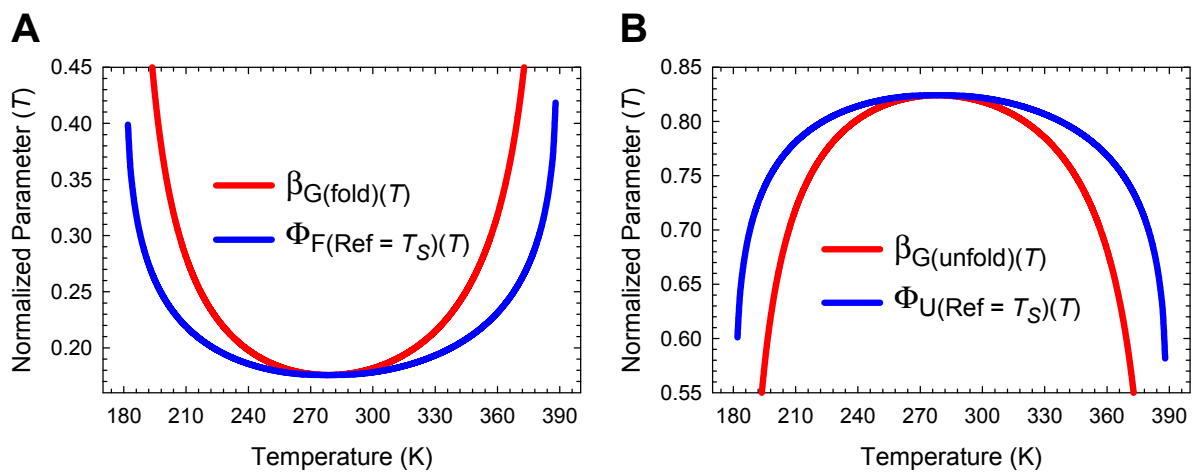

**Figure 21−figure supplement 3.**

**Comparison of Leffler $\beta_{G(T)}$ and Fersht $\Phi_{(internal)(T)}$ when the protein at $T_S$ is defined as the wild type.**

**(A)** $\beta_{G(fold)(T)}$ is almost identical to $\Phi_{F(internal)(T)}$ around the temperature of maximum stability, but as the temperature deviates from $T_S$, $\beta_{G(fold)(T)}$ increases far more steeply than $\Phi_{F(internal)(T)}$, such that for $T \neq T_S$ we have $\beta_{G(fold)(T)} > \Phi_{F(internal)(T)}$. **(B)** $\beta_{G(unfold)(T)}$ is almost identical to $\Phi_{U(internal)(T)}$ around the temperature of maximum stability, but as the temperature deviates from $T_S$, $\beta_{G(unfold)(T)}$ decreases far more steeply than $\Phi_{U(internal)(T)}$, such that for $T \neq T_S$ we have $\beta_{G(unfold)(T)} < \Phi_{U(internal)(T)}$.

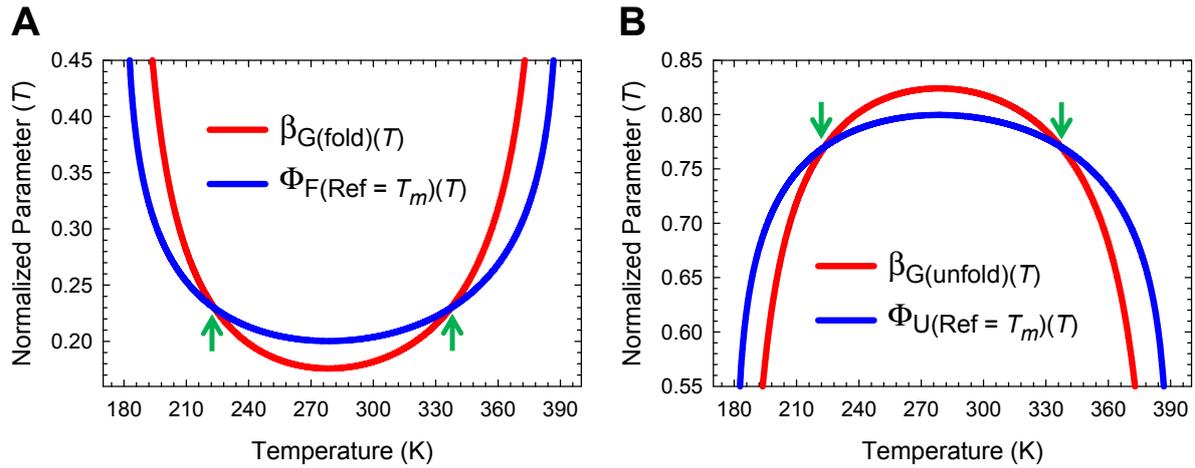

**Figure 21−figure supplement 4.**

**Comparison of the Leffler $\beta_{G(T)}$ and Fersht $\Phi_{(internal)(T)}$ when the protein at $T_m$ is defined as the wild type.**

**(A)** $\beta_{G(fold)(T)} < \Phi_{F(internal)(T)}$ for $T_c < T < T_m$ and $\beta_{G(fold)(T)} > \Phi_{F(internal)(T)}$ for $T < T_c$ and $T > T_m$.
**(B)** $\beta_{G(unfold)(T)} > \Phi_{U(internal)(T)}$ for $T_c < T < T_m$ and $\beta_{G(unfold)(T)} < \Phi_{U(internal)(T)}$ for $T < T_c$ and $T > T_m$. Note that $\Phi_{F(internal)(T)}$ and $\Phi_{U(internal)(T)}$ are undefined for $T_c$ and $T_m$ (green pointers).

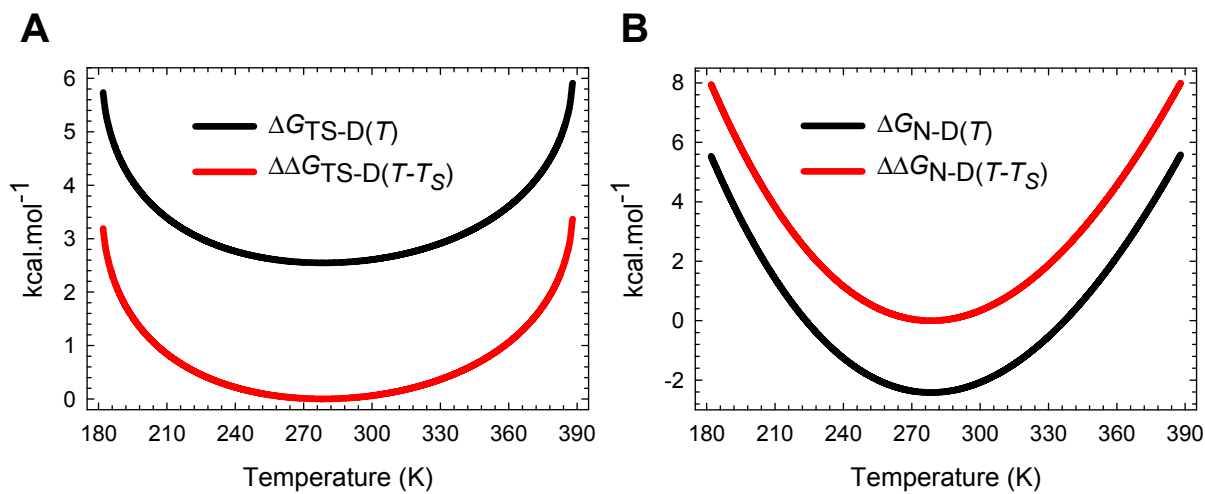

**Figure 21−figure supplement 5.**

**Transformation of $\Delta G_{TS\text{-}D(T)}$ and $\Delta G_{N\text{-}D(T)}$ to generate $\Phi_{F(internal)(T)}$ when the protein at $T_S$ is defined as the wild type.**

**(A)** The transformation $\Delta G_{TS\text{-}D(T)} - \Delta G_{TS\text{-}D(T_S)}$ (the numerator in Eq. (44)) lowers the $\Delta G_{TS\text{-}D(T)}$ function such that the value of $\Delta\Delta G_{TS\text{-}D(T\text{-}T_S)}$ is zero at the reference temperature. **(B)** The transformation $\Delta G_{N\text{-}D(T)} - \Delta G_{N\text{-}D(T_S)}$ (the denominator in Eq. (44)) raises the $\Delta G_{N\text{-}D(T)}$ function such that the value of $\Delta\Delta G_{N\text{-}D(T\text{-}T_S)}$ is zero at the reference temperature. The unmodified and the transformed curves are shown in black and red, respectively.

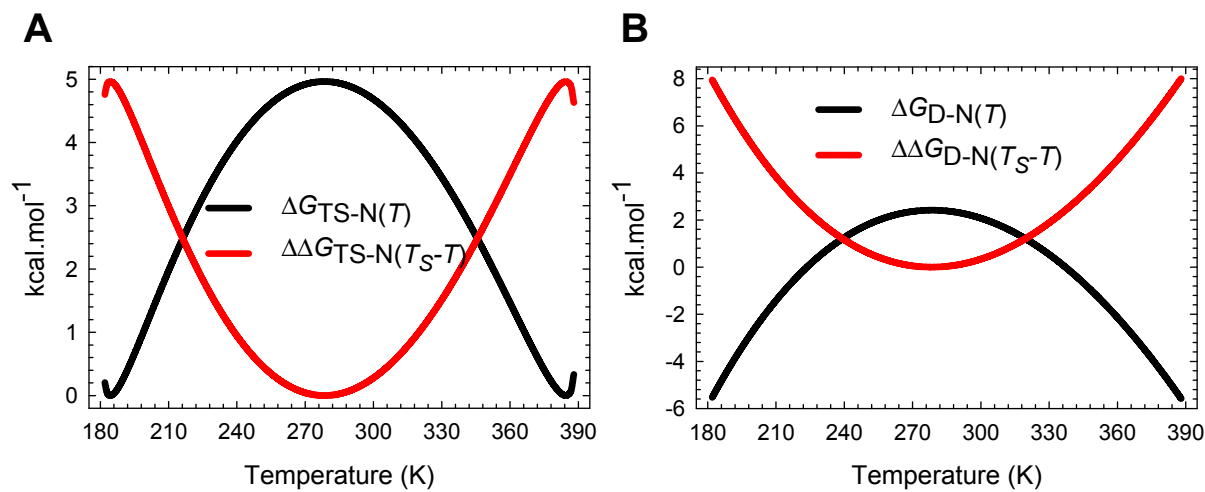

**Figure 21−figure supplement 6.**

**Transformation of $\Delta G_{TS\text{-}N(T)}$ and $\Delta G_{D\text{-}N(T)}$ to generate $\Phi_{U(internal)(T)}$ when the protein at $T_S$ is defined as the wild type.**

**(A)** The transformation $\Delta G_{TS\text{-}N(T_S)} - \Delta G_{TS\text{-}N(T)}$ (the numerator in Eq. (45)) flips the $\Delta G_{TS\text{-}N(T)}$ function vertically and concomitantly shifts it along the ordinate such that the value of $\Delta\Delta G_{TS\text{-}N(T_S\text{-}T)}$ at the reference temperature is zero. **(B)** The transformation $\Delta G_{D\text{-}N(T_S)} - \Delta G_{D\text{-}N(T)}$ (the denominator in Eq. (45)) flips the $\Delta G_{D\text{-}N(T)}$ function vertically and concomitantly shifts it along the ordinate such that the value of $\Delta\Delta G_{D\text{-}N(T_S\text{-}T)}$ at the reference temperature is zero. The unmodified and the transformed curves are shown in black and red, respectively.

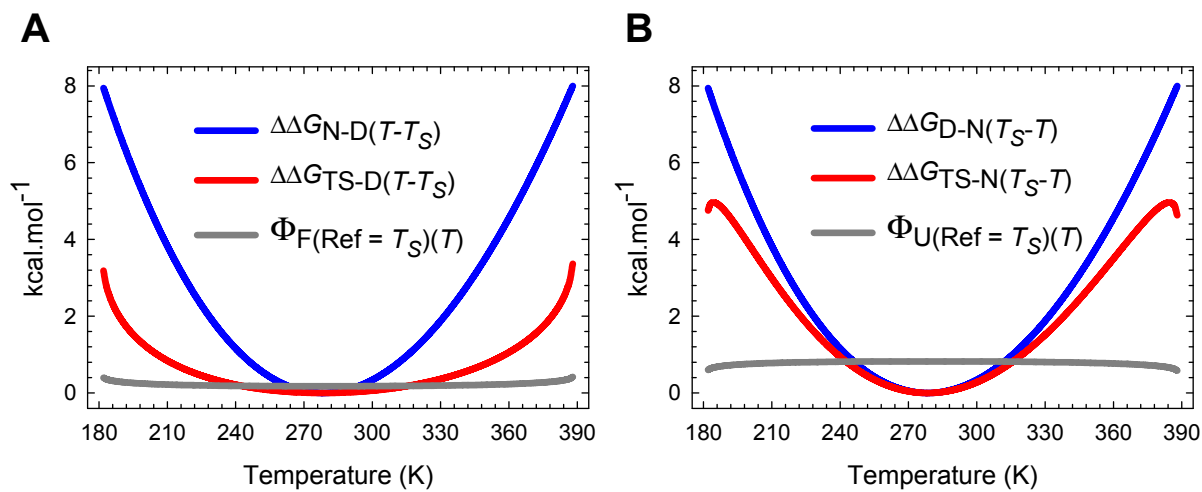

**Figure 21−figure supplement 7.**

**An overlay of transformed curves and $\Phi_{(internal)(T)}$ when the protein at $T_S$ is defined as the wild type.**

**(A)** Dividing $\Delta\Delta G_{TS-D(T-T_S)}$ by $\Delta\Delta G_{N-D(T-T_S)}$ generates $\Phi_{F(internal)(T)}$ with its minimum at $T_S$. **(B)** Dividing $\Delta\Delta G_{TS-N(T_S-T)}$ by $\Delta\Delta G_{D-N(T_S-T)}$ generates $\Phi_{U(internal)(T)}$ with its maximum at $T_S$. Note that the dimensions of the ordinate apply only to the red and the blue curves since $\Phi_{(internal)(T)}$ is dimensionless.

Page **118** of **129**

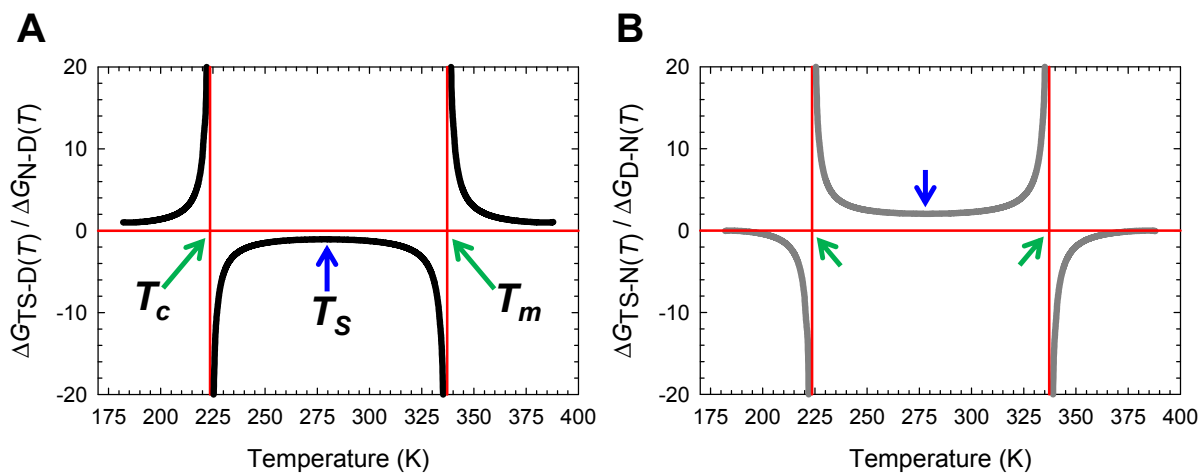

**Figure 21−figure supplement 8.**

**Temperature-dependence of the ratio of the Gibbs activation energies to stability.**

**(A)** The ratio $\Delta G_{TS-D(T)}/\Delta G_{N-D(T)}$ is negative for $T_c < T < T_m$ and positive for $T < T_c$ and $T > T_m$. **(B)** The ratio $\Delta G_{TS-N(T)}/\Delta G_{D-N(T)}$ is positive for $T_c < T < T_m$ and negative for $T < T_c$ and $T > T_m$. The vertical asymptotes are a consequence of $\Delta G_{D-N(T)} = -\Delta G_{N-D(T)}$ approaching zero as the temperature approaches $T_c$ and $T_m$. Note that the ordinate is dimensionless, and that $\left(\Delta G_{TS-D(T)}/\Delta G_{N-D(T)}\right) + \left(\Delta G_{TS-N(T)}/\Delta G_{D-N(T)}\right) = 1$ for a two-state system.

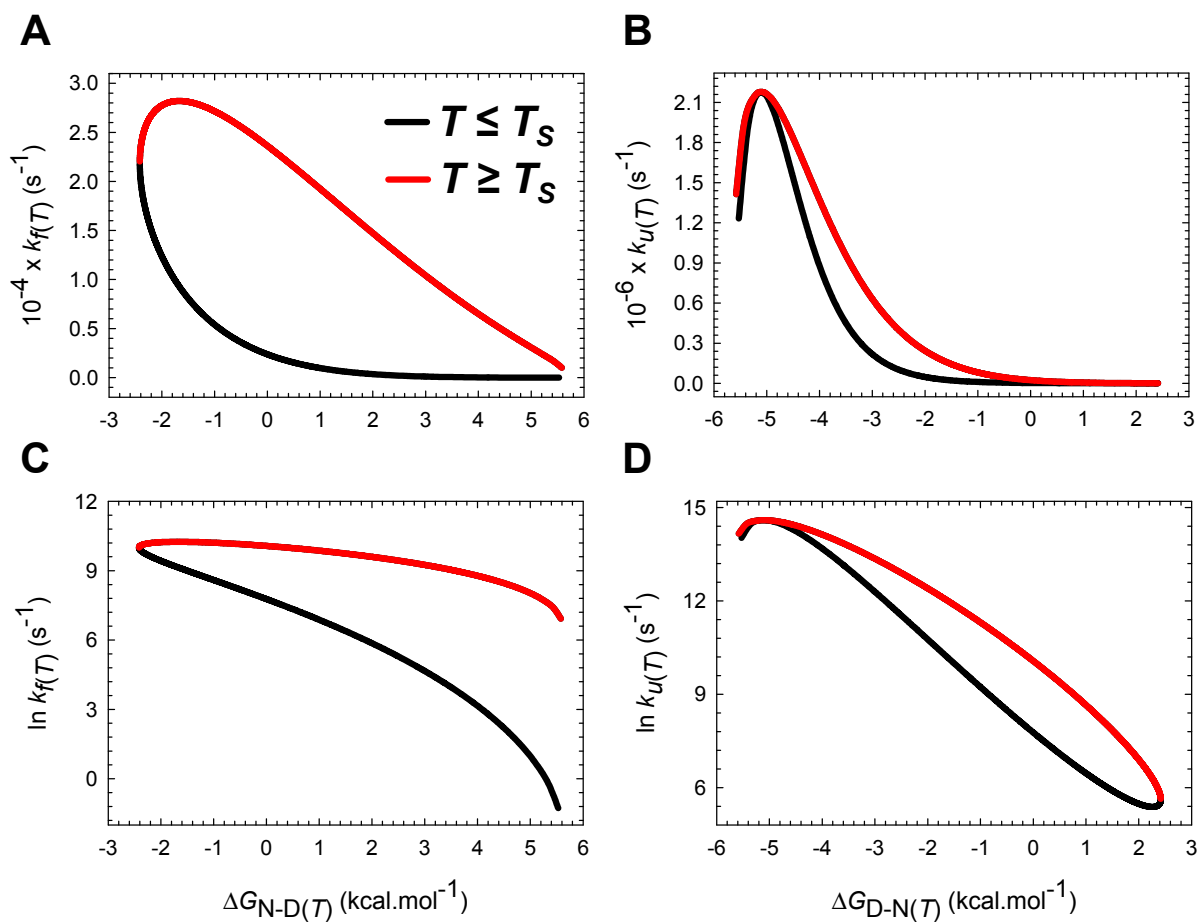

**Figure 21−figure supplement 9.**

**The complex non-linear relationship between the rate constants and the difference in Gibbs energies between the ground states.**

(A) $k_{f(T)}$ vs the Gibbs energy of folding at equilibrium. The slope of this plot equals $-k_{f(T)} \Delta H_{TS-D(T)} / \Delta S_{N-D(T)} RT^2$. (B) $k_{u(T)}$ vs the Gibbs energy of unfolding at equilibrium. The *Marcus-inverted-regimes* which occur at very low and high temperatures are towards the extreme left. The slope of this plot is given by $-k_{u(T)} \Delta H_{TS-N(T)} / \Delta S_{D-N(T)} RT^2$. (C) Natural logarithm of $k_{f(T)}$ vs the Gibbs energy of folding at equilibrium (slope = $-\Delta H_{TS-D(T)} / \Delta S_{N-D(T)} RT^2$). (D) Natural logarithm of $k_{u(T)}$ vs the Gibbs energy of unfolding at equilibrium (slope = $-\Delta H_{TS-N(T)} / \Delta S_{D-N(T)} RT^2$). The abscissae for plots A and C, and plots B and D are identical. The colour-code is identical for all the four plots.

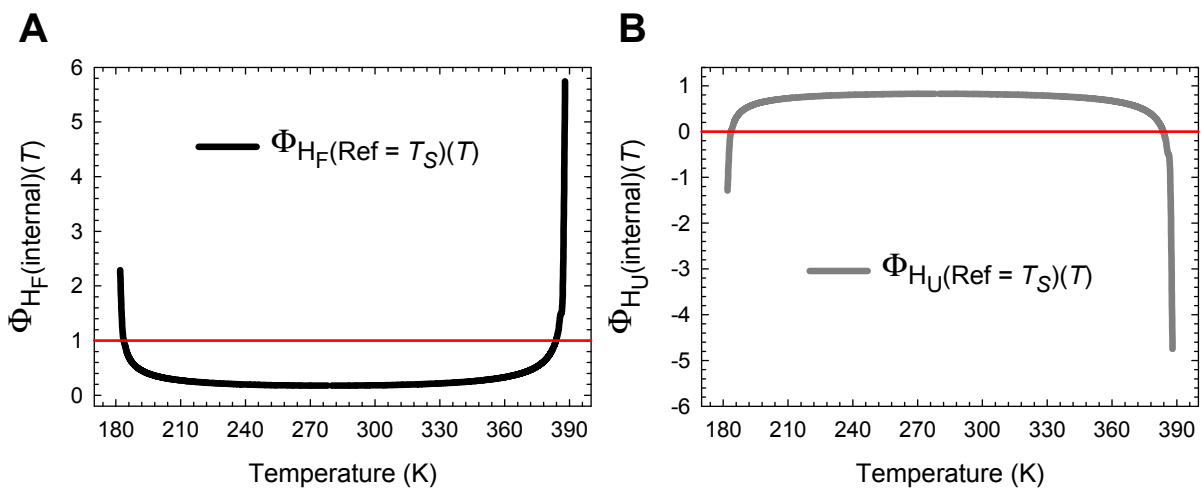

**Figure 22.**

**Temperature-dependence of $\Phi_{H(internal)(T)}$ when the protein at $T_S$ is defined as the wild type.**

**(A)** Temperature-dependence of $\Phi_{H_F(internal)(T)}$. **(B)** Temperature-dependence of $\Phi_{H_U(internal)(T)}$. Note that both these curves are undefined at $T_S$. Although the algebraic sum of $\Phi_{H_F(internal)(T)}$ and $\Phi_{H_U(internal)(T)}$ is unity for all temperatures, they need not necessarily be are not restricted to a canonical range of $0 \leq \Phi \leq 1$. The parameters $\Phi_{H_F(internal)(T)}$ and $\Phi_{H_U(internal)(T)}$ are the "*enthalpic analogues*" of $\Phi_{F(internal)(T)}$ and $\Phi_{U(internal)(T)}$, respectively (the subscript "H" indicates we are using enthalpy instead of Gibbs energy). Consequently, this figure is the enthalpic equivalent of **Figure 21**.

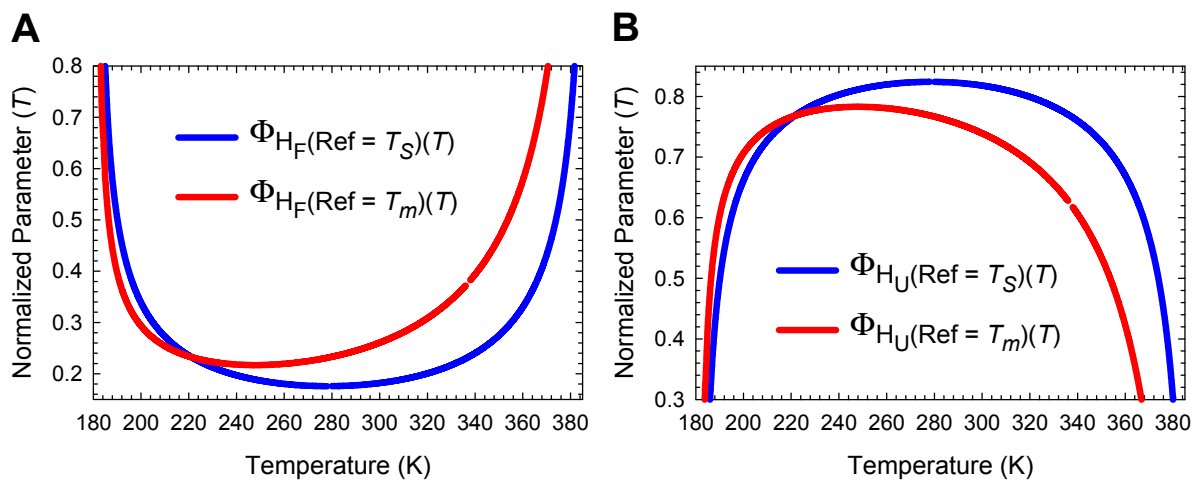

**Figure 22−figure supplement 1.**

**The magnitude of $\Phi_{H(internal)(T)}$ is dependent on the definition of the wild type.**

**(A)** A comparison of $\Phi_{H_F(internal)(T)}$ calculated using proteins at $T_S$ and $T_m$ as the wild types. **(B)** A comparison of $\Phi_{H_U(internal)(T)}$ calculated using proteins at $T_S$ and $T_m$ as the wild types. For the blue curves we have $\Delta H_{(wt)} \equiv \Delta H_{(T_S)}$, and for the red curves, we have $\Delta H_{(wt)} \equiv \Delta H_{(T_m)}$. This notation applies to both equilibrium and activation enthalpies. The blue curves are undefined (0/0) at $T_S$, and the red curves are undefined at $T_c$ and $T_m$. This figure is the enthalpic equivalent of **Figure 21−figure supplement 1**.

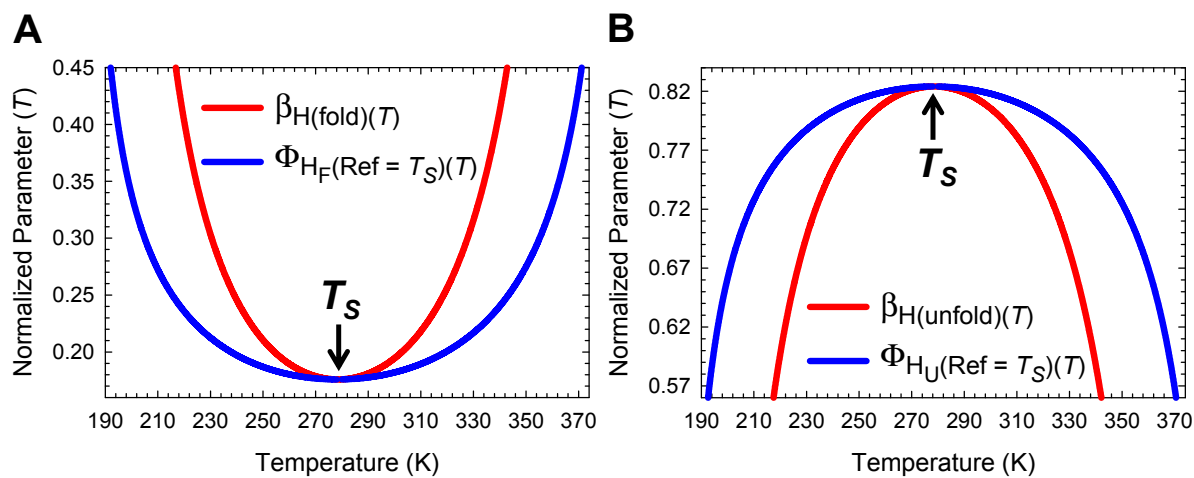

**Figure 22−figure supplement 2.**

**Comparison of the Leffler $\beta_{H(T)}$ and $\Phi_{H(internal)(T)}$ when the protein at $T_S$ is defined as the wild type.**

**(A)** $\beta_{H(fold)(T)}$ is almost identical to $\Phi_{H_F(internal)(T)}$ around the temperature of maximum stability, but as the temperature deviates from $T_S$, $\beta_{H(fold)(T)}$ increases far more steeply than $\Phi_{H_F(internal)(T)}$, such that for $T \neq T_S$ we have $\beta_{H(fold)(T)} > \Phi_{H_F(internal)(T)}$. **(B)** $\beta_{H(unfold)(T)}$ is almost identical to $\Phi_{H_U(internal)(T)}$ around the temperature of maximum stability, but as the temperature deviates from $T_S$, $\beta_{H(unfold)(T)}$ decreases far more steeply than $\Phi_{H_U(internal)(T)}$, such that for $T \neq T_S$ we have $\beta_{H(unfold)(T)} < \Phi_{H_U(internal)(T)}$. Note that the parameters $\Phi_{H_F(internal)(T)}$ and $\Phi_{H_U(internal)(T)}$ are the Fersht-analogues of the Leffler $\beta_{H(fold)(T)}$ and $\beta_{H(unfold)(T)}$, respectively (see heat capacity RC). Consequently, this figure is analogous to a comparison of Leffler $\beta_{G(fold)(T)}$ and Fersht $\Phi_{F(internal)(T)}$, and Leffler $\beta_{G(unfold)(T)}$ and Fersht $\Phi_{U(internal)(T)}$ (see **Figure 21−figure supplement 3**).

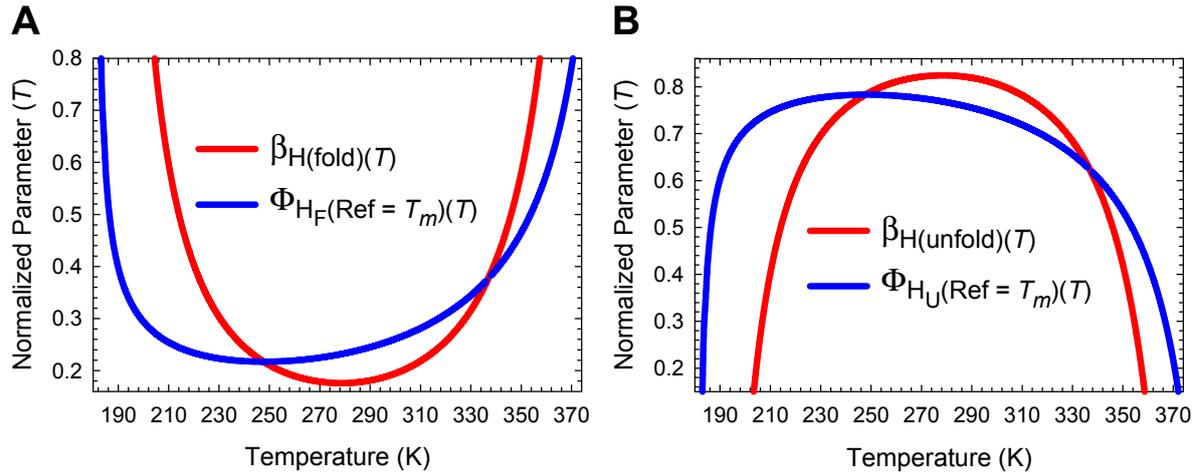

**Figure 22−figure supplement 3.**

**Comparison of the Leffler $\beta_{H(T)}$ and $\Phi_{H(internal)(T)}$ when the protein at $T_m$ is defined as the wild type.**

Changing the definition of the wild type from $T_S$ (see previous figure) to $T_m$ has a dramatic effect on the relationship between the Leffler-$\beta_{H(T)}$ and the Fersht-$\Phi_{H(internal)(T)}$. **(A)** $\beta_{H(fold)(T)} < \Phi_{H_F(internal)(T)}$ for ~248 K $< T <T_m$, $\beta_{H(fold)(T)} > \Phi_{H_F(internal)(T)}$ for $T <$ ~248 K and $T > T_m$, and identical when $T =$ ~248 K. **(B)** $\beta_{H(unfold)(T)} > \Phi_{H_U(internal)(T)}$ for ~248 K $< T < T_m$, $\beta_{H(unfold)(T)} < \Phi_{H_U(internal)(T)}$ for $T <$ ~248 K and $T > T_m$, and identical when $T =$ ~248 K. Note that $\Phi_{H_F(internal)(T)}$ and $\Phi_{H_U(internal)(T)}$ are undefined at $T_m$ and the discontinuity in the functions is apparent upon close inspection. This figure is the enthalpic equivalent of **Figure 21−figure supplement 4**.

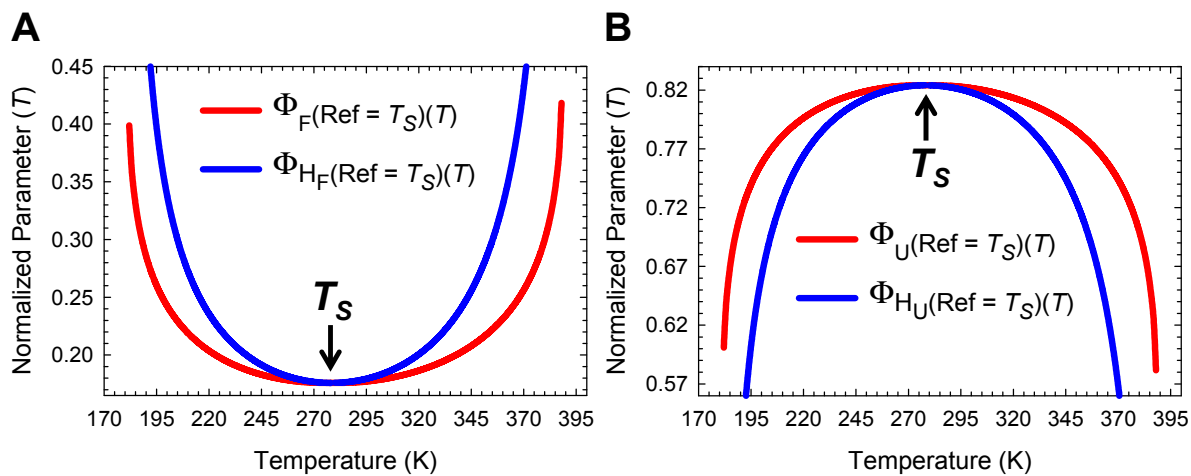

**Figure 22−figure supplement 4.**

**Comparison of $\Phi_{(internal)(T)}$ and $\Phi_{H(internal)(T)}$ when the protein at $T_S$ is defined as the wild type.**

**(A)** The normalized Gibbs parameter $\Phi_{F(internal)(T)}$ is almost identical to the normalized enthalpic parameter $\Phi_{H_F(internal)(T)}$ around the temperature of maximum stability, but as the temperature deviates from $T_S$, $\Phi_{H_F(internal)(T)}$ increases far more steeply than $\Phi_{F(internal)(T)}$, such that for $T \neq T_S$ we have $\Phi_{H_F(internal)(T)} > \Phi_{F(internal)(T)}$. **(B)** The normalized Gibbs parameter $\Phi_{U(internal)(T)}$ is almost identical to the normalized enthalpic parameter $\Phi_{H_U(internal)(T)}$ around the temperature of maximum stability, but as the temperature deviates from $T_S$, $\Phi_{H_U(internal)(T)}$ decreases far more steeply than $\Phi_{U(internal)(T)}$, such that for $T \neq T_S$ we have $\Phi_{H_F(internal)(T)} < \Phi_{U(internal)(T)}$. Since $\Phi_{(internal)(T)}$ and $\Phi_{H(internal)(T)}$ are the Fersht-analogues of Leffler $\beta_{G(T)}$ and $\beta_{H(T)}$, respectively, this figure is analogous to comparing the temperature-dependent position of the TSE along the entropic and heat capacity RCs (i.e., a comparison of the temperature-dependence of $\beta_{G(T)}$ and $\beta_{H(T)}$; see **Figure 17**).

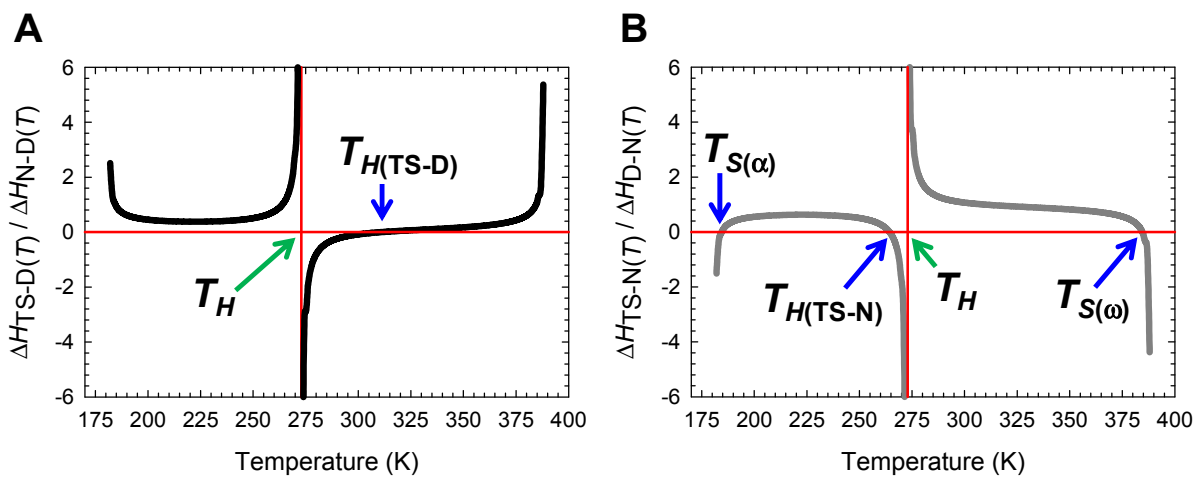

**Figure 22−figure supplement 5.**

**Temperature-dependence of the ratio of the activation enthalpies to equilibrium enthalpies.**

**(A)** The ratio $\Delta H_{\text{TS-D}(T)}/\Delta H_{\text{N-D}(T)}$ is positive for $T_\alpha \leq T < T_H$ and $T_{H(\text{TS-D})} < T \leq T_\omega$, negative for $T_H < T < T_{H(\text{TS-D})}$ and zero at $T_{H(\text{TS-D})}$. **(B)** The ratio $\Delta H_{\text{TS-N}(T)}/\Delta H_{\text{D-N}(T)}$ is positive for $T_{S(\alpha)} < T < T_{H(\text{TS-N})}$ and $T_H < T < T_{S(\omega)}$, negative for $T_\alpha \leq T < T_{S(\alpha)}$, $T_{H(\text{TS-N})} < T < T_H$, and $T_{S(\omega)} < T \leq T_\omega$, and zero at $T_{S(\alpha)}$, $T_{H(\text{TS-N})}$, and $T_{S(\omega)}$. The vertical asymptotes are a consequence of $\Delta H_{\text{D-N}(T)} = -\Delta H_{\text{N-D}(T)}$ approaching zero as $T \to T_H$. Note that the ordinate is dimensionless, and that $\left(\Delta H_{\text{TS-D}(T)}/\Delta H_{\text{N-D}(T)}\right) + \left(\Delta H_{\text{TS-N}(T)}/\Delta H_{\text{D-N}(T)}\right) = 1$ for a two-state system.

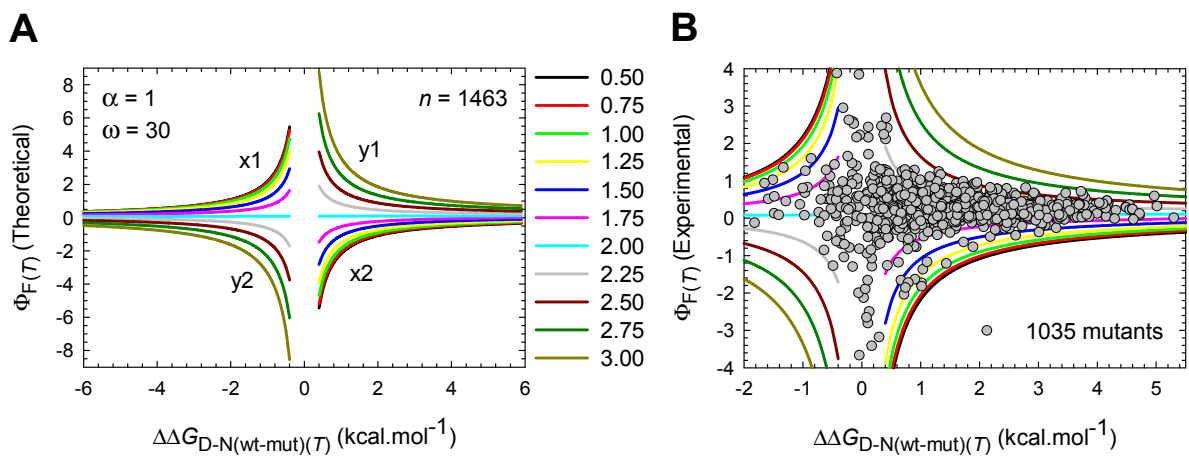

**Figure 23.**

**Comparison of the theoretical and experimental $\Phi_{F(T)}$ values (structural perturbation).**

(A) Theoretical limits of $\Phi_{F(T)}$ values according to parabolic approximation where all the 1463 theoretical mutants have the following common parameters: $\alpha$ = 1 M$^2$.mol.kcal$^{-1}$ (DSE-parabola); $\omega$ = 30 M$^2$.mol.kcal$^{-1}$ (NSE-parabola). The wild type was arbitrarily chosen to be the one with parameters $\Delta G_{D-N(T)}$ = 6 kcal.mol$^{-1}$ and $m_{D-N}$ = 2 kcal.mol$^{-1}$.M$^{-1}$. The legend indicates the variation in $m_{D-N}$ values in kcal.mol$^{-1}$.M$^{-1}$. The quadrants labelled $x1$ and $x2$ are for mutants whose $m_{D-N} < m_{D-N(wt)}$ (i.e., a contraction of the RC) and the quadrants labelled $y1$ and $y2$ are for those mutants whose $m_{D-N} > m_{D-N(wt)}$ (i.e., an expansion of the RC). Close inspection shows that for those mutants whose stabilities have changed but not their $m_{D-N}$ values, the $\Phi_{F(T)}$ values are positive but very close to zero (shown in cyan). Theoretical $\Phi_F$ values corresponding to $\Delta\Delta G_{D-N(wt-mut)(T)}$ = 0.0 ± 0.4 kcal.mol$^{-1}$ have been excluded for clarity. This corresponds to about 6.7% error on the wild type $\Delta G_{D-N(wt)(T)}$. (B) An overlay of theoretical $\Phi_{F(T)}$ and experimental $\Phi_{F(T)}$ values in water for 1035 mutants from 31 two-state systems. Data used to calculate the experimental $\Phi_{F(T)}$ values were taken from published literature (detailed information is given elsewhere). The vertical asymptotes are a consequence of $\Delta\Delta G_{D-N(wt-mut)(T)}$ approaching zero.

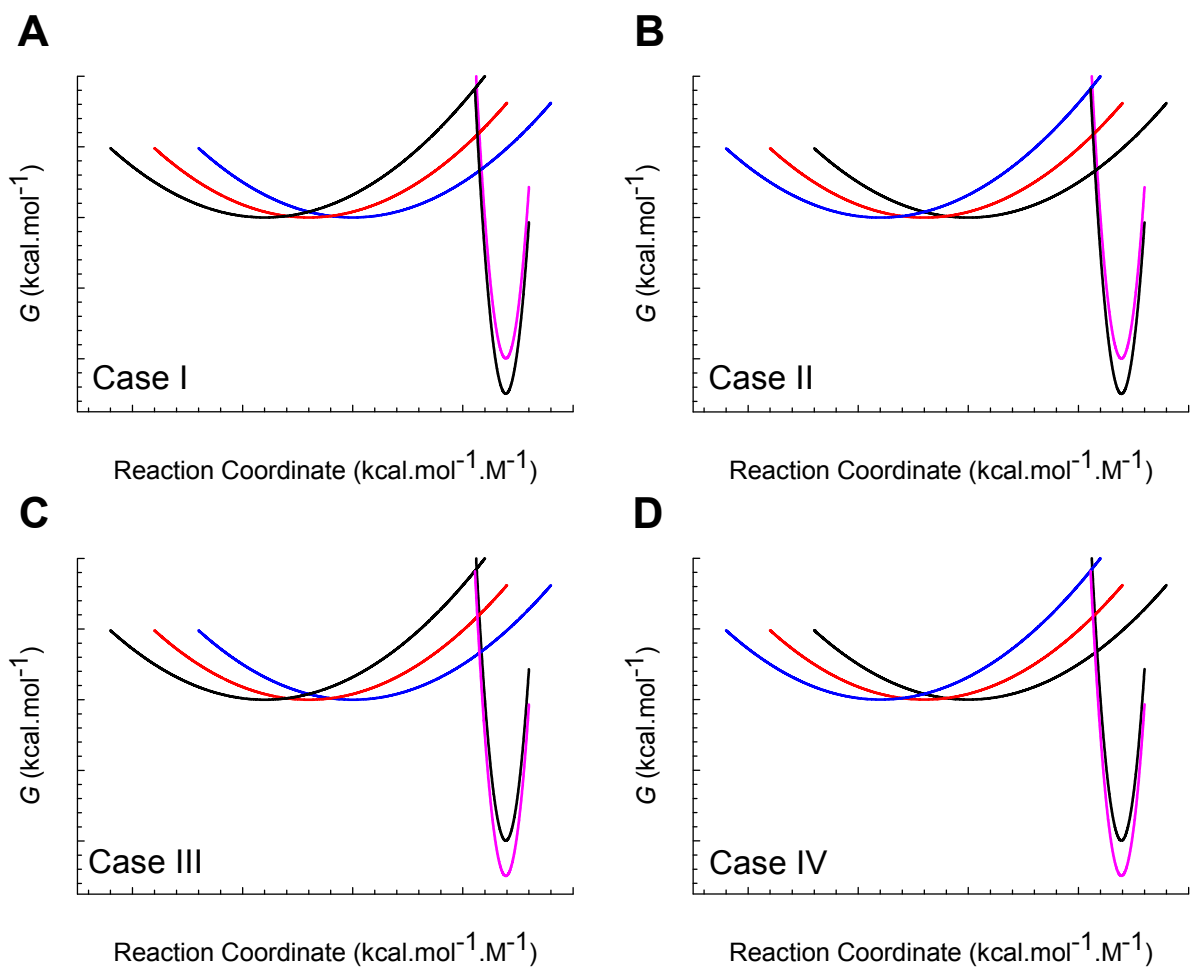

**Figure 23−figure supplement 1.**

**Parabolic Gibbs energy curves to illustrate the effect of concomitant changes in $\Delta G_{D-N(T)}$ and $m_{D-N}$ on the position of the TSE along the abscissa and ordinate.**

The parabolas corresponding to the DSE and NSE of the wild type are shown in black in all the four plots. The mutant DSE-parabolas are shown in blue and red while the mutant NSE-parabolas are shown in magenta. **(A)** The introduced mutation causes a concomitant decrease in $\Delta G_{D-N(T)}$ and the mean-length of the RC. **(B)** The introduced mutation causes a decrease in $\Delta G_{D-N(T)}$ but an increase in the mean-length of the RC. **(C)** The introduced mutation stabilizes the mutant but causes a concomitant decrease in the mean-length of the RC. **(D)** The introduced mutation stabilizes the protein but concomitantly causes an increase in the mean-length of the RC. The curvatures of all the DSE-parabolas ($\alpha = 1$ $M^2.mol.kcal^{-1}$) and all the NSE-parabolas ($\omega = 30$ $M^2.mol.kcal^{-1}$) are identical.

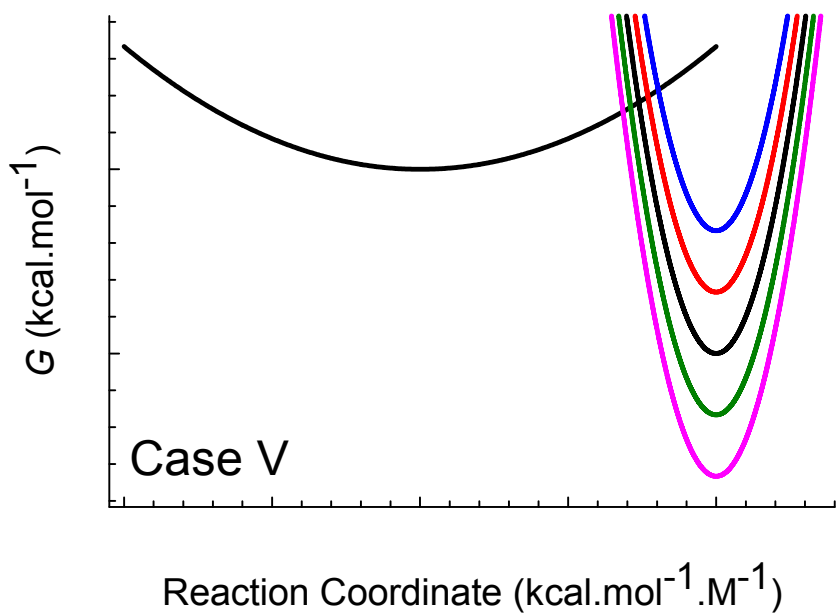

Reaction Coordinate (kcal.mol$^{-1}$.M$^{-1}$)

**Figure 23−figure supplement 2.**

**Parabolic Gibbs energy curves to illustrate the effect of a change in $\Delta G_{D-N(T)}$ on the position of the TSE along the abscissa and ordinate.**

The wild type DSE and NSE-parabolas are shown in black, the destabilized mutants are shown in red and blue, and the stabilized mutants are shown in green and magenta. As the protein is increasingly destabilized the *curve-crossing* along the RC shifts closer to the vertex of the NSE-parabola, and can be due to a stabilized DSE or a destabilized NSE, or both. Conversely, as the protein is increasingly stabilized, the *curve-crossing* along the RC shifts away from the vertex of NSE-parabola and this can be due to a destabilized DSE or a stabilized NSE, or both. The force constant for the DSE-parabola is $\alpha = 1$ M$^2$.mol.kcal$^{-1}$. The curvatures of all the NSE-parabolas are identical ($\omega = 30$ M$^2$.mol.kcal$^{-1}$).